\documentclass[thesis,letterpaper,12pt]{utthesis} 
\title{Explicit Asymptotic Solutions of $\nu_e + e^-$ Neutrino Networks for Large Sets of Partial Differential Equations in Core-Collapse Supernovae}	       
\author{Raghav Chari}                
\copyrightYear{2024}            
\graduationMonth{April}         
\majorProfessor{Dr. Mike Guidry}	    
\viceProvost{Christine Nattrass} 
\major{Physics}					
\degree{Bachelor of Science in Physics}	    			
\college{Arts and Sciences}               
\dept{Physics and Astronomy}				
\university{The University  of Tennessee, Knoxville}	
\numberOfCommitteeMembers{3} 
\committeeMemberA {Dr. Mike Guidry}	
\committeeMemberB {Dr. Sean Lindsay}	
\committeeMemberC {Dr. Vassilios Mewes}	
\usepackage{nomencl}    
\usepackage{pgfplots}
\usepackage{float}                      
\usepackage{natbib}                     
\usepackage{booktabs} 
\usepackage{longtable} 
\usepackage{graphicx}					
\graphicspath{ {figures/}{figures/eps/}{figures/pdf/} }%
\usepackage{epstopdf}
\usepackage{fancyhdr}                   
\usepackage{url}                        
\usepackage[inactive]{srcltx}		 	
\usepackage{relsize}                    
\usepackage{booktabs}                   
\usepackage[config, labelfont={bf}]{caption,subfig} 
\usepackage{mathrsfs}
\usepackage[linesnumbered,ruled,vlined]{algorithm2e}
\usepackage{bm}
\begin{document}
    \pagenumbering{roman}
    \setcounter{page}{1}
    %
    %
    
    \addToPDFBookmarks{0}{Front Matter}{rootNode} 
    \addToPDFBookmarks{1}{Title}{i} 
    \makeTitlePage 
    %
    %
    \makeCopyrightPage 
    \addToPDFBookmarks{1}{Dedication}{b} 
    \chapter*{}
\begin{center}
\it Dedication ...

To my friends and family, for their unwavering support, love, and the belief that I could pursue my passion. Special thanks to the endless support my Mother, Father and Savannah Bedell. 
Most of all to my older brother, Rohith Chari, for his guidance, support, and
belief in me.

\end{center} 
    %
    \addToPDFBookmarks{1}{Acknowledgements}{c} 
    \chapter*{Acknowledgements}

I would like to thank the University of Tennessee, Knoxville's Department of Physics \& Astronomy for their support and to all the outstanding professors and mentors within the Department. I am particularly thankful to Dr. Sean Lindsay, whose immense support and guidance has been paramount to my education. I would also like to thank Dr. Eirik Endeve and Dr. Vassilios Mewes from Oak Ridge National Laboratories for their support and guidance throughout my research. I am also deeply grateful to Ph.D. students Adam Cole and Nick Brey for their support and mentorship over the three years of this project. Their contributions have been instrumental to my growth and success. Additionally, I wish to acknowledge Olivia Clark, whose assistance in data analysis has played a critical role in the completion of this project. Most of all I would like to thank Dr. Mike Guidry for the opportunity to work on this project and for his support and guidance as my advisor. This work was also made possible through generous funding from several sources, which have been instrumental to the advancement of this research:


\footnotesize{
\begin{itemize}
    \item[] Chari R., Guidry M. (2024). \textit{Enhancing Astrophysical Modeling: Integrating WEAKLIB with Fast Explicit Neutrino Networks for Advanced Large Scale Neutrino Electron Scattering}, Advanced Undergraduate Research Activity (AURA).
    \item[] Chari R., Guidry M. (2021). \textit{New Approaches to Astrophysical Nucleosynthesis and Neutrino Transport}, Fellowship, Department of Physics and Astronomy, University of Tennessee, Knoxville.
    \item[] Chari R. (2022, 2023). Undergraduate Research \& Fellowships Travel Grants, University of Tennessee, Knoxville.
\end{itemize}
}

    %
    \addToPDFBookmarks{1}{Quote}{d} 
    29 mtime=1698379370.79967556
72 LIBARCHIVE.xattr.com.apple.quarantine=MDA4Mzs2NTNhZDA3OTtPdXRsb29rOw
60 SCHILY.xattr.com.apple.quarantine=0083;653ad079;Outlook;
    %
    \addToPDFBookmarks{1}{Abstract}{e} 
    \chapter*{Abstract}
\label{ch:abstract}

In physics, accurately modeling large-scale phenomena such as core-collapse supernovae, (CCSN), and neutron star mergers
are computationally challenging and require solving large sets of partial and ordinary differential equations. Traditional methods used widely in the scientific community are predominantly implicit, which are approximations that often require drastic simplifications and can be computationally inefficient. This thesis presents results on a new software suite titled ``Fast Explicit Neutrino Networks'' or ``FENN'', that introduces a suite of algebraically stabilized
explicit methods known as explicit asymptotic for modeling Neutrino Electron Scattering, (NES), presenting a novel approach that combines the stability of
traditional methods with enhanced computational efficiency. Initial results show that FENN can deliver accurate solutions for neutrino networks at improved computational speeds. This thesis further covers new results for scaled networks beyond the constraints of standard energy groupings, as well as the dynamics of neutrino interactions such as the scattering of various neutrino flavors---electron neutrinos ($\nu_e$), electron anti-neutrinos ($\bar{\nu}_e$), and muon/tau neutrinos ($\nu_{\mu,\tau}$) as well as their anti-particles ($\bar{\nu}_{\mu,\tau}$)---off electrons.

    %
    \addToPDFBookmarks{0}{Table of Contents}{f}
    \tableofcontents 
    \addToTOC{List of Tables} 
    \listoftables 

    \addToTOC{List of Figures} 
    \listoffigures 
    \makenomenclature 
    \addToPDFBookmarks{0}{Nomenclature}{g} 
    \printnomenclature[1.25in] 
    \newpage
    \pagenumbering{arabic}
    \setcounter{page}{1}
    72 LIBARCHIVE.xattr.com.apple.quarantine=MDA4Mzs2NTNhZDA3OTtPdXRsb29rOw
60 SCHILY.xattr.com.apple.quarantine=0083;653ad079;Outlook;
    \chapter{Introduction} \label{ch:introduction}

Realistic simulations of core-collapse supernovae and stellar explosions involve solving sets of non-linear partial and ordinary differential equations related to hydrodynamics, radiation transport, and thermonuclear reactions. This gives stiff systems that are computationally challenging and requires careful choice of numerical methods to make them efficient. \citep{oran05, gear71, lam91, press92}.

Explicit numerical methods, therefore, are usually simple as they calculate based on the present states without information about future states. Alternatively, implicit methods are typically based on predictions of the states to be made in future iterations and require complex matrix inversions that are computationally expensive.

When it comes to realistic astrophysical systems, explicit methods such as forward Euler struggle with stiff systems of differential equations. This is due to the stabnility constraints of the method that impose very small timesteps, which are impractical in large-scale simulations.

In a series of papers, the explicit asymptotic method was applied to thermonuclear networks \citep{guidJCP,guidAsy,guidPE,guidQSS,guid2023a,guid2016,haid2015,haid2016,brey2022,chup2008, fege2011}. 

Specifically, in \citep{guidJCP} we can consider the following equations:

\begin{align}
 \frac{d y_i}{d t} &= F_i^{+}-F_i^{-} \nonumber \\
&= \left(f_1^{+}+f_2^{+}+\ldots\right)_i - \left(f_1^{-}+f_2^{-}\right)_i \nonumber \\
&= \left(f_1^{+}-f_1^{-}\right)_i + \left(f_2^{+}-f_2^{-}\right)_i + \ldots = \sum_j\left(f_j^{+}-f_j^{-}\right)_i
\label{ThermoEquation}
\end{align}
where $y_i(i=1 \ldots N)$ are species, and $t$ is time. We can denote the fluxes between species $i$ and $j$ by $\left(f_j^{\pm}\right)_i$, and the sum for each variable $i$ for all variables $j$ coupled to $i$ by a non-zero flux $\left(f_j^{\pm}\right)_i$.

Therefore, for an $N$-species network there will be $N$ equations in the populations $y_i$.
Here \citep{guidJCP} applies an explicit approximation known as the explicit asymptotic to method to Eq. \ref{ThermoEquation}. 
Given the differential equations in the form
\begin{equation}
F_i^{-}=(k_1^i+k_2^i+\ldots+k_m^i) y_i \equiv k^i y_i,
\end{equation}
where $k_j^j$ are rate parameters (with the subscript $m$ indicating the total number of these processes) for processes depleting $y_i$. $\tau_j^i=1 / k_j^i$ are the characteristic timescales defined for this set of equations, the effective total depletion rate and corresponding timescale are
\begin{equation}
k^i \equiv \frac{F_i^{-}}{y_i}, \quad \tau^i=\frac{1}{k^i}.
\end{equation}
Thus, at timestep $t_n$ the finite-difference approximation is
\begin{equation}
y_i(t_n)=\frac{F_i^{+}(t_n)}{k^i(t_n)}-\left.\frac{1}{k^i(t_n)} \frac{d y_i}{d t}\right|_{t=t_n}.
\end{equation}
The asymptotic limit where $F_i^{+} \simeq F_i^{-}$ gives the first approximation $y_i^{(1)}(t_n)$ and local error $E_n^{(1)}$,
\begin{equation}
y_i^{(1)}(t_n)=\frac{F_i^{+}(t_n)}{k^1(t_n)}, \quad E_n^{(1)} \equiv y(t_n)-y^{(1)}(t_n)=-\frac{1}{k(t_n)} \frac{d y}{d t}(t_n).
\end{equation}
For small $d y_i / d t$, the correction term is
\begin{equation}
\frac{d y}{d t}(t_n)=\frac{1}{\Delta t}(y_i(t_n)-y_i(t_{n-1}))+\frac{1}{\Delta t}(E_n^{(1)}-E_{n-1}^{(1)})+O(\Delta t).
\end{equation}
Here, \(O(\Delta t)\) represents the higher-order terms in the expansion, which are proportional to \(\Delta t\). These terms are significant when \(\Delta t\) is large, but as \(\Delta t\) approaches zero, their contribution diminishes.
This leads to:
\begin{equation}
y_n^{(2)}=\frac{1}{1+k_n \Delta t}\left(y_{n-1}+F_n^{+} \Delta t\right),
\end{equation}
for large $k \Delta t$. Implementing an asymptotic algorithm requires defining a critical value $\kappa$ of $k \Delta t$:
\begin{enumerate}
    \item If $k^i \Delta t<\kappa$, update the population numerically by the explicit Euler method.
    \item If $k \Delta t \geq \kappa$, update the population algebraically using the provided approximation.
\end{enumerate}
The stability of the integration is contingent upon the condition $k^i \Delta t<1$, with $\kappa=1$ chosen for demonstration.

However, in this thesis, we will apply them to scaled neutrino networks (larger than standard network sizes) for, specifically, Neutrino Electron Scattering  ($\nu_e + e^-$). Energy from neutrinos that are released during a core-collapse supernova is important for a large portion of the supernova. Further, it is believed that the neutrinos released during the core collapse contribute to the shock wave that propagates through the star. While most codes focus on the full neutrino matter interaction such as electron-positron pair creation and annihilation, emission and absorption on nucleons and nuclei, and scattering on nucleons and nuclei, the Neutrino Electron Scattering portion is a large enough portion of this that the results from these methods can be applied to the more general formulation of the neutrino transport problem.

In general, the entirety of the neutrino transport problem is difficult to simulate because of the weak interaction further described in section \ref{ch:Mathematical Formulation}. Kinetic models are required that use the Boltzmann equation that depends on position and momentum coordinates in phase space. Due to the nature of this, implicit methods such as backward Euler are traditionally used.

In \citep{PhysRevD.109.103019}, we applied the explicit asymptotic algorithms \citep{guidJCP} by introducing a novel suite of computational algorithms known as  Fast Explicit Neutrino Networks (FENN) that were shown to balance speed and accuracy for the explicit asymptotic methods at a fixed energy grid and bin size. Following this, in a subsequent study introduced for neutrino networks in \citep{2024AAS...24313505C}, we expand on the foundational work by emphasizing the importance of scalability.

As of the writing of this thesis, scalability has not been applied to this problem in the same way \citep{guidJCP} has applied them in the context of reaction networks. In this thesis, we will compare our results to validate the fixed network sizes in \citep{PhysRevD.109.103019}, as well as extend them to network sizes up to 180 species.

    \chapter{Mathematical Formulation} \label{ch:Mathematical Formulation}
\section{Neutrino Electron Scattering}

\citep{bru85} provides an extensive mathematical framework for the transport of neutrinos inside astrophysical core-collapse supernovae. The formulation in FENN will focus on Neutrino Electron Scattering (NES), which we now summarize.

\subsection{Formulation of Neutrino-Electron Scattering}

Let's begin with the basics,

\begin{equation}
    \lambda^{(t)}(\omega) \rightarrow \Lambda^{(t)}(\omega) = \frac{3 \lambda^{(t)}(\omega)}{3+\lambda^{(t)}(\omega) \left| \nabla \psi^{(0)}(\omega) \right| / \psi^{(0)}(\omega)},
\end{equation}
Here $\lambda^{(t)}(\omega)$ is the modification of the mean free path, where the focus is on the right-hand side of the equation, which describes neutrino-electron scattering and thermal production terms. This can be simplified by expanding the kernels of these terms into a Legendre series and truncating after the first two terms

\begin{equation}
    R_{\mathrm{NES}}^{\mathrm{in} / \text{out}}\left(\omega, \omega^{\prime}, \cos \theta \right) = \frac{1}{2} \sum_l (2 l + 1) \Phi_{l, \mathrm{NES}}^{\mathrm{in} / \text{out}}\left(\omega, \omega^{\prime} \right) P_l(\cos \theta),
\end{equation}
Using this expansion, the quantities $A_{\mathrm{NES}}^{(\alpha)}(\omega)$, $B_{\mathrm{NES}}^{(\alpha)}(\omega)$, and $C_{\mathrm{NES}}^{(\alpha)}(\omega)$ are expressed as:

\begin{align}
    A_{\mathrm{NES}}^{(0)}(\omega) & = -\frac{2 \pi}{c(2 \pi \hbar c)^3} \int_0^{\infty} \omega^{\prime 2} d \omega^{\prime} \left\{ \Phi_{0, \mathrm{NES}}^{\mathrm{in}}\left(\omega, \omega^{\prime}\right) \psi^{(0)}\left(\omega^{\prime}\right) + \Phi_{0, \mathrm{NES}}^{\text{out}}\left(\omega, \omega^{\prime}\right) \left[1 - \psi^{(0)}\left(\omega^{\prime}\right)\right] \right\}, \\
    B_{\mathrm{NES}}^{(0)}(\omega) & = -\frac{2 \pi}{3 c(2 \pi \hbar c)^3} \int_0^{\infty} \omega^{\prime 2} d \omega^{\prime} \left[ \Phi_{1, \mathrm{NES}}^{\mathrm{in}}\left(\omega, \omega^{\prime}\right) - \Phi_{1, \mathrm{NES}}^{\text{out}}\left(\omega, \omega^{\prime}\right) \right] \psi^{(1)}\left(\omega^{\prime}\right), \\
    C_{\mathrm{NES}}^{(0)}(\omega) & = \frac{2 \pi}{c(2 \pi \hbar c)^3} \int_0^{\infty} \omega^{\prime 2} d \omega^{\prime} \Phi_{0, \mathrm{NES}}^{\mathrm{in}}\left(\omega, \omega^{\prime}\right) \psi^{(0)}\left(\omega^{\prime}\right).
\end{align}
For the isoenergetic scattering kernel, a similar expansion yields
\begin{equation}
    R_{\mathrm{IS}}^0(\omega, \omega, \cos \theta) \approx \frac{1}{2} \Phi_{0, \mathrm{IS}}(\omega) + \frac{3}{2} \Phi_{1, \mathrm{IS}}(\omega) \cos \theta.
\end{equation}
Finally, the $B_{\mathrm{IS}}^{(1)}(\omega)$ is given by:
\begin{equation}
    B_{\mathrm{IS}}^{(1)}(\omega) = \frac{2 \pi}{c(2 \pi \hbar c)^3} \omega^2 \left[ \Phi_{1, \mathrm{IS}}(\omega) - \Phi_{0, \mathrm{IS}}(\omega) \right].
\end{equation}
NES involves the exchange of charged ($M_w$) and neutral ($M_z$) bosons, leading to distinct matrix elements for electron-type neutrinos formulated later in this chapter. 

\subsection{Weak Interaction}
The scattering process involving the exchange of charged ($M_w$) and neutral ($M_z$) bosons is described by

\begin{equation}
M_w=\frac{G}{\sqrt{2}}\left[\bar{u}_\nu(q') \gamma^\mu(1-\gamma_5) u_e(p_e)\right]\left[\bar{u}_e(p_e') \gamma_\mu(1-\gamma_5) u_\nu(q)\right],
\end{equation}
\begin{equation}
M_z=\frac{G}{\sqrt{2}}\left[\bar{u}_\nu(q') \gamma^\mu(1-\gamma_5) u_\nu(q)\right]\left[\bar{u}_e(p_e') \gamma_\mu(a-b \gamma_5) u_e(p_e)\right],
\end{equation}
 where $a=-\frac{1}{2}+2 \sin^2 \theta_{\mathrm{w}}$, $b=-\frac{1}{2}$, and $G$ is the weak interaction coupling constant. Combining these interactions, we obtain the total matrix element $M$ as:
\begin{equation}
M=\frac{G}{\sqrt{2}}\left[\bar{u}_\nu(q') \gamma^\mu(1-\gamma_5) u_\nu(q)\right]\left[\bar{u}_e(p_e') \gamma_\mu(C_V-C_A \gamma_5) u_e(p_e)\right],
\end{equation}
with $C_V=a+1$ and $C_A=b+1$. The transition rate for this process is given by

\begin{align}
r = & \frac{G^2}{\omega \omega' E_e E_e'}(2 \pi)^4 \delta^4(q+p_e-q'-p_e') \times \nonumber \\
& \left[(C_V+C_A)^2 p_e \cdot q p_e' \cdot q' + (C_V-C_A)^2 p_e' \cdot q p_e \cdot q' - M_e^2(C_V^2-C_A^2) q' \cdot q\right],
\end{align}
assuming the last term is negligible for relativistic electrons.
The in and out scattering rates, \( \Phi_{l, \mathrm{NES}}^{\{\text{in}, \text{out}\}} \), are derived from phase-space integration, yielding:
\begin{align}
\Phi_{l, \mathrm{NES}}^{\left\{\begin{array}{c}
\text{in} \\
\text{out}
\end{array}\right.} = & \frac{G^2}{\pi \omega^2 \omega'^2} \int dE_e F_e(E_e)\left[1-F_e(E_e+\omega-\omega')\right]\left\{\begin{array}{c}
\exp[-\beta(\omega-\omega')] \\
1
\end{array}\right\} \\
& \times\left[(C_V+C_A)^2 H_l^{\mathrm{I}}(\omega, \omega', E_e) + (C_V-C_A)^2 H_l^{\mathrm{II}}(\omega, \omega', E_e)\right],
\end{align}
where \( H_l^{\mathrm{I}} \) and \( H_l^{\mathrm{II}} \) are functions detailed in \citep{Yueh1976}.

The expression for \( B_{\mathrm{NES}}^{(1)}(\omega) \) is then given by:
\[
B_{\mathrm{NES}}^{(1)}(\omega)=-\frac{2 \pi}{c(2 \pi \hbar c)^3} \int_0^{\infty} \omega^2 d \omega^{\prime}\left\{\Phi_{0, \mathrm{NES}}^{\mathrm{in}}\left(\omega, \omega^{\prime}\right) \psi^{(0)}\left(\omega^{\prime}\right)+\Phi_{0, \mathrm{NES}}^{\text{out }}\left(\omega, \omega^{\prime}\right)\left[1-\psi^{(0)}\left(\omega^{\prime}\right)\right]\right\} ,
\]
where \( \omega^{\prime}, \omega, l, T, \eta \) are parameters, with \( \eta = \mu_e / k_B T \). To interpolate the NES opacity on a consistent energy grid for \( \omega \) and \( \omega^{\prime} \) within a given state \( \left(\rho_i, T_i, Y_{e i}\right) \), it is necessary to interpolate for \( \eta \) first.

The NES tables are five-dimensional tables with $\left(\omega^{\prime}, \omega, l, T, \eta\right)$, where $\eta=\mu_e / k_B T$. We then interpolate the NES opacities at some $\rho$, $T$, $Y_e$ detailed in Table \ref{ch:appendix}.

\begin{figure}[H]
\centering 
\includegraphics[width=5in]{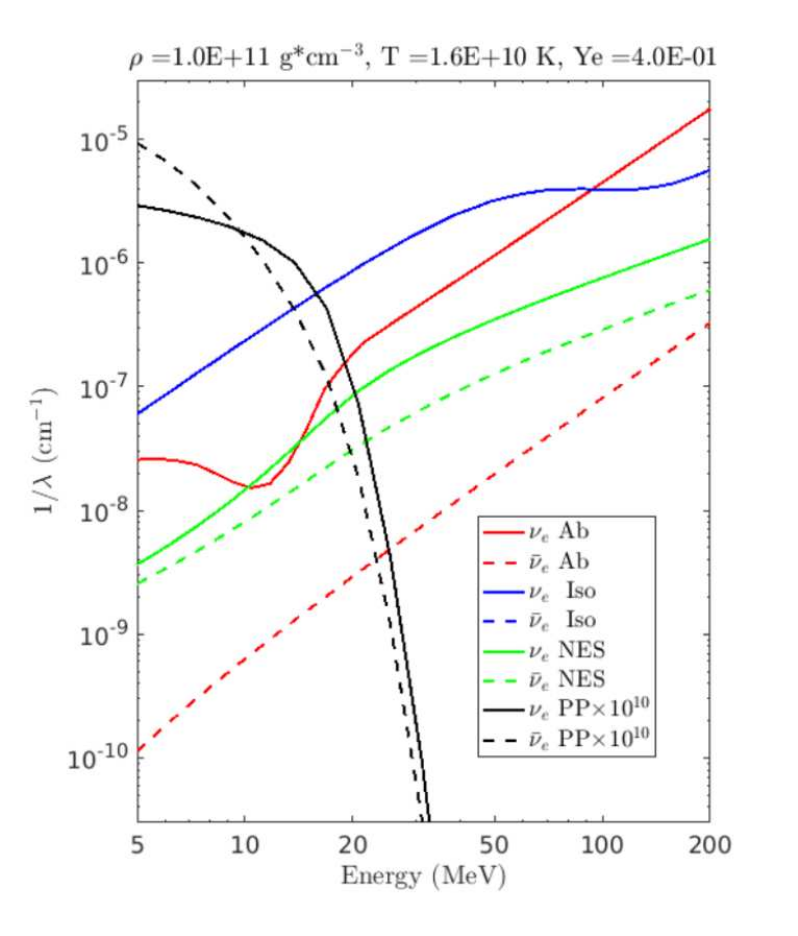}
\caption{WeakLib opacities adopted from \citep{bru85}. }\label{Opacities}
\end{figure}

\subsection{Constructing Energy Bins}

In FENN, the model describes the evolution of the spectral neutrino number distribution within discretized energy bins, each characterized by a specific volume of phase space $V$. \citep{1993ApJ...410..740M} demonstrated how realistic treatment of the NES influences the result. The underlying physics is governed by the spatially homogeneous Boltzmann equation, in this case modified to account for the inelastic neutrino-electron scattering, assuming constant background matter during the processes of interaction. This formulation can be given by,
\begin{equation}
\frac{d \mathcal{F}}{d t}= (1-\mathcal{F}) \int_{V_P} \mathcal{R}^{\text{in}}(\boldsymbol{\varepsilon}, \boldsymbol{\varepsilon}^{\prime}, \boldsymbol{n} \cdot \boldsymbol{n}^{\prime} ; \boldsymbol{u}) \mathcal{F}^{\prime} d V_{P^{\prime}}
- \mathcal{F} \int_{V_P} \mathcal{R}^{\text{out}}(\boldsymbol{\varepsilon}, \boldsymbol{\varepsilon}^{\prime}, \boldsymbol{n} \cdot \boldsymbol{n}^{\prime} ; \boldsymbol{u})(1-\mathcal{F}^{\prime}) d V_{P^{\prime}},
\end{equation}
where $\mathcal{F}$, the phase-space density, is normalized between $[0,1]$, and $\mathcal{R}^{\text{in/out}}$ denote the transition rates into or out of an energy bin, influenced by the energy before and after collision, $\varepsilon$ and $\varepsilon^{\prime}$, and the orientation of neutrino momenta through $\boldsymbol{n}$ and $\boldsymbol{n}^{\prime}$.

The scattering kernel approximation is given by \citep{smit96}
\begin{equation}
\mathscr{R}^{\text {in/out }}\left(\epsilon, \epsilon^{\prime}, \cos \alpha, \mathbf{u}\right) \approx \sum_{\ell=0}^L \Phi_{\ell}^{\text {in/out }}\left(\epsilon, \epsilon^{\prime}, \mathbf{u}\right) P_{\ell}(\cos \alpha),
\end{equation}
with the orthogonality relationship
\begin{equation}
\Phi_{\ell}^{\text {in/out }}\left(\epsilon, \epsilon^{\prime}, \mathbf{u}\right)=\frac{2 \ell+1}{2} \int_{-1}^1 \mathscr{R}^{\text {in/out }}\left(\epsilon, \epsilon^{\prime}, \cos \alpha, \mathbf{u}\right) P_{\ell}(\cos \alpha) d \cos \alpha.
\end{equation}
Integrating the phase space density over all directions yields the neutrino number density
\begin{equation}
\mathscr{N}(\epsilon, \mathbf{x}, t)=\frac{1}{4 \pi} \int_0^{2 \pi} \int_0^\pi \mathscr{F} \sin \vartheta d \vartheta d \phi.
\end{equation}
The isotropic scattering approximation leads to the following equation for the evolution of neutrino number density
\begin{equation}
\begin{aligned}
\frac{d \mathscr{N}}{d t} = & (1-\mathscr{N}) \int_{\mathbb{R}^{+}} \mathcal{R}^{\text{in}}\left(\epsilon, \epsilon^{\prime}\right) \mathscr{N}\left(\epsilon^{\prime}\right) d V_{\epsilon^{\prime}} \\
& - \mathscr{N} \int_{\mathbb{R}^{+}} \mathcal{R}^{\text{out}}\left(\epsilon, \epsilon^{\prime}\right)\left(1-\mathscr{N}\left(\epsilon^{\prime}\right)\right) d V_{\epsilon^{\prime}}.
\end{aligned}
\label{neutrinonumberdenisty}
\end{equation}
Ensuring conservation and equilibrium, we have
\begin{equation}
\mathcal{R}^{\text {in }}\left(\epsilon, \epsilon^{\prime}\right)=\mathcal{R}^{\text {out }}\left(\epsilon^{\prime}, \epsilon\right),
\end{equation}
\begin{equation}
\mathcal{R}^{in}\left(\epsilon, \epsilon^{\prime}\right)=\mathcal{R}^{\text {out }}\left(\epsilon^{\prime}, \epsilon\right) e^{\theta\left(\epsilon^{\prime}-\epsilon\right)}.
\end{equation}
And the equilibrium phase space density is
\begin{equation}
\mathscr{N}_{E q}(\epsilon, \mathbf{u})=\frac{1}{1+e^{\beta\left(\epsilon-\mu_\nu\right)}}.
\end{equation}

The discretization of the problem is formulated in \citep{lac2020}. In summary, we can discretize the energy domain into $N_b$ energy bins for computational purposes. Each bin is centered at

\begin{equation}
\epsilon_i = \frac{\epsilon_{i-1/2} + \epsilon_{i+1/2}}{2}
\end{equation}
and the volume of each energy bin is given by:
\begin{equation}
\Delta V_i^\epsilon = \int_{\epsilon_{i-1/2}}^{\epsilon_{i+1/2}} dV_\epsilon = \frac{4\pi}{3}(\epsilon_{i+1/2}^3 - \epsilon_{i-1/2}^3).
\label{EnergyBinVolume}
\end{equation}
Using the finite-volume approach to discretize the governing equation, we approximate
\begin{equation}
\frac{d \mathscr{N}}{d t} \approx (1-\mathscr{N}) \int_{D^\epsilon} \mathcal{R}^{in}(\epsilon, \epsilon') \mathscr{N}(\epsilon') dV_{\epsilon'} - \mathscr{N} \int_{D^\epsilon} \mathcal{R}^{\text{out}}(\epsilon, \epsilon')(1-\mathscr{N}(\epsilon')) dV_{\epsilon'}.
\end{equation}

For the $i^{th}$ bin with a corresponding $\mathscr{N}_i$ number density, the discretized form becomes
\begin{equation}
\begin{aligned}
\frac{d \mathscr{N}_i}{d t} = & \sum_{k=1}^{N_b} (1-\mathscr{N}_i) \int_{\epsilon_{k-1/2}}^{\epsilon_{k+1/2}} \mathcal{R}_{ik}^{in}(\epsilon, \epsilon') \mathscr{N}_k(\epsilon') dV_{\epsilon'} \\
& - \sum_{k=1}^{N_b} \mathscr{N}_i \int_{\epsilon_{k-1/2}}^{\epsilon_{k+1/2}} \mathcal{R}_{ik}^{\text{out}}(\epsilon, \epsilon')(1-\mathscr{N}_k(\epsilon')) dV_{\epsilon'}.
\end{aligned}
\end{equation}
Assuming constant scattering rates within each bin allows us to simplify further
\begin{equation}
\mathscr{N}_i(t) = \frac{1}{\Delta V_i^\epsilon} \int_{\epsilon_{i-1/2}}^{\epsilon_{i+1/2}} \mathscr{N}(\epsilon, t) dV_\epsilon,
\end{equation}
which represents the volume-averaged particle density in each energy bin. This leads us to the expression for the rate of change of neutrinos in the $i^{th}$ energy bin
\begin{equation}
\frac{d \mathscr{N}_i}{d t} = (1-\mathscr{N}_i) \sum_{k=1}^{N_b} \hat{\mathcal{R}}_{ik}^{in} \mathscr{N}_k - \mathscr{N}_i \sum_{k=1}^{N_b} \hat{\mathcal{R}}_{ik}^{\text{out}} (1-\mathscr{N}_k),
\label{RateChanges}
\end{equation}
with $\hat{\mathcal{R}}^{in/out} = \mathcal{R}^{\text{in/out}} \Delta V_k^\epsilon$.

Here we describe the rate of change of neutrino populations. In \citep{PhysRevD.109.103019} the $i^{th}$ energy bin had 40 equations so $i$ ranged from 1 to 40, but here they are determined by the network size, so for the results presented in this thesis, they take on a value between 1 and 180. 

Further simplification gives
\begin{equation}
\frac{d \mathscr{N}_i}{d t} = \sum_{k=1}^{N_b} \hat{\mathcal{R}}_{ik}^{in} \mathscr{N}_k - \mathscr{N}_i \sum_{k=1}^{N_b} \left[\hat{\mathcal{R}}_{ik}^{\text{out}} + (\hat{\mathcal{R}}_{ik}^{in} - \hat{\mathcal{R}}_{ik}^{\text{out}}) \mathscr{N}_k \right].
\end{equation}
Letting $F_i^{+} = \sum_{k=1}^{N_b} \hat{\mathcal{R}}_{ik}^{in} \mathscr{N}_k$ and $\kappa_i = \sum_{k=1}^{N_b} \hat{\mathcal{R}}_{ik}^{\text{out}}$, and defining
\begin{equation}
\tilde{\kappa}_i = \sum_{k=1}^{N_b} \left[\delta_{ik} \kappa_k + (\hat{\mathcal{R}}_{ik}^{in} - \hat{\mathcal{R}}_{ik}^{\text{out}}) \mathscr{N}_k \right],
\end{equation}
we obtain
\begin{equation}
\frac{d \mathscr{N}_i}{d t} = F_i^{+} - \tilde{\kappa}_i \mathscr{N}_i = C_i,
\end{equation}
where $\delta_{ik}$ is the Kronecker delta. This encapsulates the flux of neutrinos associated with the $i^{th}$ bin and the rate parameter for neutrino-electron collisions. An exact solution for the differential equation is provided by
\begin{equation}
\mathscr{N}_i(t) = \mathscr{N}_0 e^{-\tilde{\kappa}_i t} + \frac{F_i^{+}}{\tilde{\kappa}_i} (1 - e^{-\tilde{\kappa}_i t}),
\end{equation}
where $\mathscr{N}_0$ is the initial number density. The mean free path of neutrinos, assumed to travel at light speed, is
\begin{equation}
\ell = \frac{c}{\tilde{\kappa}_i}.
\end{equation}

\subsection{Matrix Formulation of the NES Model}

The discretized NES model culminates in a matrix formulation. This formulation is structured around the spectral number density of neutrinos, $\mathcal{N}$, within the discretized energy bins, which transforms the Boltzmann equation into a system of linear differential equations represented in Eq. \ref{RateChanges} where $\mathcal{N}_i$ and $\mathcal{N}_k$ represent the neutrino number densities in bins $i$ and $k$, respectively, and $N_b$ is the total number of bins. The transition rates $\mathcal{R}_{ik}^{\text{in/out}}$ encapsulate the physics of neutrino scattering between these bins.

Converting this equation into matrix form, we introduce a vector $\mathbf{\mathcal{N}}$ representing the neutrino number densities across all energy bins and a matrix $\mathbf{M}$ encoding the transition rates between bins, such that

\begin{equation}
\frac{d\mathbf{\mathcal{N}}}{dt} = \mathbf{M}(\mathbf{\mathcal{N}}) \cdot \mathbf{\mathcal{N}},
\end{equation}
where $\mathbf{M}$ is an $N_b \times N_b$ collision matrix, embodying the interaction rates between different energy bins.

The elements of the collision matrix, $M_{ik}$, are defined as
\begin{equation}
    M_{ik} = \hat{\mathcal{R}}_{ik}^{\text{in}} - \delta_{ik}\tilde{\kappa}_k,
\end{equation}
indicating the rate at which neutrinos scatter into ($\hat{\mathcal{R}}^{\text{in}}$) and out of ($\tilde{\kappa}$) the energy bins. This formulation captures both the influx and efflux of neutrinos in each bin, modulated by the in-scattering and out-scattering rates, respectively.

Defining $\dot{\mathcal{N}}_i \equiv d \mathcal{N}_i / d t$, we have explicitly for the ``Collision Matrix''

\begin{equation}
\left(\begin{array}{c}
\dot{\mathcal{N}}_1 \\
\dot{\mathcal{N}}_2 \\
\dot{\mathcal{N}}_3 \\
\vdots \\
\dot{\mathcal{N}}_{N_b}
\end{array}\right)=\left(\begin{array}{ccccc}
M_{11} & M_{12} & M_{13} & \ldots & M_{1 N_b} \\
M_{21} & M_{22} & M_{23} & \ldots & M_{2 N_b} \\
M_{31} & M_{32} & M_{33} & \ldots & M_{3 N_b} \\
\vdots & \vdots & \vdots & \ldots & \vdots \\
M_{N_b 1} & M_{N_b 2} & M_{N_b 3} & \ldots & M_{N_b N_b}
\end{array}\right)\left(\begin{array}{c}
\mathcal{N}_1 \\
\mathcal{N}_2 \\
\mathcal{N}_3 \\
\vdots \\
\mathcal{N}_{N_b}
\end{array}\right)
\label{CollisionMatrix}
\end{equation}
The results presented in \cite{PhysRevD.109.103019} used a $40 \times 40$ matrix, but here we can incorporate matrices up to $180 \times 180$. Standard explicit methods for solving the matrix differential equations in stiff systems often require non-competitive, extremely small time steps to maintain stability. However, it will be shown that algebraically stabilized explicit methods can achieve competitive time steps. Explicit integration methods are used at each time step to simulate the dynamical neutrino distributions from well-defined initial conditions.

On the other hand, implicit schemes require much more costly matrix inversions at every step. While it is true that implicit methods can demand multiple matrix inversions within one time step, especially during strong interaction phases, the primary concern is the non-linear impact of matrix inversions on computing time. Typically, the dependence of computing time on the size of the matrix is quadratic or worse. By using less costly matrix-vector multiplications, we can reduce the computational load and take more cost-effective time steps.

    \chapter{Numerical Integration}
\label{NumericalIntegration}
\section{Numerical Integration Methods for Neutrino Distribution}

The two dominant algorithms used in numerical simulations for the distributions of neutrinos over astrophysical events in this thesis consist of the explicit asymptotic method and the forward Euler algorithm. In this regard, such methods play an active role in stabilizing the evolution of the neutrino distribution, hence keeping at bay most of the computational difficulties sometimes induced by stiff differential equations.

\subsection{Explicit Asymptotic Method}

As described in more depth for neutrinos in \citep{lac2020}, as well as in thermonuclear networks in \citep{guidJCP, guidAsy}, the explicit asymptotic method approximates the future state of neutrino number densities. More specifically, the explicit asymptotic method approximates the future state of the number density of neutrinos by iterating solutions for $\mathcal{N}_i^{n+1}$. This calculates the number density at the next time step $n+1$, based on the current time step $n$. The derivative of this expansion is given by
\begin{equation}
\mathscr{N}_i = \frac{1}{\tilde{\kappa}_i}\left(F_i^{+} - \frac{d \mathscr{N}_i}{d t}\right),
\end{equation}
and in the asymptotic limit where $\frac{d \mathcal{N}_i}{d t} \rightarrow 0$ implying $F_i^{+} \approx \tilde{\kappa}_i$, the finite-difference approximation yields:
\begin{equation}
\frac{d \mathscr{N}_i^n}{d t} = \frac{\mathscr{N}_i^{n+1} - \mathscr{N}_i^n}{\Delta t} - \frac{\Delta t}{2} \frac{d^2 \mathscr{N}_i^n}{d t^2} + \ldots,
\end{equation}
truncating at the first term for small derivatives. This leads to an approximate expression for $\mathscr{N}_i^{n+1}$:
\begin{equation}
\mathscr{N}_i^{n+1} = \mathscr{N}_i^n + \frac{\Delta t C_i^n}{1 + \tilde{\kappa}_i \Delta t}.
\end{equation}
Particle number conservation check for the explicit asymptotic (EA) method involves,
\begin{equation}
N_{\text{tot}}^{n+1} = N_{\text{tot}}^n + \Delta t \sum_{i=1}^{N_b} \frac{C_i^n}{1 + \tilde{\kappa}_i \Delta t} \Delta V_i^\epsilon,
\end{equation}
with the Taylor Series expansion of the denominator
\begin{equation}
\frac{1}{1 + \tilde{\kappa}_i \Delta t} = \sum_{m=0}^{\infty}(-1)^m(\tilde{\kappa}_i \Delta t)^m,
\end{equation}
resulting in
\begin{equation}
N_{\text{tot}}^{n+1} = N_{\text{tot}}^n + \Delta t \sum_{i=1}^{N_b} C_i^n \Delta V_i^\epsilon + \sum_{i=1}^{N_b} \sum_{m=1}^{\infty} C_i^n \Delta V_i^\epsilon(-1)^m(\tilde{\kappa}_i)^m(\Delta t)^{m+1},
\end{equation}
where $C_i^n$ is directly derived from the rate of change of 
 $\mathcal{N}_i^n$. This approach, based on the first term of Taylor series expansion, has the significant advantage of providing a clear formula that allows one to avoid the explicit inversion of matrices and is thus particularly efficient for large-scale simulation.

The tolerance condition for conservation of particle number is defined as:
\begin{equation}
\left| \frac{N_{\text{tot}}^{n+1} - N_{\text{tot}}^n}{N_{\text{tot}}^n} \right| \leq \text{tolC},
\label{TolC}
\end{equation}
and an additional tolerance condition for the accuracy in density between time steps is introduced
\begin{equation}
\max \left[ \frac{\left| \mathscr{N}^{n+1} - \mathscr{N}^n \right|}{\max \left( \mathscr{N}^n, 10^{-8} \right)} \right] \leq \text{tolN}.
\end{equation}

\subsection{Explicit Forward Euler Algorithm}
The Forward Euler Algorithm is a simplistic explicit numerical integration requiring small time steps for stability in stiff systems. We can apply it directly to the neutrino distribution evolution as
\begin{equation}
\mathcal{N}_i^{n+1}=\mathcal{N}_i^n+\Delta t C_i^n,
\label{ForwardEuler}
\end{equation}
with the change rate $C_i^n$ defined as
\begin{equation}
C_i^n \equiv \frac{d \mathcal{N}_i^n}{d t}=\eta_i-\tilde{\kappa}_i \mathcal{N}_i^n.
\end{equation}
The total particle number at a specific time $t$ is approximated by the sum over all energy bins
\begin{equation}
N_{\text{tot}}(t) \approx \sum_{i=1}^{N_b} \mathscr{N}_i(t) \Delta V_i^\epsilon.
\end{equation}
By multiplying Eq. \ref{ForwardEuler} by $\Delta V_i^\epsilon$ and summing over all bins, we derive the expression for the total particle number at the next time step
\begin{equation}
N_{\text{tot}}^{n+1} = N_{\text{tot}}^n + \Delta t \sum_{i=1}^{N_b} C_i^n \Delta V_i^\epsilon.
\end{equation}
Evaluating the summation of $C_i^n \Delta V_i^\epsilon$ leads to the conclusion that the total particle number remains conserved
\begin{equation}
\sum_{i=1}^{N_b} C_i^n \Delta V_i^\epsilon = \sum_{i=1}^{N_b} \sum_{k=1}^{N_b} \left(\mathscr{R}_{ik}^{\text{in}} - \mathscr{R}_{ki}^{\text{out}}\right) (1-\mathscr{N}_i) \mathscr{N}_k \Delta V_i^\epsilon \Delta V_k^\epsilon = 0,
\end{equation}
which holds due to the symmetry of the scattering kernels, ensuring $N_{\text{tot}}^{n+1} = N_{\text{tot}}^n$, thus confirming that the forward Euler method conserves particle number at each step to machine precision.

This scheme ensures stability for time steps $\Delta t$ smaller than a critical threshold $\tau_{\mathrm{c}}$, prescribed by the inverse of the fastest rate in the system. It has an advantage in being a simple and computationally effective way for small time steps, which could fully integrate the system dynamics.

Therefore the decision to use either the forward Euler or the explicit asymptotic method relies on the comparative size of the integration time step $\Delta t$ and the critical time step $\tau_{\mathrm{c}}$. For $\Delta t < \tau_{\mathrm{c}}$, the forward Euler method is stable and preferred due to its simplicity. On the other hand, for $\Delta t \geq \tau_{\mathrm{c}}$, when the forward Euler scheme can become unstable, the use of the explicit asymptotic method is far more stable and shows us better results without instability for larger values of the time step.

This approach ensures that the integration done in FENN remains stable and efficient, allowing for accurate simulations of neutrino dynamics across varying timescales and interaction intensities.

\subsection{Backward Euler Method}

The backward Euler (BE) method, an implicit method used for comparison with the explicit asymptotic (EA) method, is applied as follows:
\begin{equation}
\mathscr{N}_i^{n+1} = \mathscr{N}_i^n + \Delta t C_i^{n+1},
\end{equation}
where it is noted that BE conserves particle number within a specified tolerance and requires solving an algebraic equation for future density values $\mathscr{N}_i^{n+1}$. This is approached using the Newton-Raphson method with a defined tolerance parameter, tolBE. The algebraic equation to solve is
\begin{equation}
\mathscr{N}_i^{n+1}(1 + \tilde{\kappa}_i(\mathscr{N}^{n+1}) \Delta t) - \Delta t F_i^{+}(\mathscr{N}^{n+1}) - \mathscr{N}_i^n = 0,
\end{equation}
or concisely, $f(\mathscr{N}_i^{n+1}) = 0$. The iterative formula for the Newton-Raphson method is:
\begin{equation}
\mathscr{N}_{i, k+1}^{n+1} = \mathscr{N}_{i, k}^{n+1} - \frac{f(\mathscr{N}_{i, k}^{n+1})}{f'(\mathscr{N}_{i, k}^{n+1})},
\end{equation}
with the termination condition
\begin{equation}
\frac{\left| f(\mathscr{N}_{i, k}^{n+1}) / f'(\mathscr{N}_{i, k}^{n+1}) \right|}{\mathscr{N}_{i, k+1}^{n+1}} < \text{tolBE}.
\end{equation}
The Newton-Raphson method's vector form, requiring matrix inversions, is
\begin{equation}
\mathscr{N}_{k+1}^{n+1} = \mathscr{N}_k^{n+1} - \frac{f(\mathscr{N}_k^{n+1})}{f'(\mathscr{N}_k^{n+1})}.
\end{equation}
    \chapter{Integration of WeakLib and FENN} \label{ch:Algorithm}
\section{Pipeline Algorithms}

WeakLib is a Fortran based computational library developed at Oak Ridge Labs, that allows the computation of 5 dimensional tables as referenced in section \ref{ch:Mathematical Formulation}. To incorporate the neutrino scattering opacities ($\Phi_\text{In}$), tabulated within WeakLib, as demonstrated in the above formalism, into FENN, a dedicated pipeline algorithm was developed. This algorithm reads the tabulated data, applies necessary physical constants such as the speed of light, and seamlessly integrates this data into the FENN simulation environment. A greatly simplified pipeline was used in \citep{PhysRevD.109.103019}, however, a modified pipeline is required for integration with the WeakLib library that can read an arbitrary $\rho$, $T$, $Y_e$ grid. The new pipeline first introduced in \citep{2024AAS...24313505C} is outlined in the following sections.

\begin{figure}[H]
\centering 
\includegraphics[width=5in]{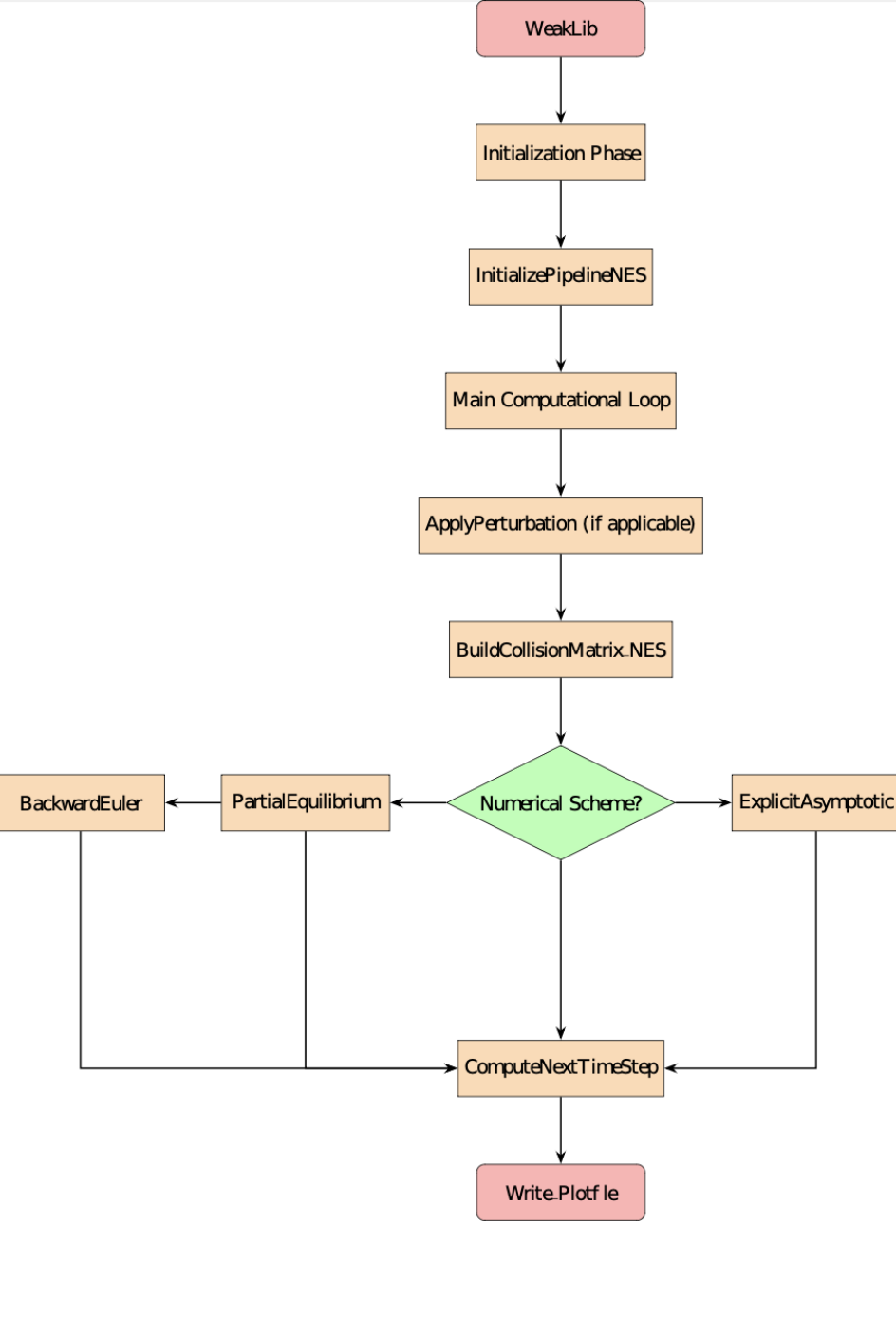}
\caption{WeakLib to FENN Integration.}\label{Flowchart}
\end{figure}

\subsection{Data Reading and Transformation}

Given a file path \texttt{filePath} and dataset name \texttt{datasetName}, the algorithm reads multidimensional data for neutrino scattering opacities $\Phi_\text{In}$.

\begin{figure}[H]
    \centering
    \includegraphics[width=5in]{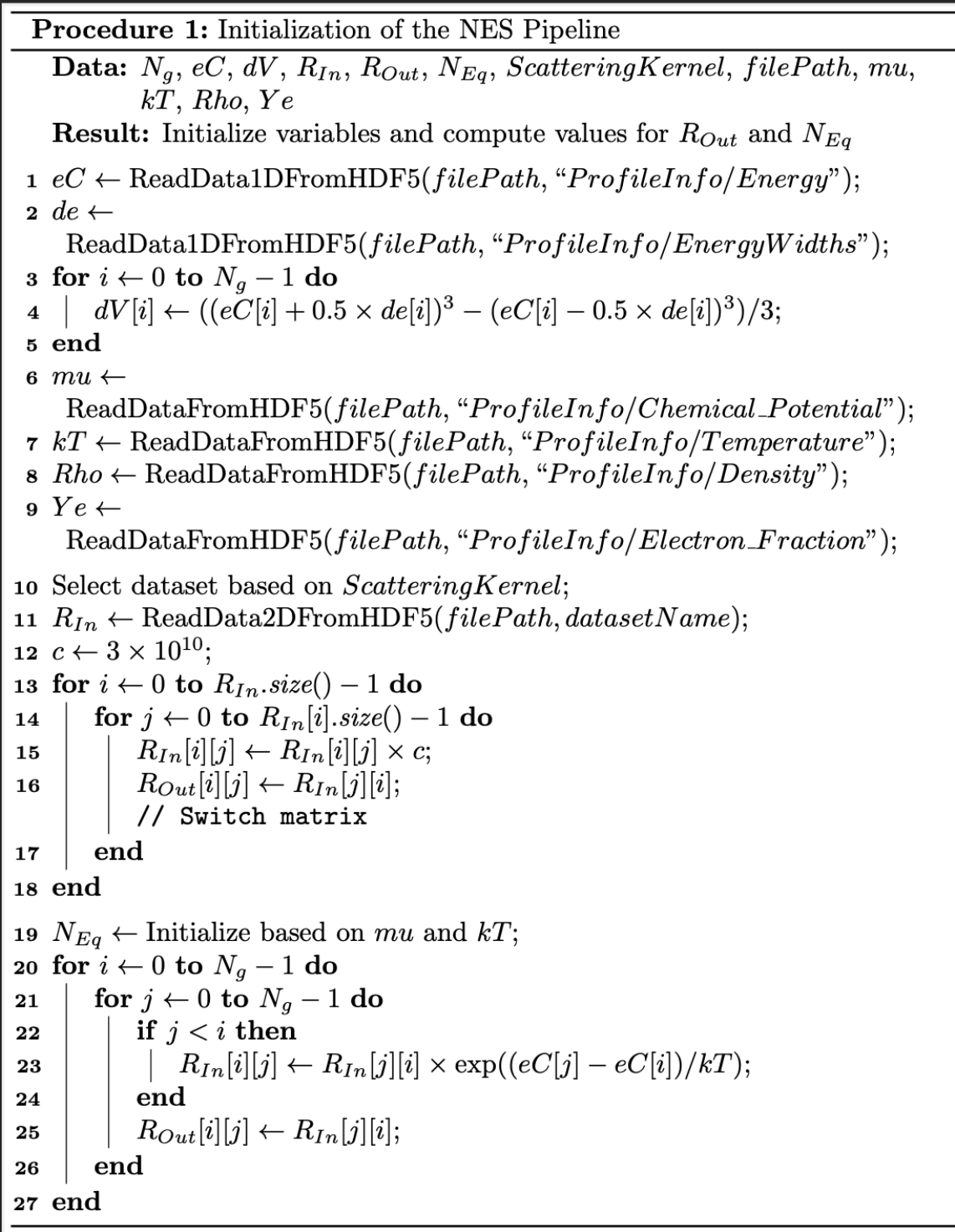}
    \caption{InitializePipelineNES algorithm}
    \label{fig:InitializePipelineNES}
\end{figure}

\subsection{Treatment of $\Phi_\text{In}$}

The speed of light $c$ is used to scale the $\Phi_\text{In}$ scattering rates read from the datasets:

\begin{equation}
    \mathbf{R}_{\text{In}} = \mathbf{R_\text{In}} \cdot c
\end{equation}
Given energy bin centers $\mathbf{eC}$ and widths $\mathbf{de}$, the volume elements $\mathbf{dV}$ are calculated from Eq. \ref{EnergyBinVolume} for $\quad i = 1, \ldots, N_g$. 
For a selected scattering kernel, the input and output rates $\mathbf{R}_{\text{In}}$ and $\mathbf{R}_{\text{Out}}$ are adjusted to reflect physical constraints and detailed balance:

\begin{equation}
    R_{\text{Out}, ij} = R_{\text{In}, ji} \exp\left(\frac{eC_j - eC_i}{kT}\right), \quad \forall i, j
\end{equation}

\subsection{Equilibrium Distribution Initialization}

The equilibrium distribution $\mathbf{N}_{\text{Eq}}$ is initialized based on the chemical potential $\mu$ and temperature $kT$:

\begin{equation}
    N_{\text{Eq}, i} = \frac{1}{\exp\left(\frac{eC_i - \mu}{kT}\right) + 1}, \quad i = 1, \ldots, N_g
\end{equation}

\section{Newton-Raphson Method Implementation for Neutrino Scattering Simulations}

Newton-Raphson is the primary solver for our backward Euler calculations. This results in costly matrix inversions at each time step, while explicit asymptotic has less costly matrix-vector multiplication, requiring more steps that are more cost effective. The backward Euler mathematical formulation is described more in depth in \citep{lac2020}.

\subsection{Jacobian Matrix Computation}

The Jacobian matrix, $\bm{J}$, crucial for the Newton-Raphson iteration, is computed based on finite differences, incorporating the input and output scattering rates, $\bm{R}_{\text{In}}$ and $\bm{R}_{\text{Out}}$, across discrete energy groups.

\begin{equation}
    \bm{J}(\bm{N}) = -\left( \bm{R}_{\text{In}} \odot (\bm{1} - \bm{N}) + \bm{R}_{\text{Out}} \odot \bm{N} \right) + \text{Diag}\left( -(\bm{R}_{\text{In}} \bm{N} + \bm{R}_{\text{Out}} (\bm{1} - \bm{N})) \right),
\end{equation}
where $\odot$ denotes the element-wise multiplication, and $\text{Diag}(\cdot)$ creates a diagonal matrix from the vector argument. The diagonal components are adjusted separately to account for the conservation laws and the specific dynamics of neutrino scattering.

\subsection{Right-Hand Side Computation}

The right-hand side (RHS) of the Newton-Raphson method encapsulates the balance between the scattering into and out of each energy state:

\begin{equation}
    \bm{F}(\bm{N}) = (\bm{1} - \bm{N}) \odot (\bm{R}_{\text{In}} \bm{N}) - \bm{N} \odot (\bm{R}_{\text{Out}} (\bm{1} - \bm{N})).
\end{equation}

\subsection{Newton-Raphson Iteration}

The Newton-Raphson iterative process for updating the neutrino occupation numbers is detailed as follows:

\begin{enumerate}
    \item Compute the Jacobian $\bm{J}(\bm{N})$ and RHS $\bm{F}(\bm{N})$ at the current estimate $\bm{N}$.
    \item Solve the linear system $\left(\bm{I} - \Delta t \bm{J}(\bm{N})\right) \Delta \bm{N} = (\bm{N}_{\text{old}} - \bm{N}) + \Delta t \bm{F}(\bm{N})$ for $\Delta \bm{N}$.
    \item Update the solution $\bm{N} \leftarrow \bm{N} + \Delta \bm{N}$.
    \item Check for convergence using the norm of $\Delta \bm{N}$ relative to $\bm{N}$; if not converged, return to step 1.
\end{enumerate}
The algorithm proceeds until the solution converges to within a specified defined tolerance.

\subsection{Timestep Functions}
Additionally, we can adjust our timestep settings by varying the Tolerance Conditions from Eq. \ref{TolC}. Each timestep has a fixed $dt_{grow}$ value for a more robust timestepping algorithm. 

\begin{figure}[H]
    \centering
    \includegraphics[width=5in]{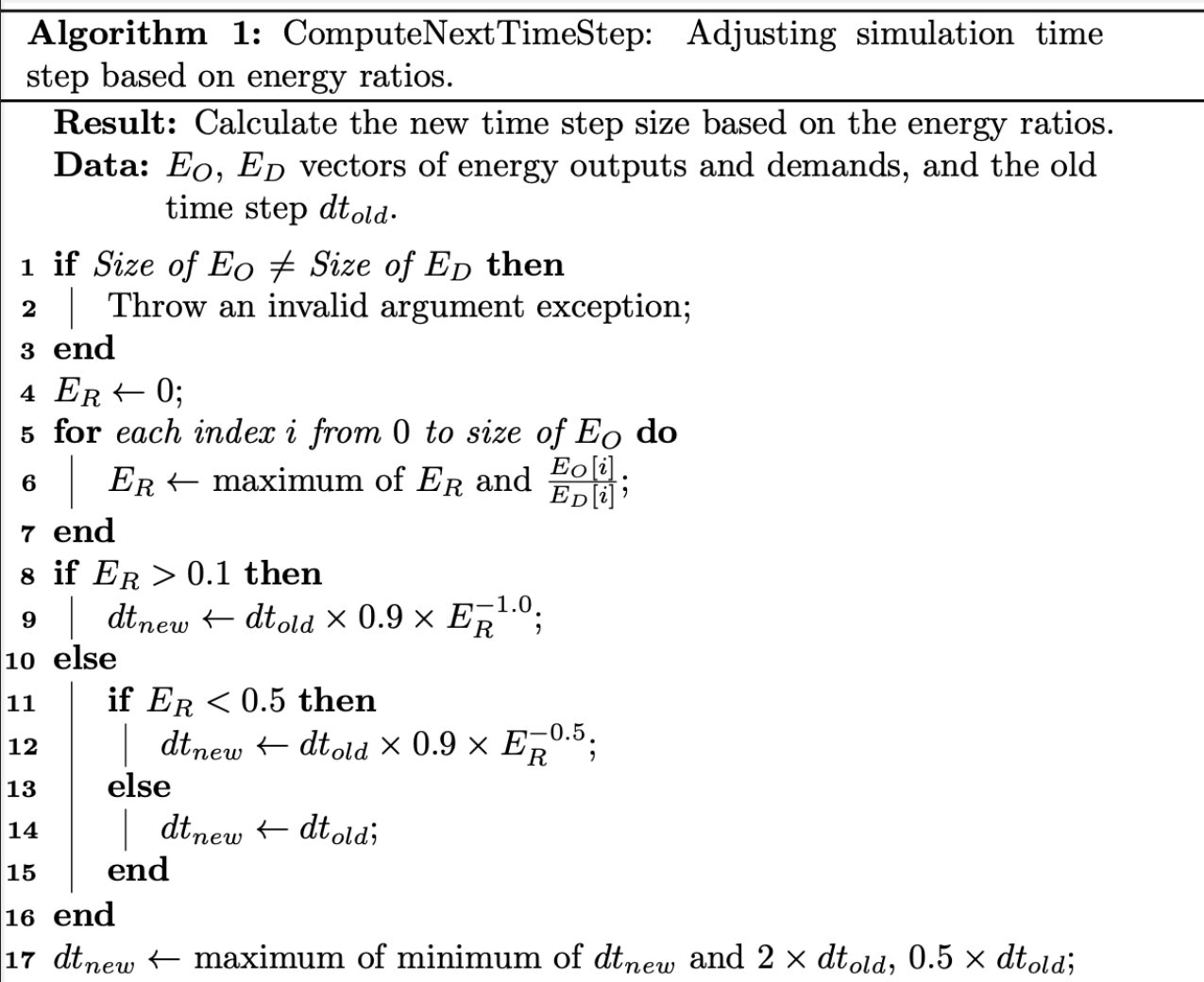}
    \caption{The ComputeNextTimeStep function modifies the time step dynamically, responding to the ratio of energy output to energy demand $(E_O \to E_D)$. It reduces the time step when this ratio is too high to preserve the accuracy of the simulation, and it expands the time step when the ratio is low to enhance efficiency.}
    \label{fig:ComputeTimeStepEA}
\end{figure}

\begin{figure}[H]
    \centering
    \includegraphics[width=5in]{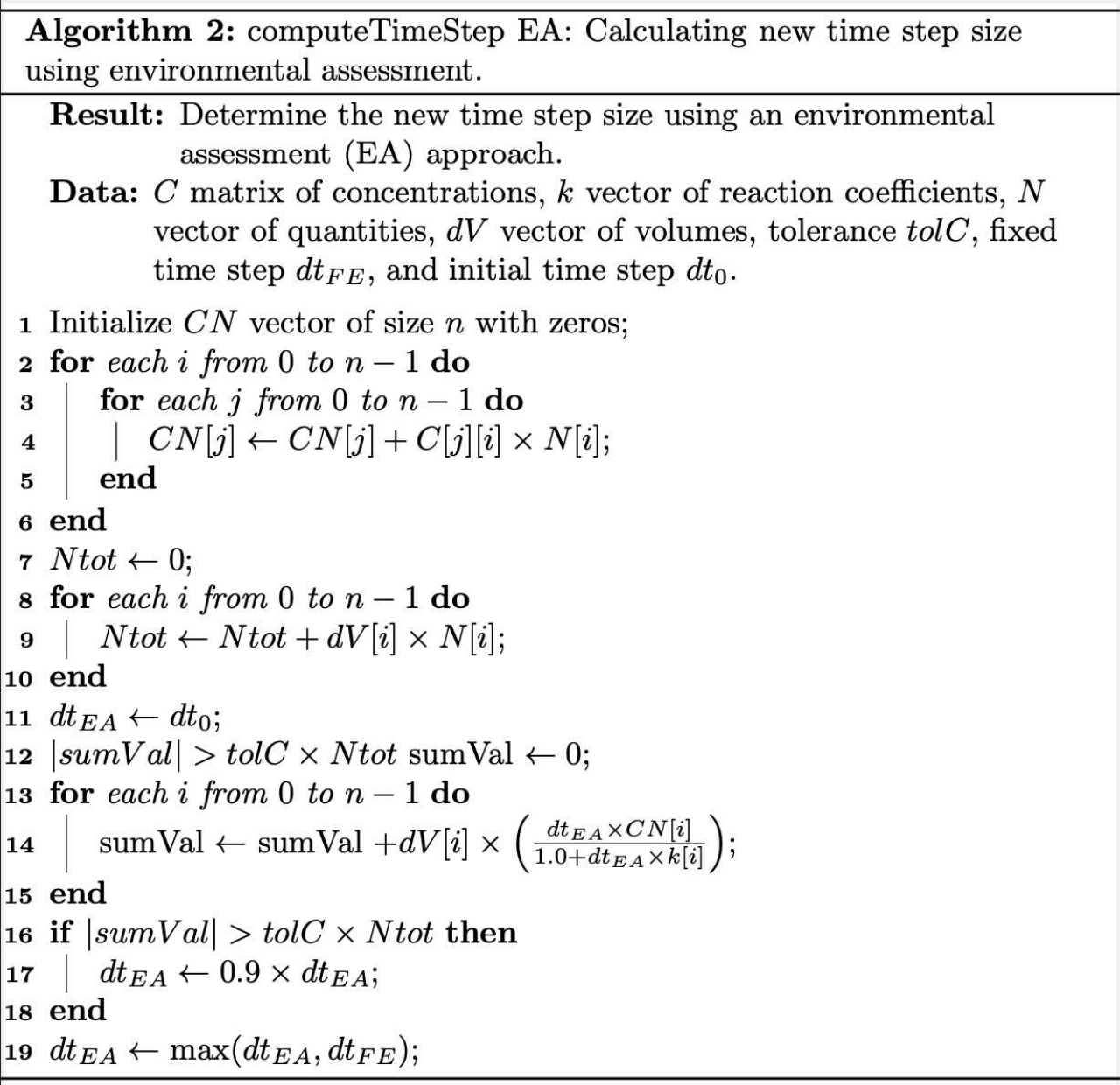}
    \caption{The computeTimeStep EA function employs an environmental assessment approach, adjusting the time step according to the total changes observed over a time step compared to a predefined tolerance. This method allows the function to modify the time step to maintain simulation precision within set limits.}
\label{fig:ComputeTimeStep}
\end{figure}
    \chapter{Tradeoff of Speed and Accuracy} \label{ch:results}

\section{Single Models}
\begin{figure}[H]
    \centering
    \includegraphics[width=5in]{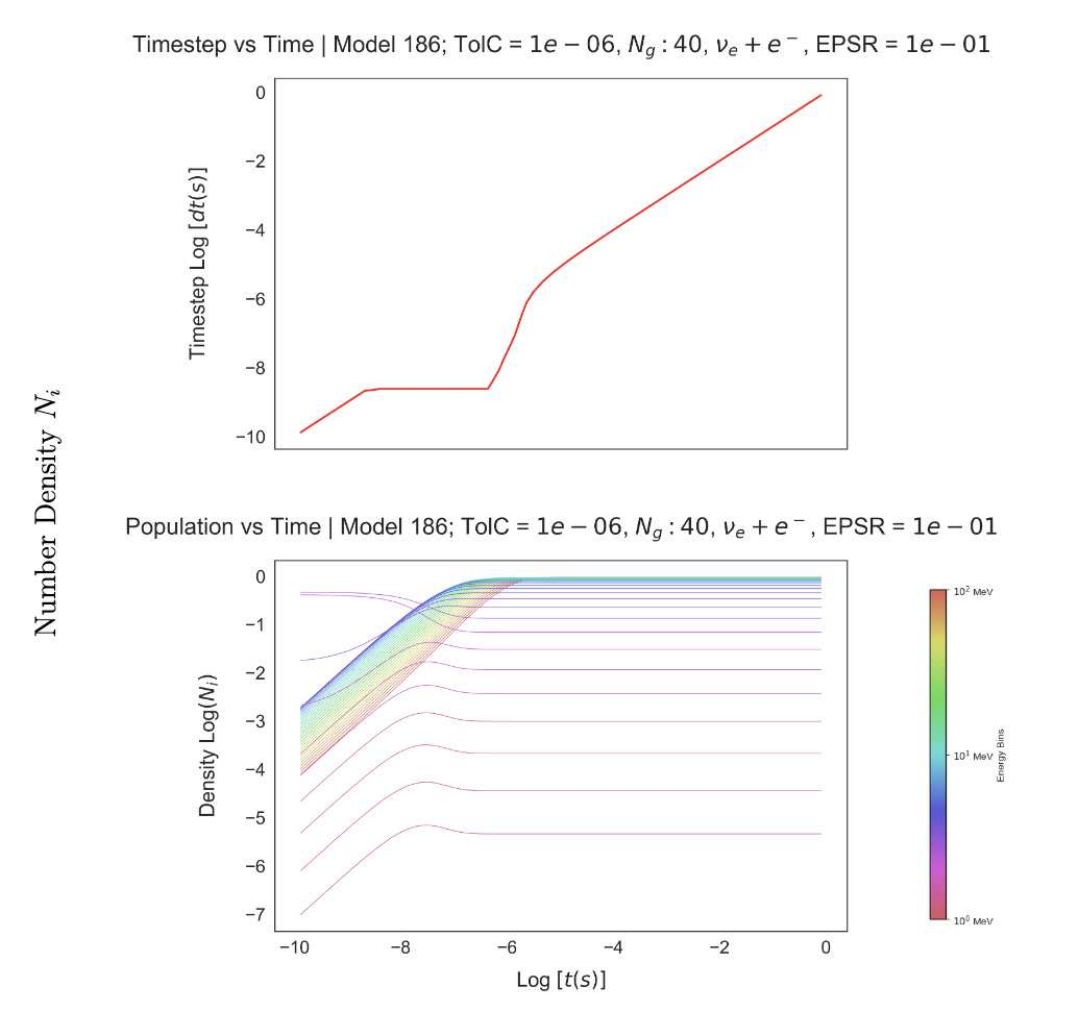}
    \caption{Dynamics of Model 186 showing Number Density $N_i$ across 40 energy bins vs time and time step vs time. The calculation referenced took 323 explicit steps.}
    \label{186PopAndTime}
\end{figure}

We can begin by presenting a representative of one set of conditions using explicit asymptotic methods. Figure \ref{186PopAndTime} shows the variation of number density $N_i$ across 40 energy bins as a function of time, as well as the progression of computational time steps in relation to the temporal axis. All corresponding model conditions are tabulated in the appendix \ref{ch:appendix}. We are using a wide range of thermodynamic conditions  expected to occur in core-collapse supernovae. 

\subsection{Heatmaps}
We can also introduce heatmaps for the scattering rates for Neutrino Electron Scattering. The following heatmaps are for a network size of 50 and reflect the rates used in the proceeding sections. 

\begin{figure}[H]
    \centering
    \includegraphics[width=5in]{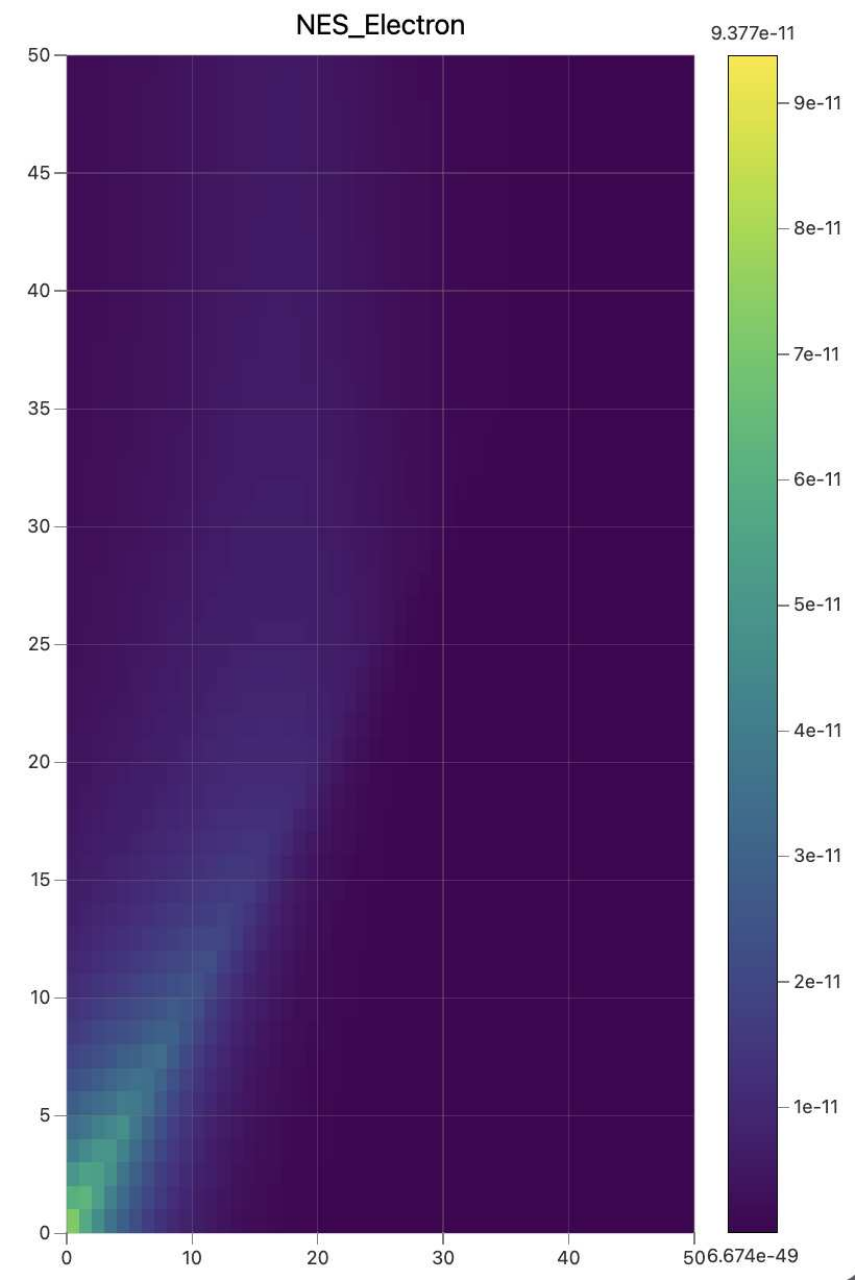}
    \caption{Model 81 Heatmap}
    \label{fig:model81Heatmap}
\end{figure}

\begin{figure}[H]
    \centering
    \includegraphics[width=5in]{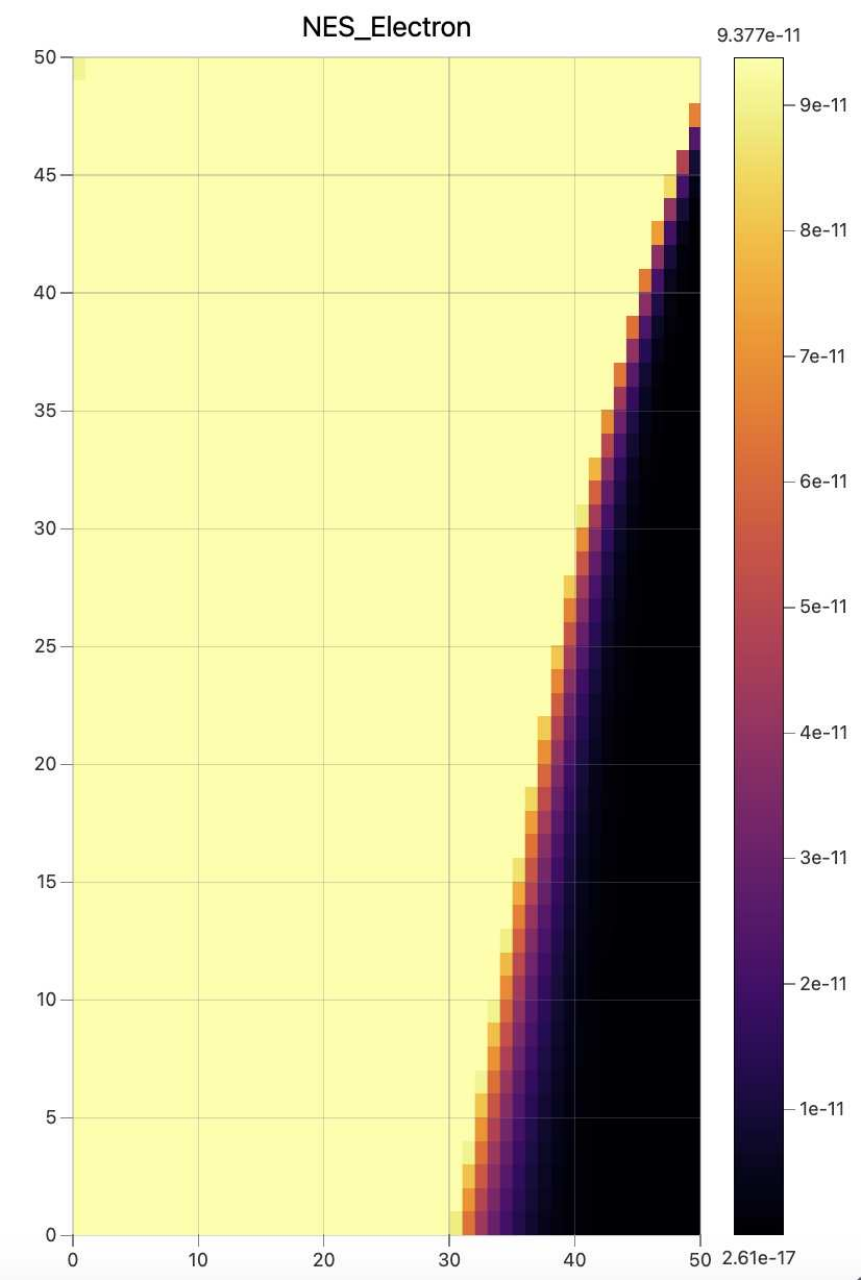}
    \caption{Model 186 Heatmap}
    \label{fig:model186Heatmap}
\end{figure}

\begin{figure}[H]
    \centering
    \includegraphics[width=5in]{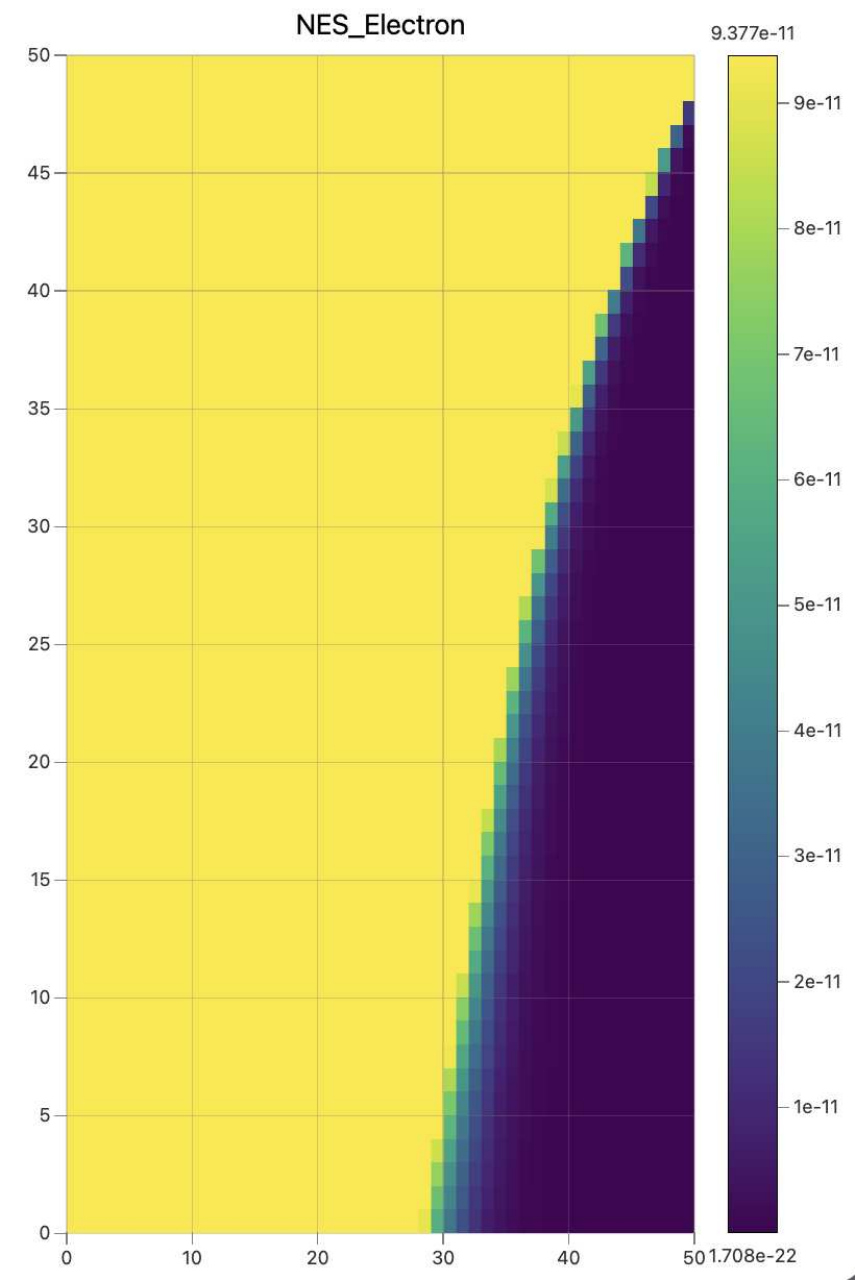}
    \caption{Model 213 Heatmap}
    \label{fig:model213Heatmap}
\end{figure}

\begin{figure}[H]
    \centering
    \includegraphics[width=5in]{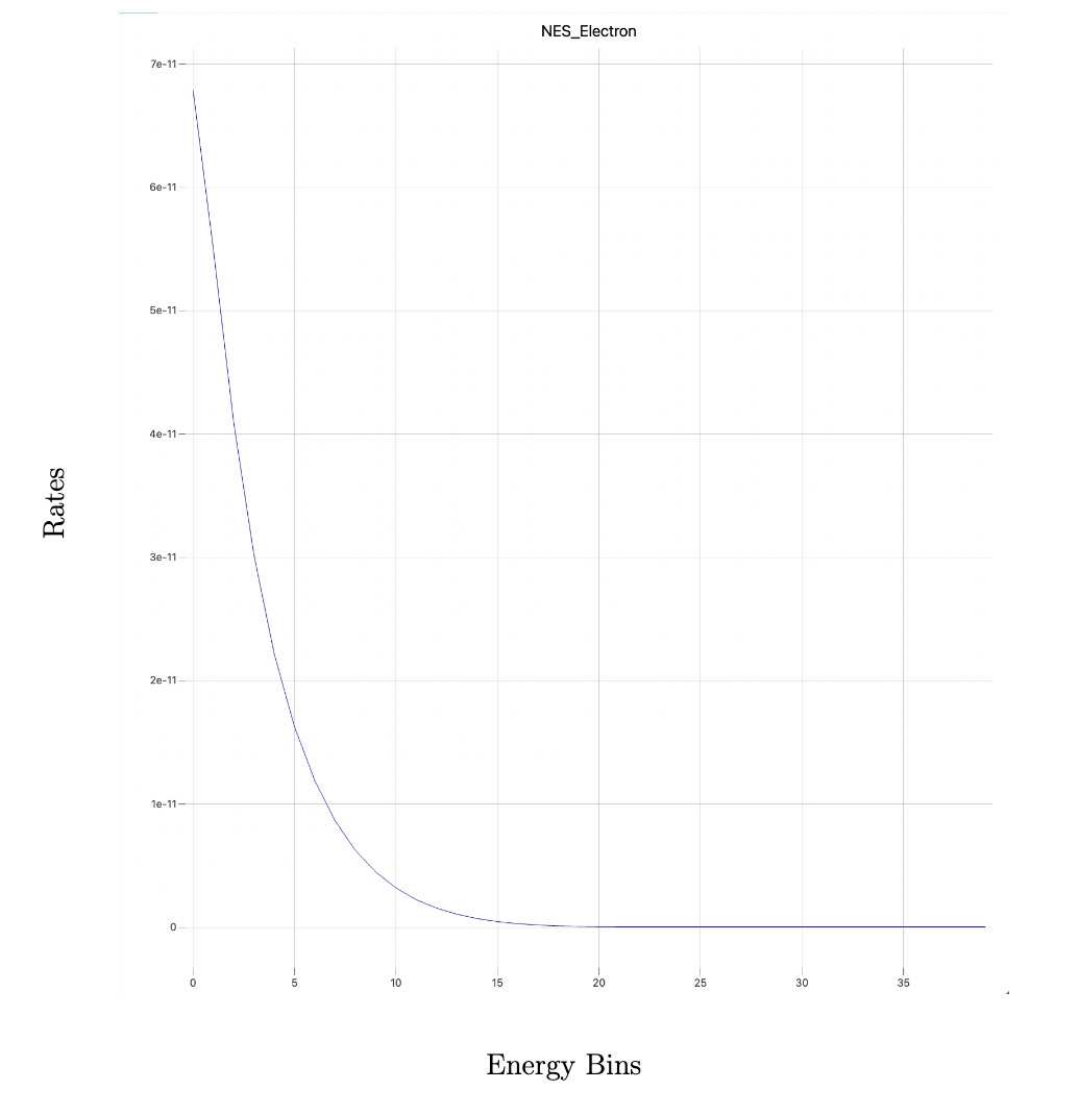}
    \caption{$\Phi_\text{In}$ for one Energy Level}
    \label{EnergyLine}
\end{figure}

Before comparing backward Euler versus explicit results we have to define a quantitative measure of accuracy, since approximations of the realistic problem introduces some level of error. More in-depth explanations are formulated in \citep{PhysRevD.109.103019}

We can characterize the accuracy of an explicit algebraic evaluation of the neutrino evolution by integrating the deviation of the neutrino number densities \(n_i\) from backward Euler results, which is our basis of comparison. We employ a root mean square calculation in number densities integrated (approximated by a sum) over all energy bins. \(R_i(t)\) is defined for each species \(i\) in the network at every time \(t\),
\begin{equation}
     R_i(t) \equiv |n_i(t) - n_i^0(t)|,
\end{equation}
where \(n_i(t)\) is the explicit asymptotic approximation for the number density in bin \(i\) at time \(t\). We can define the root mean square (RMS) error \(R(t)\) summed over all bins at time \(t\), 
\begin{equation}
    R(t) \equiv \sqrt{\sum_i R_i(t)^2} = \sqrt{\sum_i |n_i(t) - n_i^0(t)|^2},
\end{equation}
where the sum is over all energy bins and \(\sum_i n_i^0 = 1\). 

We can then approximate the total error per unit time \(\epsilon\) as
\begin{equation}
      \epsilon= \frac{1}{\delta t} \sum_{j=j_0}^{j(t_{\text{eq}})} R(t_j) \,\delta t_j,
      \label{epsilon}
\end{equation}
where \(t_j\) is the time at the \(j\)th plot output step, \(j_0\) is the first plot output step with finite \(R(t)\), \(j(t_{\text{eq}})\) is the plot output step corresponding to the equilibration time, \(\delta t_j = t_j - t_{j-1}\) is the time difference between the \(j\)th and \((j-1)\)th plot output steps, and \(\delta t \equiv t_{\text{eq}} - t(j_0)\) is the total time between the onset of finite \(R(t)\) and equilibration at \(t_{\text{eq}}\). These are represented in the \(\epsilon\) vs Time plots in the comparisons.

\subsection{A Quantitative Measure of Speed}

The explicit algebraic methods are parameterized in terms of the total number of integration steps required to complete a given simulation. This is what is known as an ``intrinsic'' measure. The total integration time is a function of the steps and the time required for each step, however, the backward Euler is doing complex matrix inversions at each step, while explicit is doing matrix-vector multiplication .

We can define a speedup factor as $F\equiv 1/(1-f)$, where $f$ is the fraction of overall computing time spent by the implicit code (backward Euler) in its linear algebra solver during for each timestep. We then expect the explicit asymptotic to be approximately $F$ times faster than backward Euler in the same computational environment.

The factor $F$ has been computed in \citep{guidAsy} using an implicit backward Euler code \textit{Xnet} \citep{hix1999}. Note that these use a dense matrix solver, so our results in a dense 40-species neutrino network are approximately valid.  

For a 40 bin calculation we can define it as:
  \begin{equation}
    \frac{\Delta t_{\rm EA}}{\Delta t_{\rm BE}} \simeq
    \frac{S_{\rm EA}}{F\cdot S_{\rm BE}} 
    \simeq 0.25 \, \left( \frac{S_{\rm EA}}{S_{\rm BE}} \right),
\label{wallClockRatio}
\end{equation}
where $\Delta t$ denotes elapsed wall clock time and $S$ is the steps. The 40-species network was employed in \citep{PhysRevD.109.103019}, however here we investigate larger networks also.

\section{Energy Grid Definitions}

\begin{figure}[H]
\centering 
\includegraphics[width=5in]{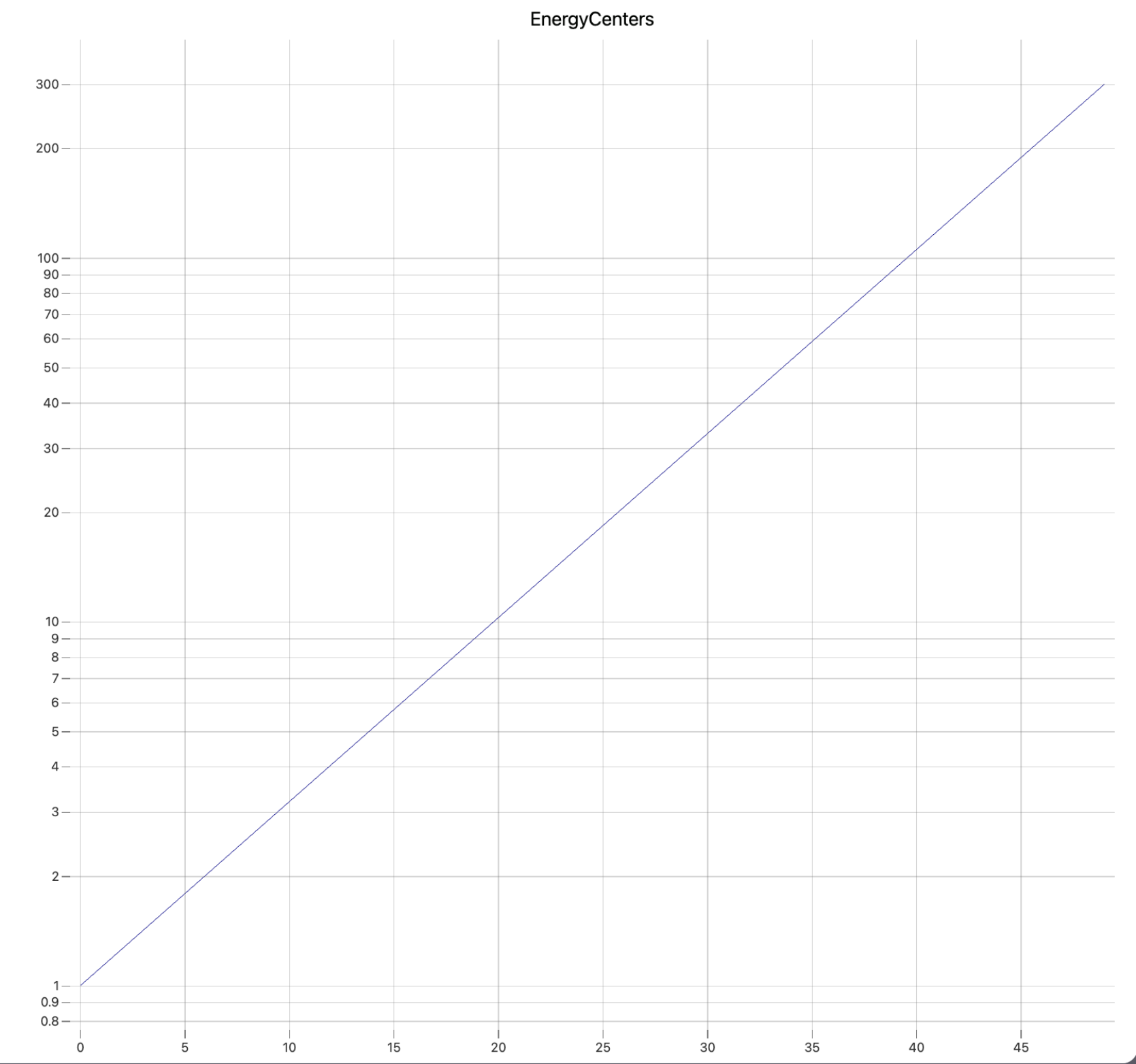}
\caption{Energy Grid for a 50-species network. Energies are logarithmically spaced from $1-300$ MeV.}
\label{Energy_Grid}
\end{figure}

All of the following calculations were done in the energy grid shown in fig. \ref{Energy_Grid}.

\begin{figure}[H]
\centering 
\includegraphics[width=5in]{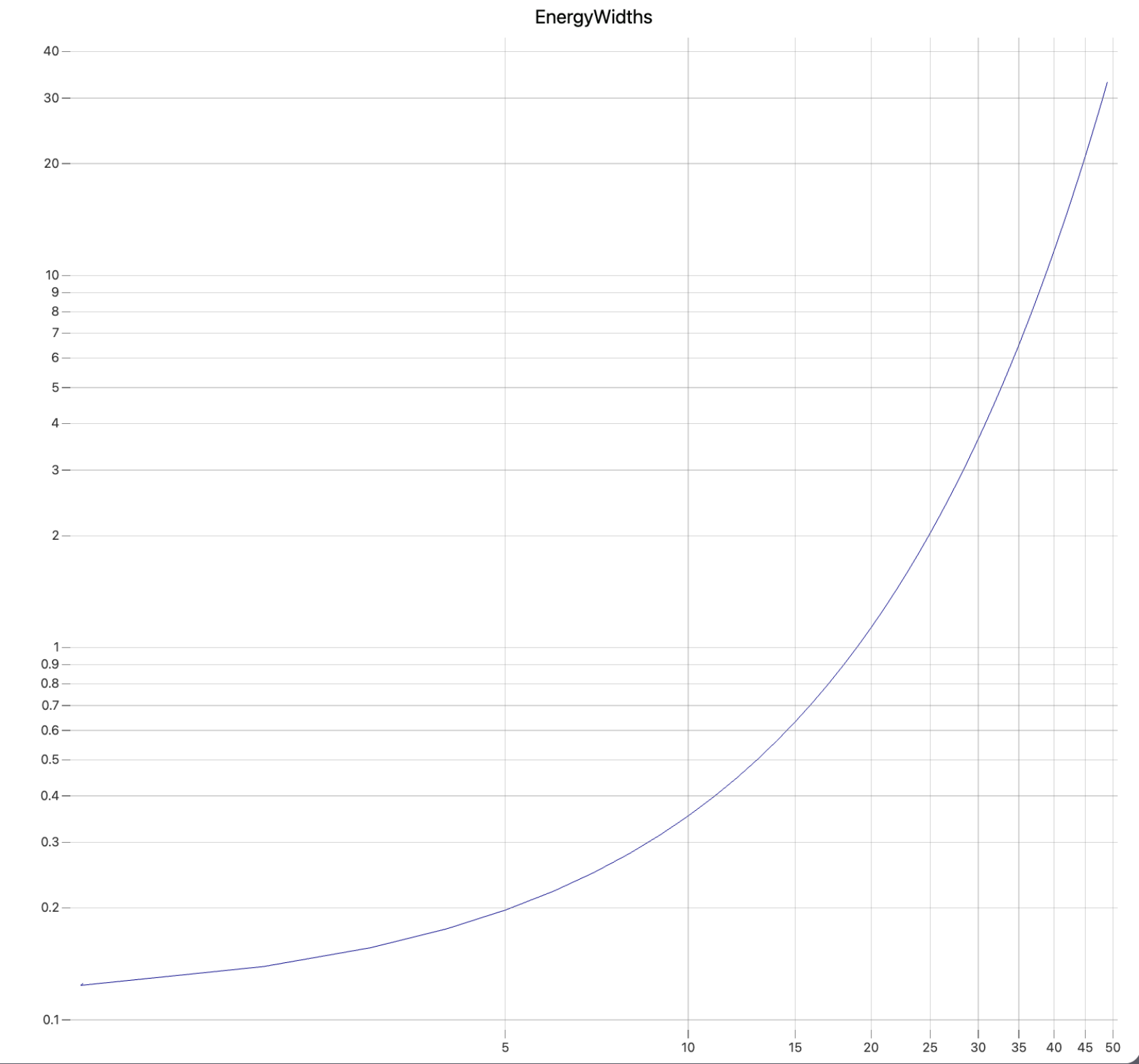}
\caption{Energy Widths}\label{Energy_Widths}
\end{figure}

Similarly for fig. \ref{Energy_Widths}. 

Data values are tabulated in the appendix \ref{ch:appendix}

\section{Model Analysis}

We can take a few of the models from appendix \ref{ch:appendix}, and do a comparative analysis of speed versus accuracy.

\subsection{Model 81 Analysis}
\label{subsec:model81}

We can begin our comparison with Model 81. We group the results by the Accurate, Intermediate, and Fast cases. These are controlled by the tolerance and are representative conditions. The following calculations were done with respect to the energy grid in fig. \ref{Energy_Grid}:

Here the time steps are represented in the Population vs Time plots as \textit{Explicit(Implicit)} $\epsilon$ is taken as a function of time calculated up until equilibrium as referenced in Eq. \ref{epsilon}.

\subsubsection{Accurate Case}
\label{subsubsec:model81accurate}

\begin{figure}[H]
    \centering
    \includegraphics[width=5in]{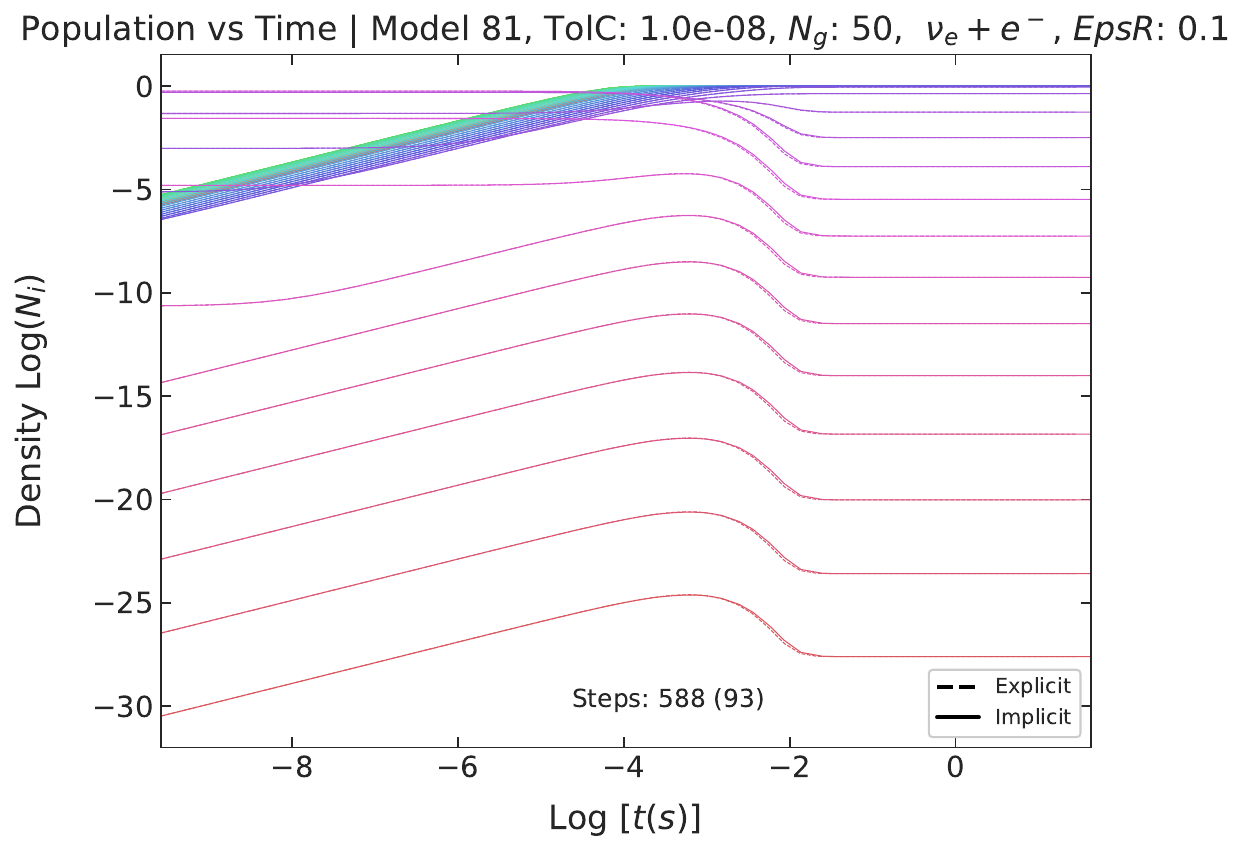}
    \caption{Population vs time for Model 81 - Accurate Case.}
    \label{fig:model81popacc}
\end{figure}

\begin{figure}[H]
    \centering
    \includegraphics[width=5in]{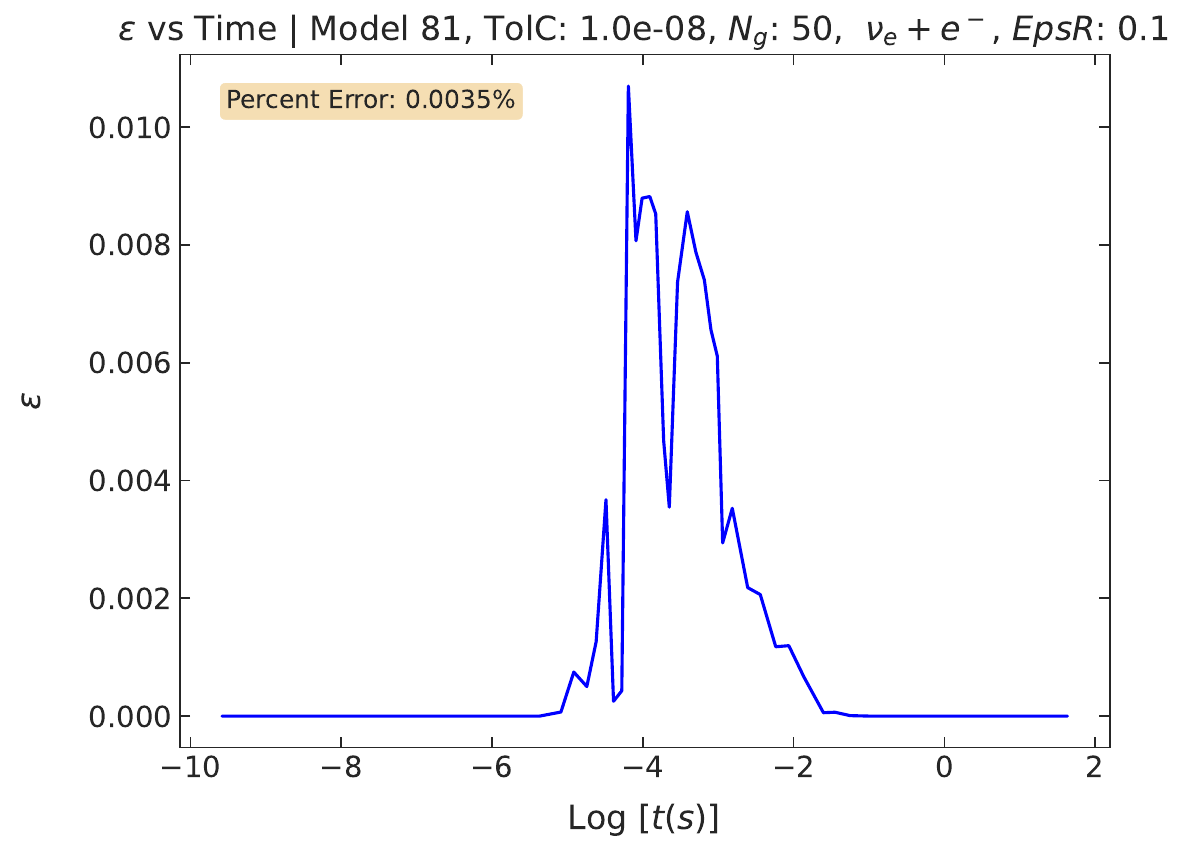}
    \caption{RMS vs time for Model 81 - Accurate Case.}
    \label{fig:model81rmsacc}
\end{figure}

\begin{figure}[H]
    \centering
    \includegraphics[width=5in]{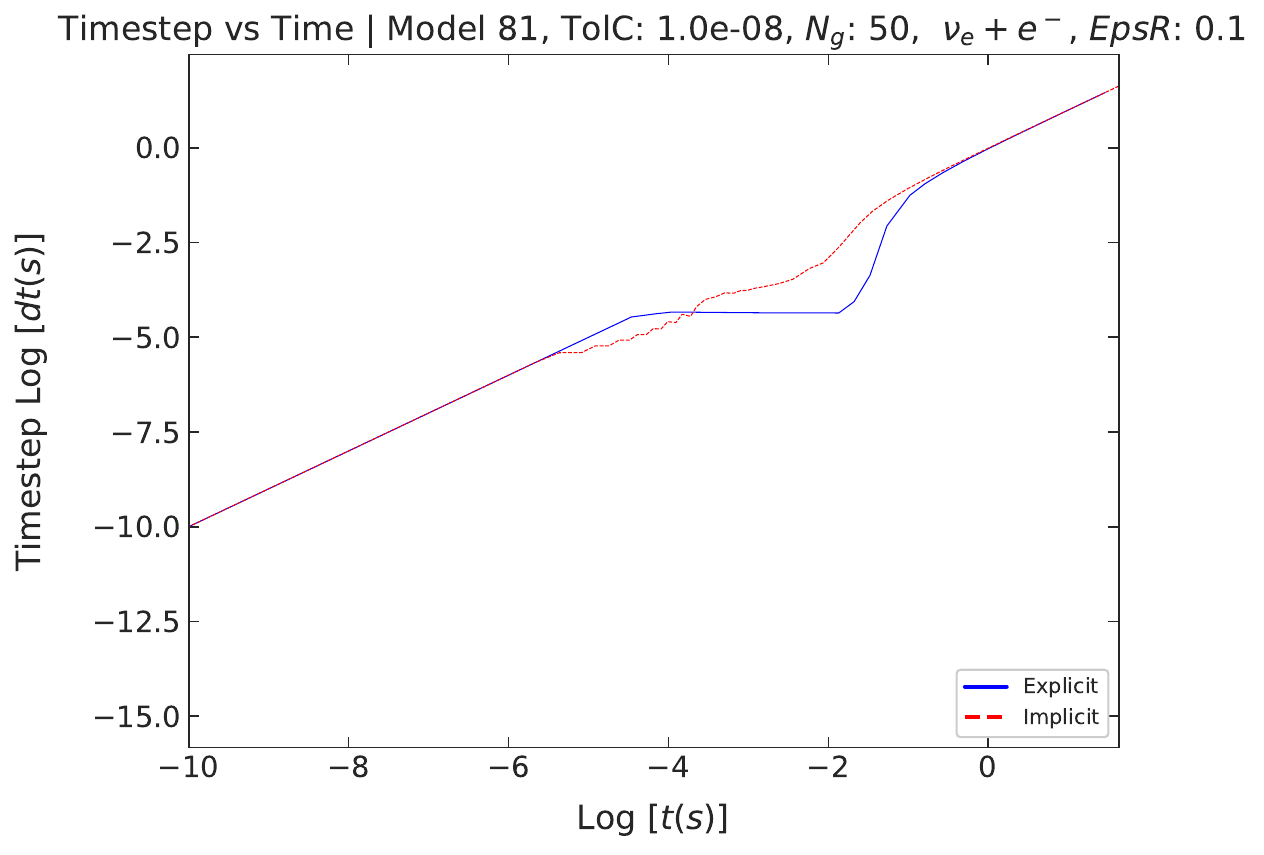}
    \caption{Time step vs time for Model 81 - Accurate Case.}
    \label{fig:model81timestepacc}
\end{figure}

Here, TolC is $10^{-8}$ which takes small time steps. The error is as a result very small.

\subsubsection{Intermediate Case}
\label{subsubsec:model81intermediate}

\begin{figure}[H]
    \centering
    \includegraphics[width=5in]{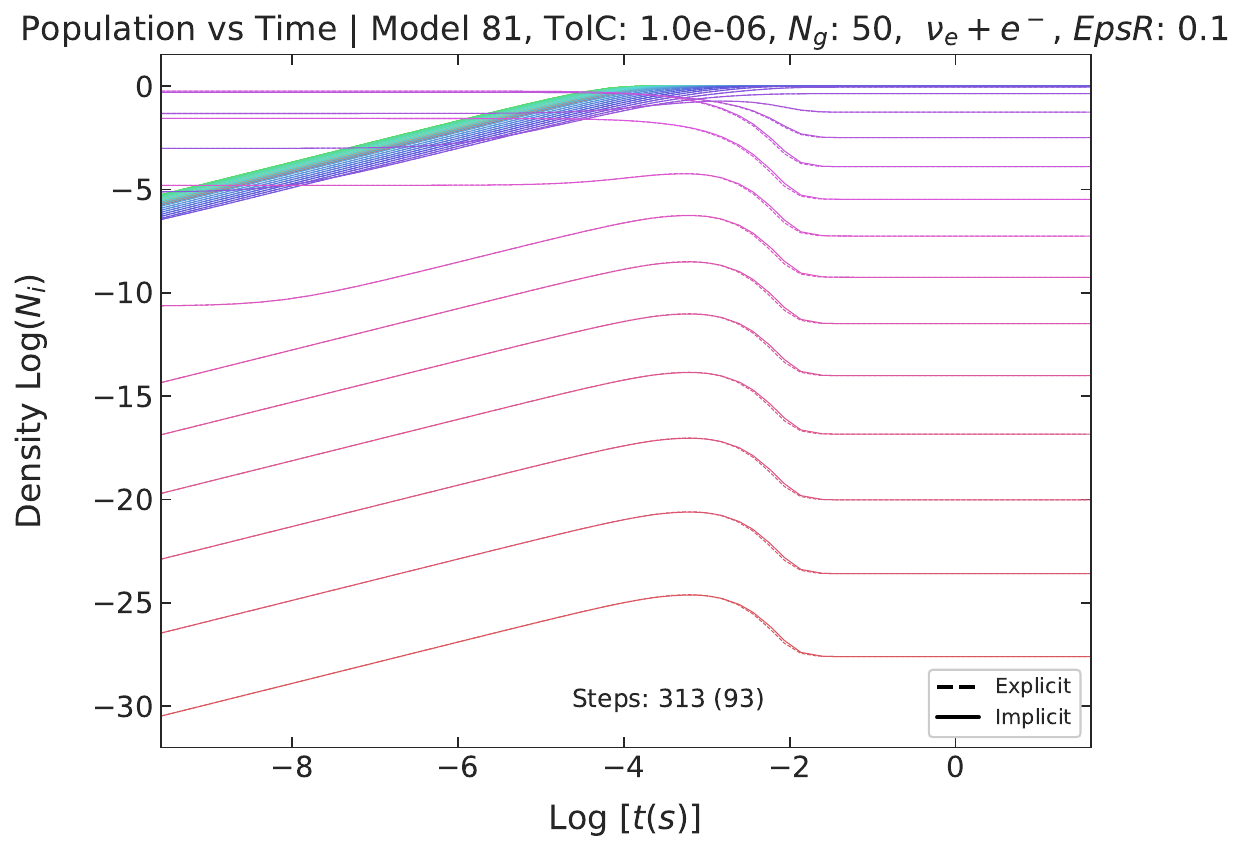}
    \caption{Population vs time for Model 81 - Intermediate Case.}
    \label{fig:model81popint}
\end{figure}

\begin{figure}[H]
    \centering
    \includegraphics[width=5in]{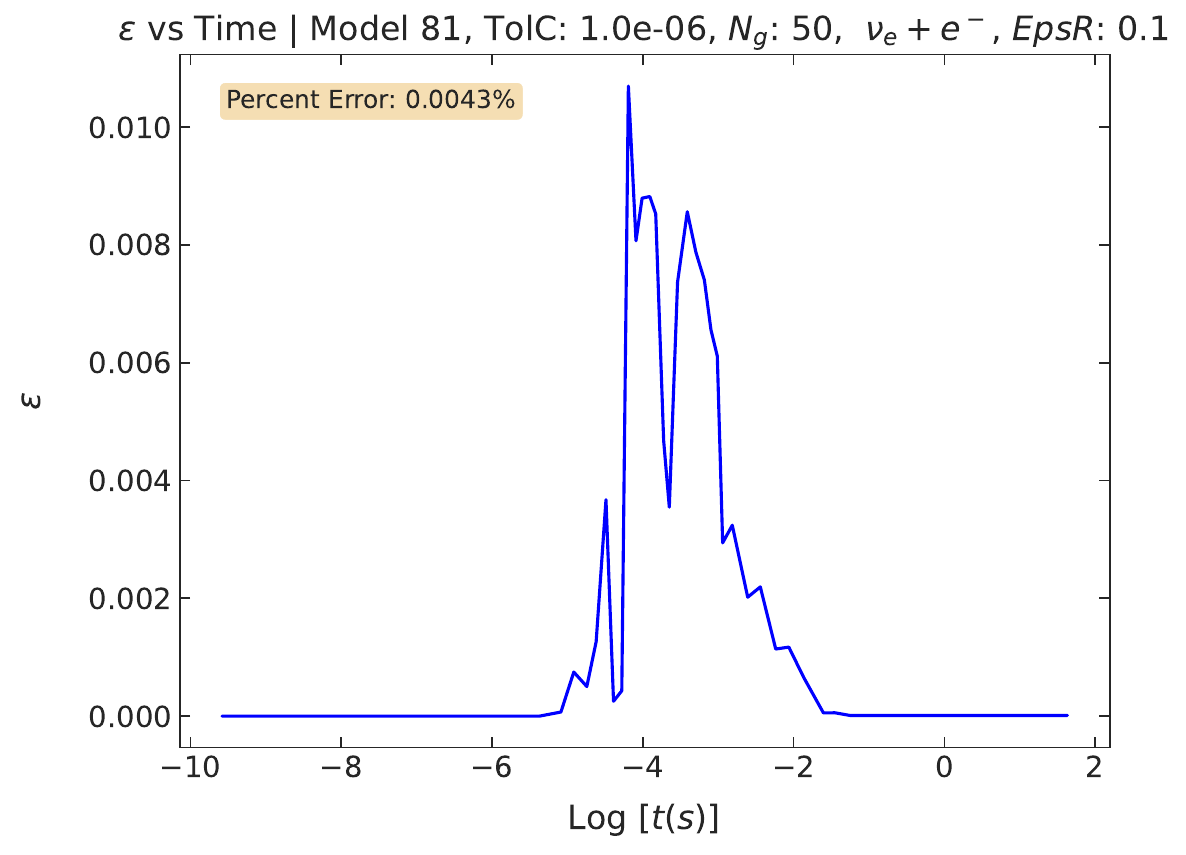}
    \caption{RMS vs time for Model 81 - Intermediate Case.}
    \label{fig:model81rmsint}
\end{figure}

\begin{figure}[H]
    \centering
    \includegraphics[width=5in]{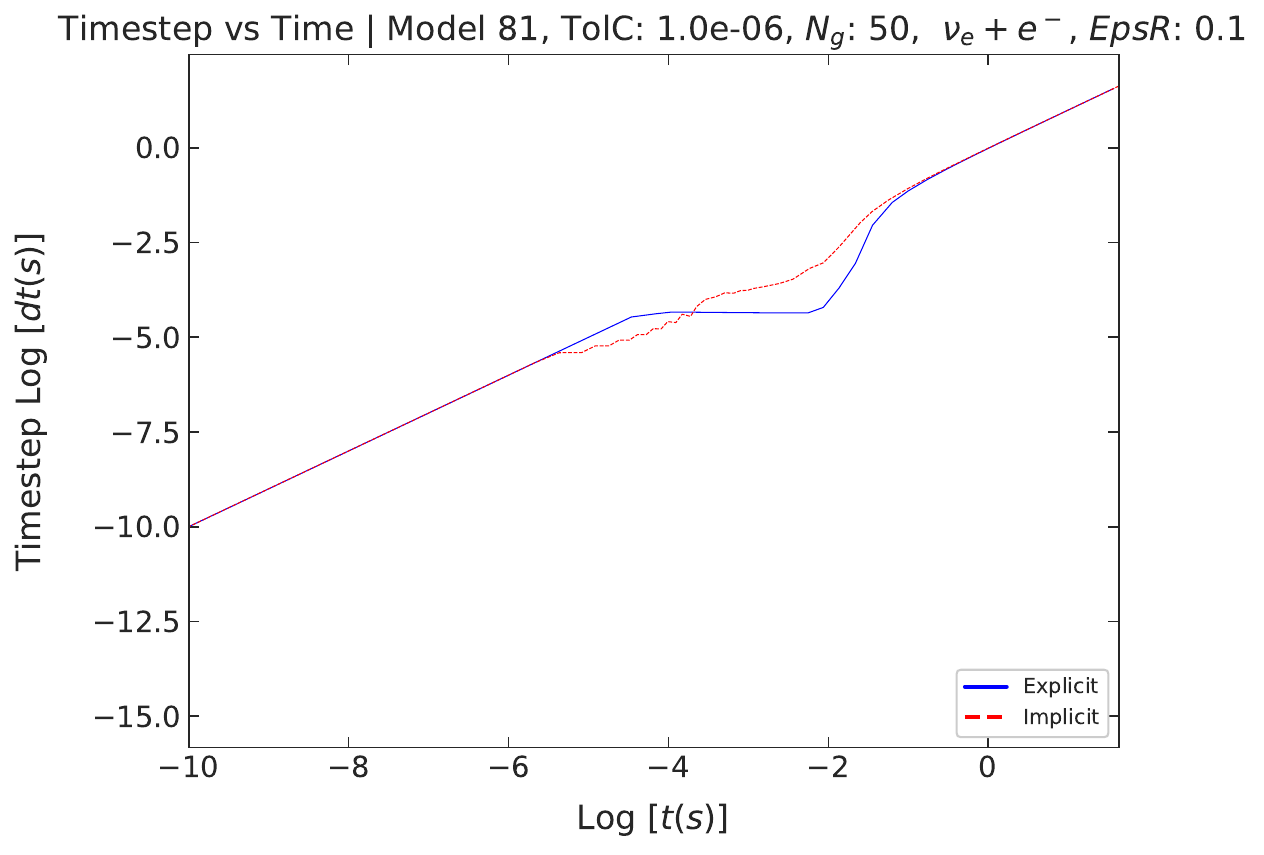}
    \caption{Time step vs time for Model 81 - Intermediate Case.}
    \label{fig:model81timestepint}
\end{figure}

For our intermediate case we can adjust TolC to $10^{-6}$ which loosens the particle conservation, taking faster time steps against the same backward Euler comparison.
\subsubsection{Fast Case}
\label{subsubsec:model81fast}

\begin{figure}[H]
    \centering
    \includegraphics[width=5in]{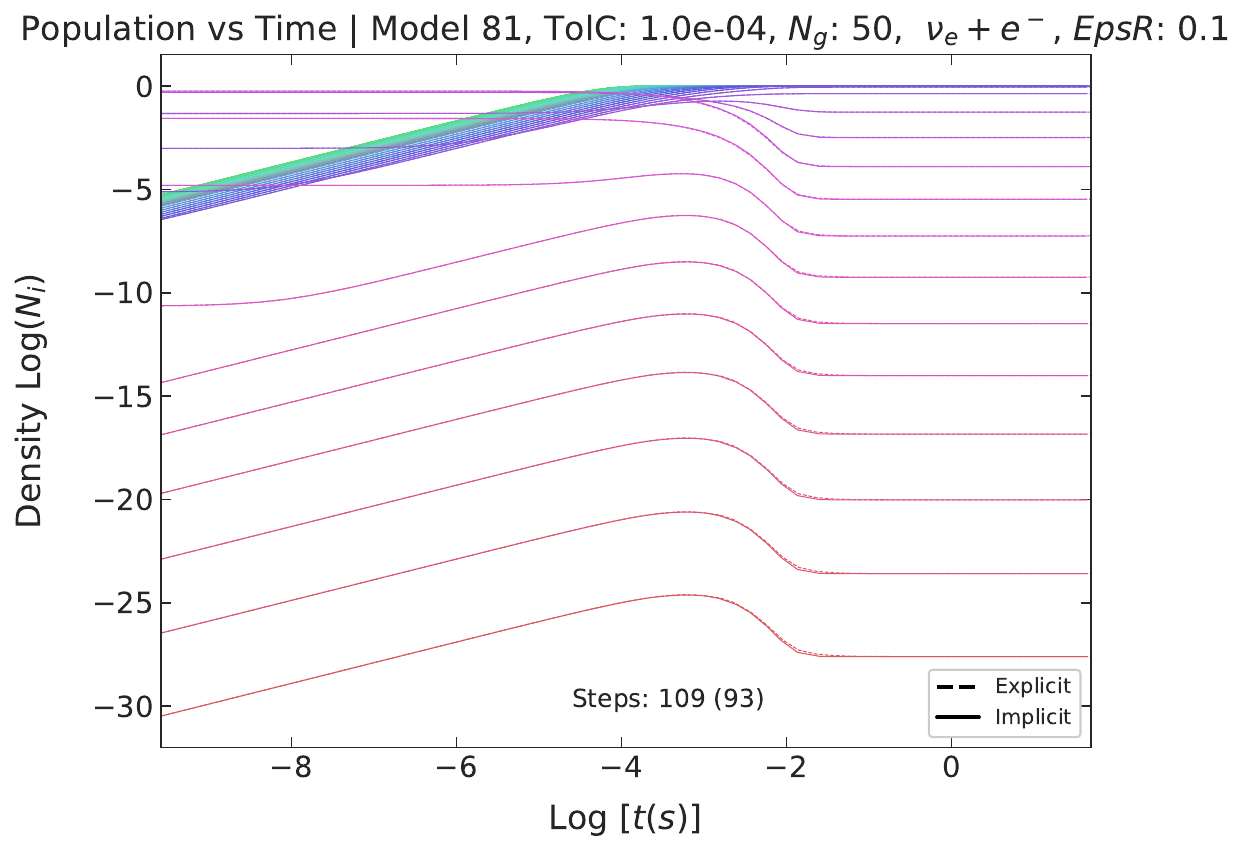}
    \caption{Population vs time for Model 81 - Fast Case.}
    \label{fig:model81popfast}
\end{figure}

\begin{figure}[H]
    \centering
    \includegraphics[width=5in]{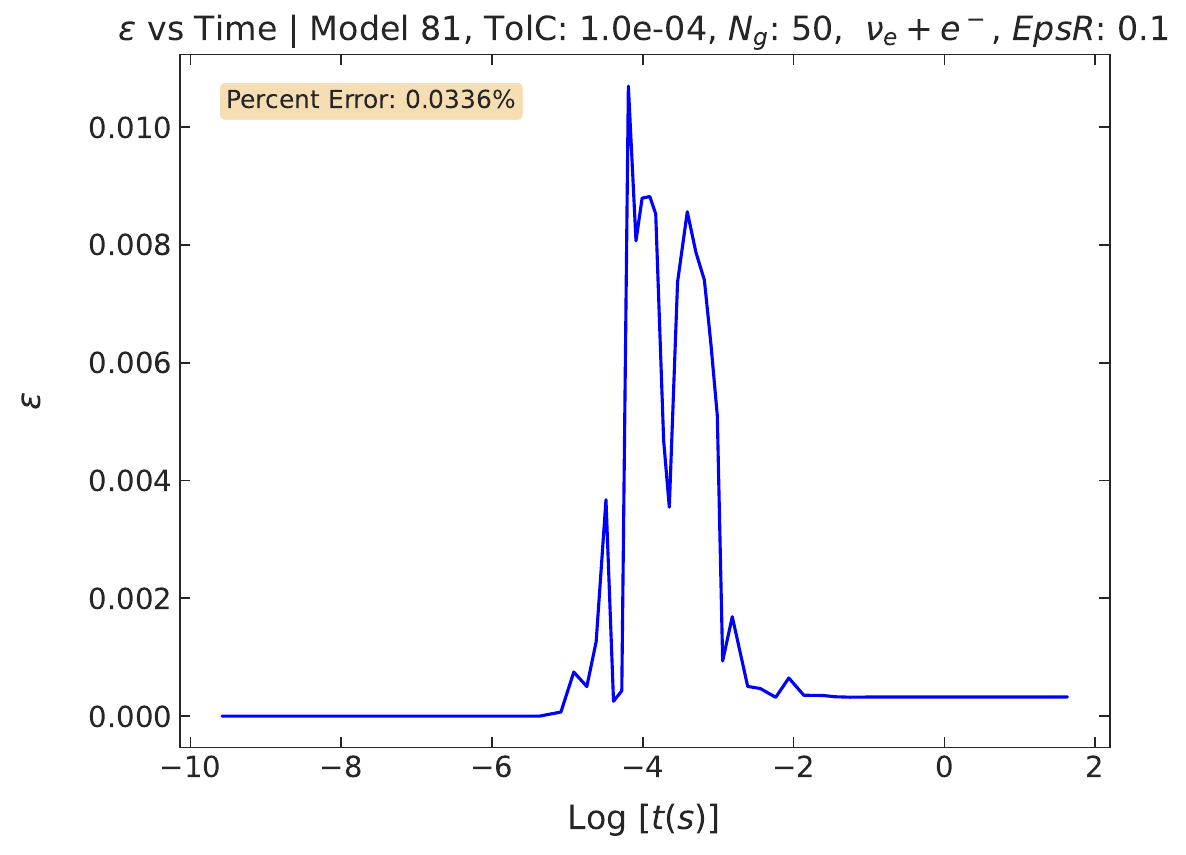}
    \caption{RMS vs time for Model 81 - Fast Case.}
    \label{fig:model81rmsfast}
\end{figure}

\begin{figure}[H]
    \centering
    \includegraphics[width=5in]{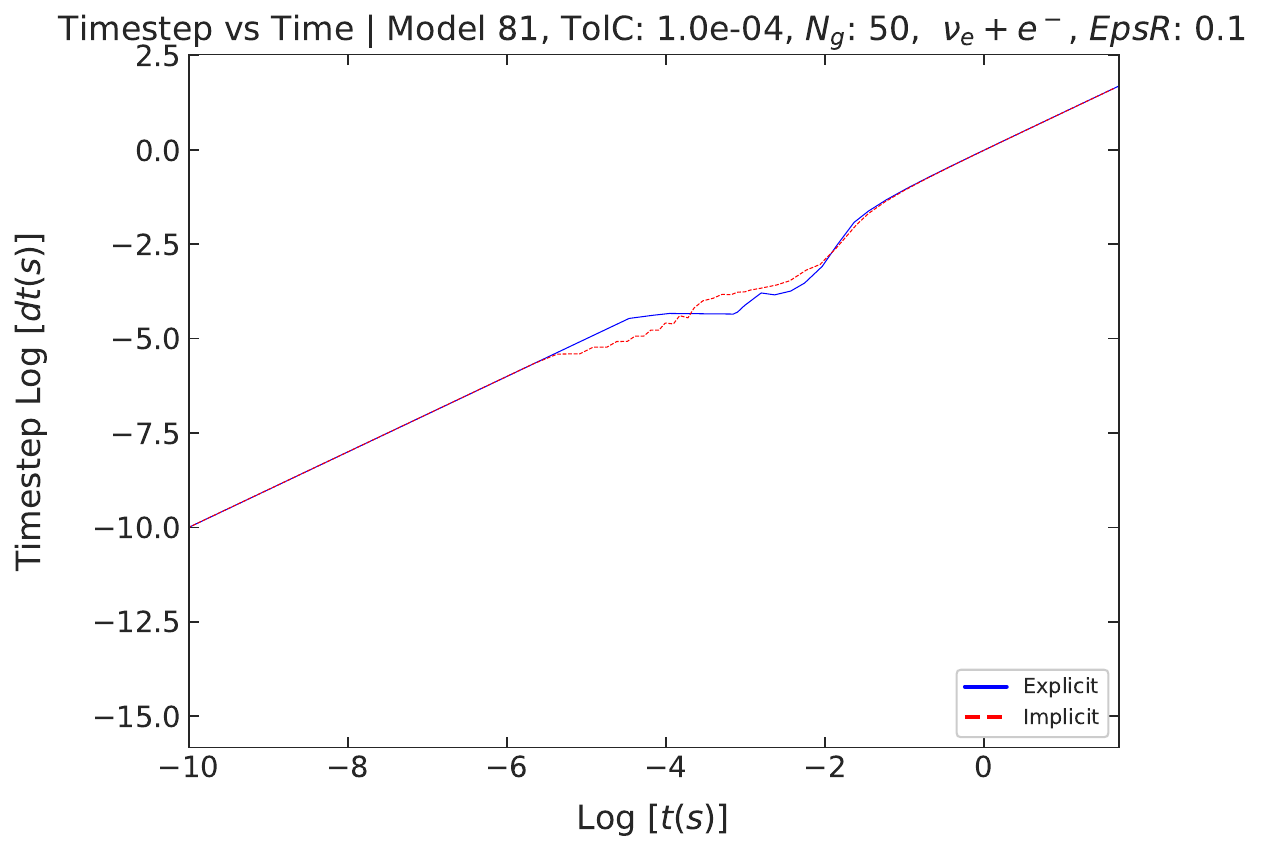}
    \caption{Time step vs time for Model 81 - Fast Case.}
    \label{fig:model81timestepfast}
\end{figure}

Similarly, we adjust TolC to $10^{-4}$ which loosens the conservation of particles even more taking only 109 time steps for explicit.

\begin{table}[H]
\centering
\begin{tabular}{lccc}
\hline
Case & Steps Explicit & Steps Implicit & Error (\%) \\
\hline
Accurate & 588 & 93 & 0.0035 \\
Intermediate & 313 & 93 & 0.0043 \\
Fast & 109 & 93 & 0.0336 \\
\hline
\end{tabular}
\caption{Summary of steps and error for Model 81.}
\label{tab:model81summary}
\end{table}
Here the error from Eq. \ref{epsilon} occurs in restricted areas of integration time. This is shown in fig. \ref{fig:model81rmsacc}, \ref{fig:model81rmsint}, and fig. \ref{fig:model81rmsfast} where the error accumulates almost entirely between $t \approx 10^{-6} -  \approx 10^{-2}$ seconds. 

 When using operator-split coupling in hydrodynamics, the continuous integration in fig. \ref{fig:model81rmsacc} can be segmented into sequential, piece wise integrations within each fluid zone. Each segment corresponds to integrating the network over one hydrodynamical timestep in the zone. Consequently, significant errors from the neutrino network for that zone are confined to a narrow range of hydroynamical integration steps, specifically between \(10^{-6}\) and \(10^{-2}\) seconds. This observation allows for potential optimization of speed versus accuracy on a per-zone basis within the hydro-kinetic system, though such optimizations are left for future exploration.

Furthermore, it's crucial to understand how neutrino network errors, as defined in Eq. \ref{epsilon} and illustrated in fig. \ref{fig:model81rmsacc} and in table \ref{tab:model81summary}, translate into overall errors in the coupled kinetic-hydrodynamical system due to specific algebraic approximations. The error calculated from Eq. \ref{epsilon} represents the RMS error per unit time across the relevant times where \(R(t)\) is non-zero. Importantly, there is no error for hydro integration steps that do not intersect with significant \(R(t)\) regions shown in fig.  \ref{fig:model81rmsacc}. For those hydro time steps that do intersect in these significant regions, the error introduced by network approximations per hydro time step \(\Delta t_{hydro}\) will depend on the integrated value of \(R(t)\) during that time step, influenced by factors such as hydro integration methods, zone sizes, and fluid characteristics at the time. This makes the error case-specific. However, the relatively low values of \(E\) chosen in fig. \ref{fig:model81rmsacc} indicate that the maximum overall error can be limited to a few percent or less while still achieving sufficient speed, when applied to specific scenarios.

\subsection{Model 186 Analysis}
\label{subsec:model186}

Similarly, we can proceed for Model 186

\subsubsection{Accurate Case}
\label{subsubsec:model186accurate}

\begin{figure}[H]
    \centering
    \includegraphics[width=5in]{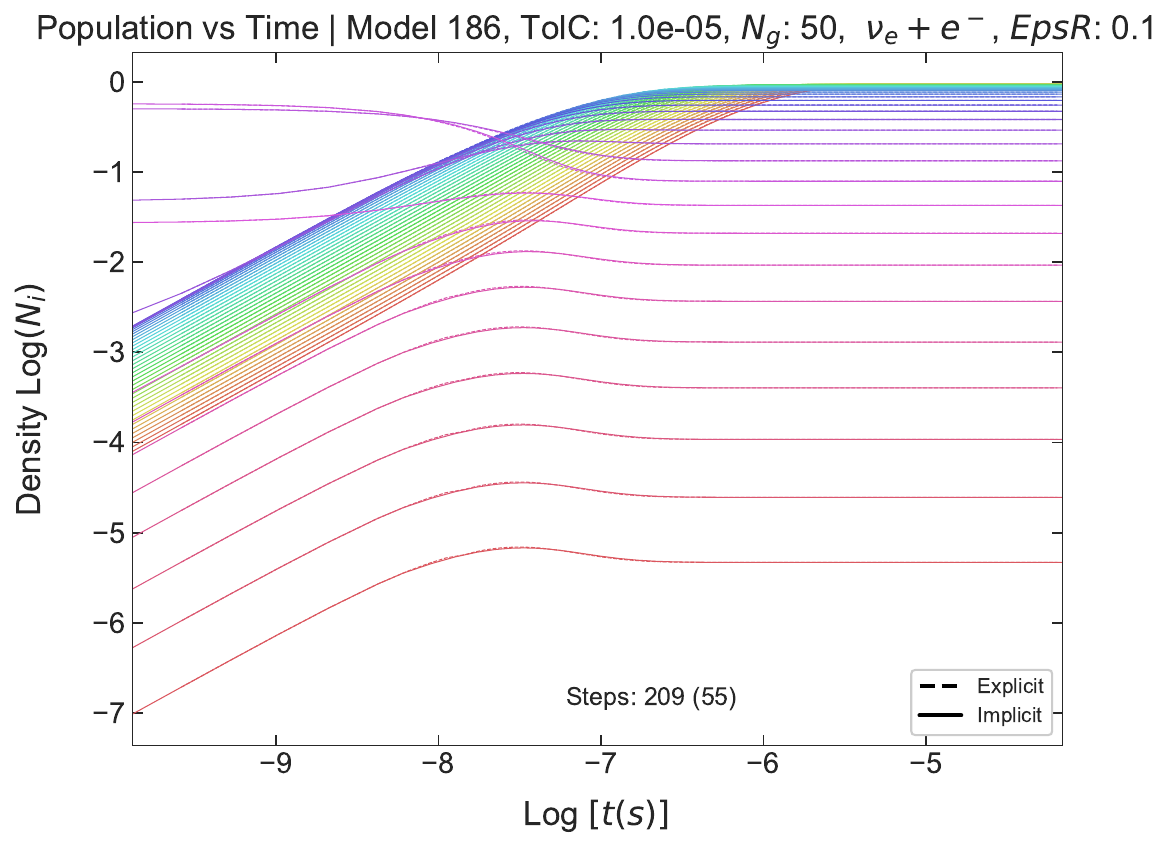}
    \caption{Population vs time for Model 186 - Accurate Case.}
    \label{fig:model186popacc}
\end{figure}

\begin{figure}[H]
    \centering
    \includegraphics[width=5in]{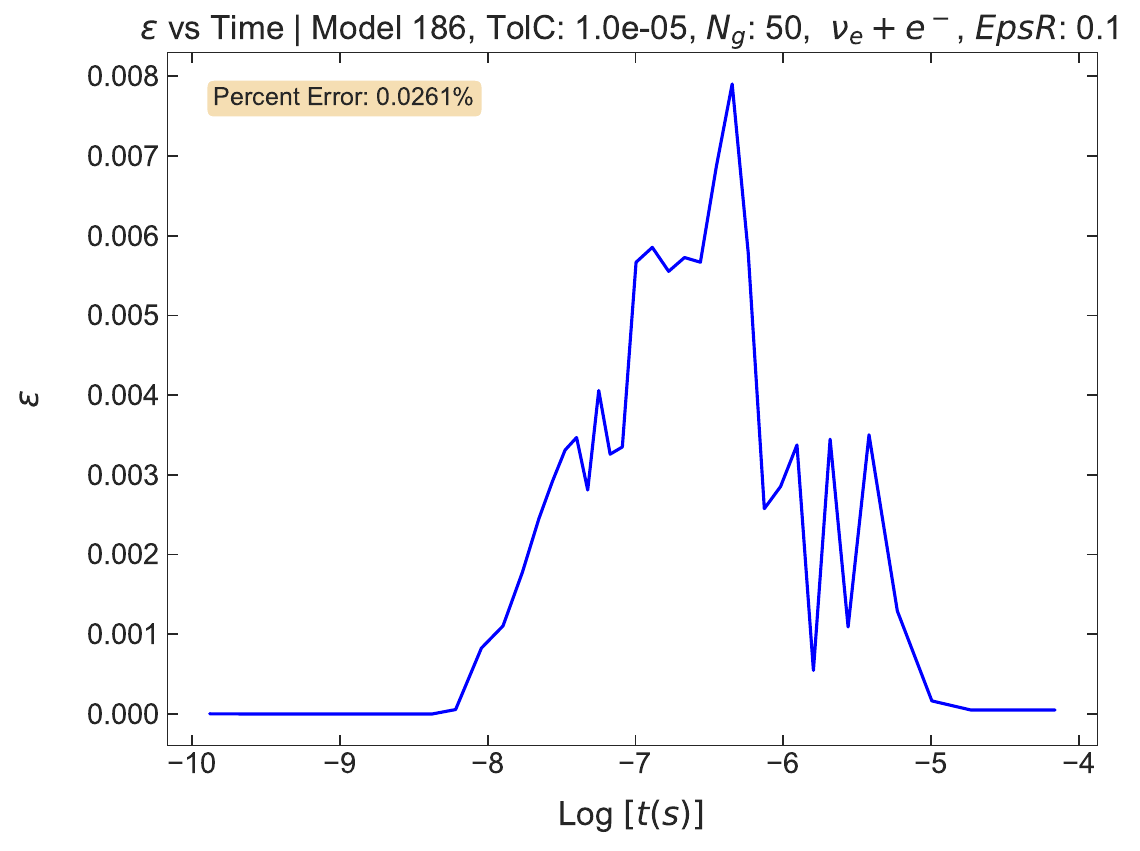}
    \caption{RMS vs time for Model 186 - Accurate Case.}
    \label{fig:model186rmsacc}
\end{figure}

\begin{figure}[H]
    \centering
    \includegraphics[width=5in]{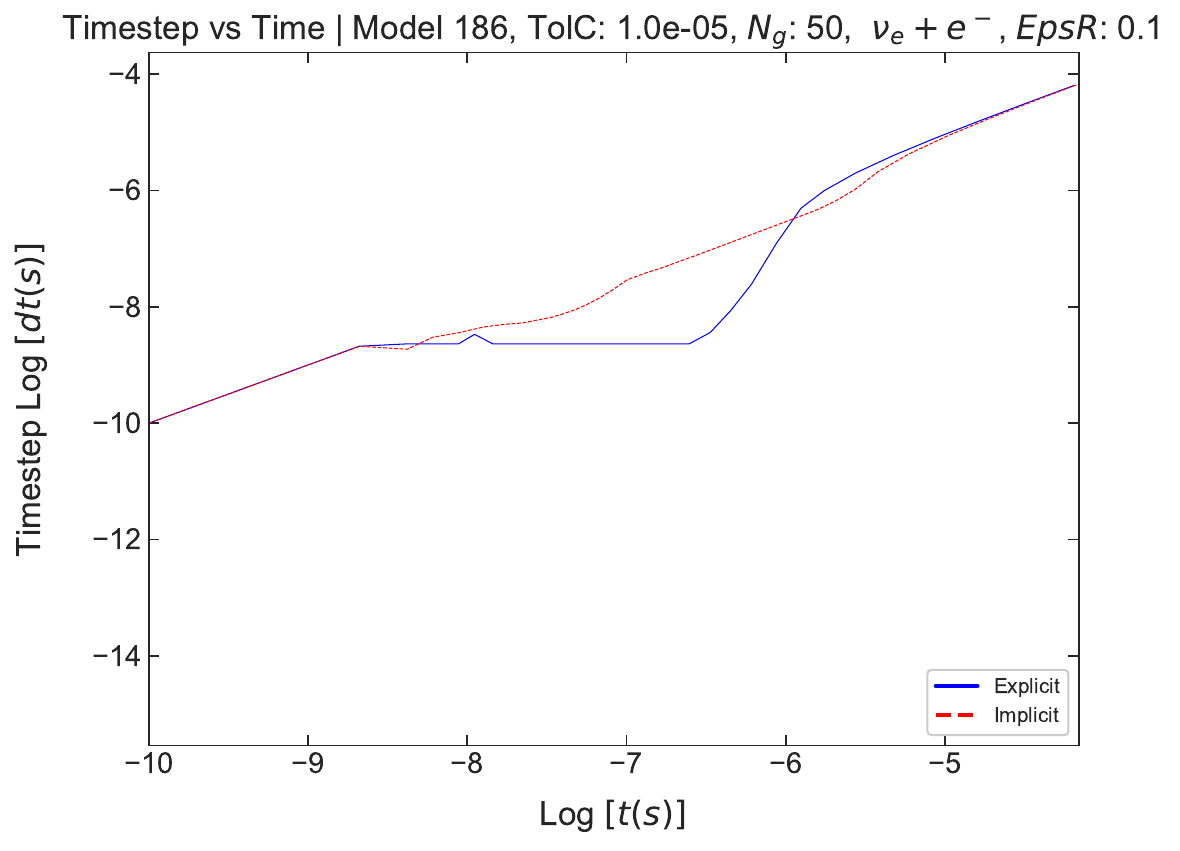}
    \caption{Timestep vs time for Model 186 - Accurate Case.}
    \label{fig:model186timestepacc}
\end{figure}

\subsubsection{Intermediate Case}
\label{subsubsec:model186intermediate}

\begin{figure}[H]
    \centering
    \includegraphics[width=5in]{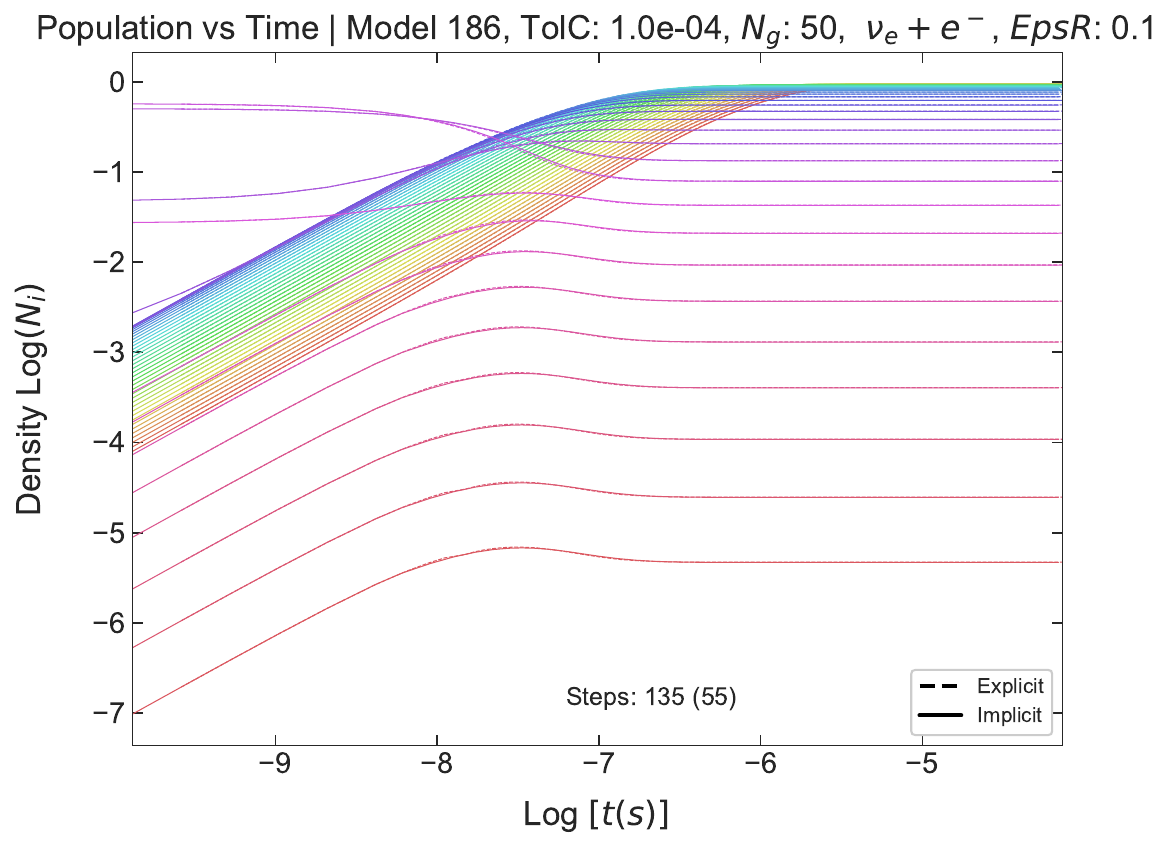}
    \caption{Population vs time for Model 186 - Intermediate Case.}
    \label{fig:model186popint}
\end{figure}

\begin{figure}[H]
    \centering
    \includegraphics[width=5in]{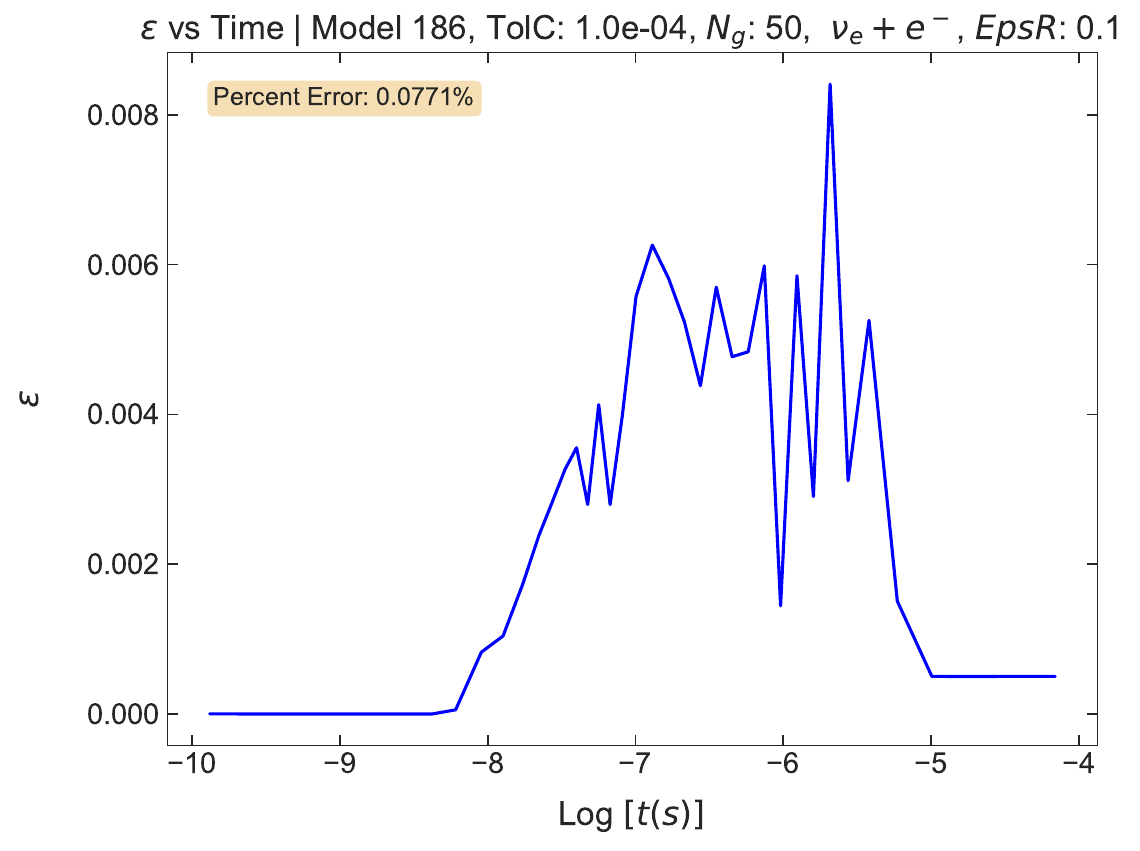}
    \caption{RMS vs time for Model 186 - Intermediate Case.}
    \label{fig:model186rmsint}
\end{figure}

\begin{figure}[H]
    \centering
    \includegraphics[width=5in]{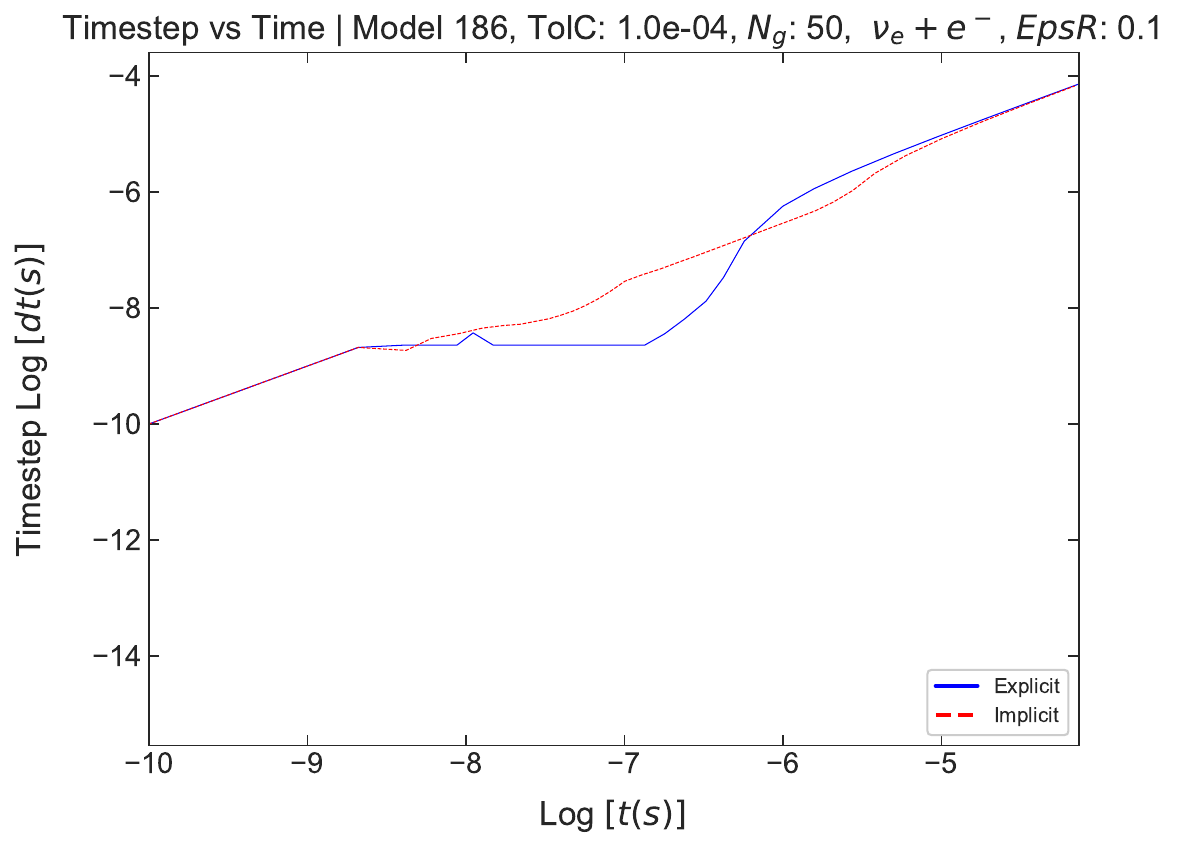}
    \caption{Timestep vs time for Model 186 - Intermediate Case.}
    \label{fig:model186timestepint}
\end{figure}

\subsubsection{Fast Case}
\label{subsubsec:model186fast}

\begin{figure}[H]
    \centering
    \includegraphics[width=5in]{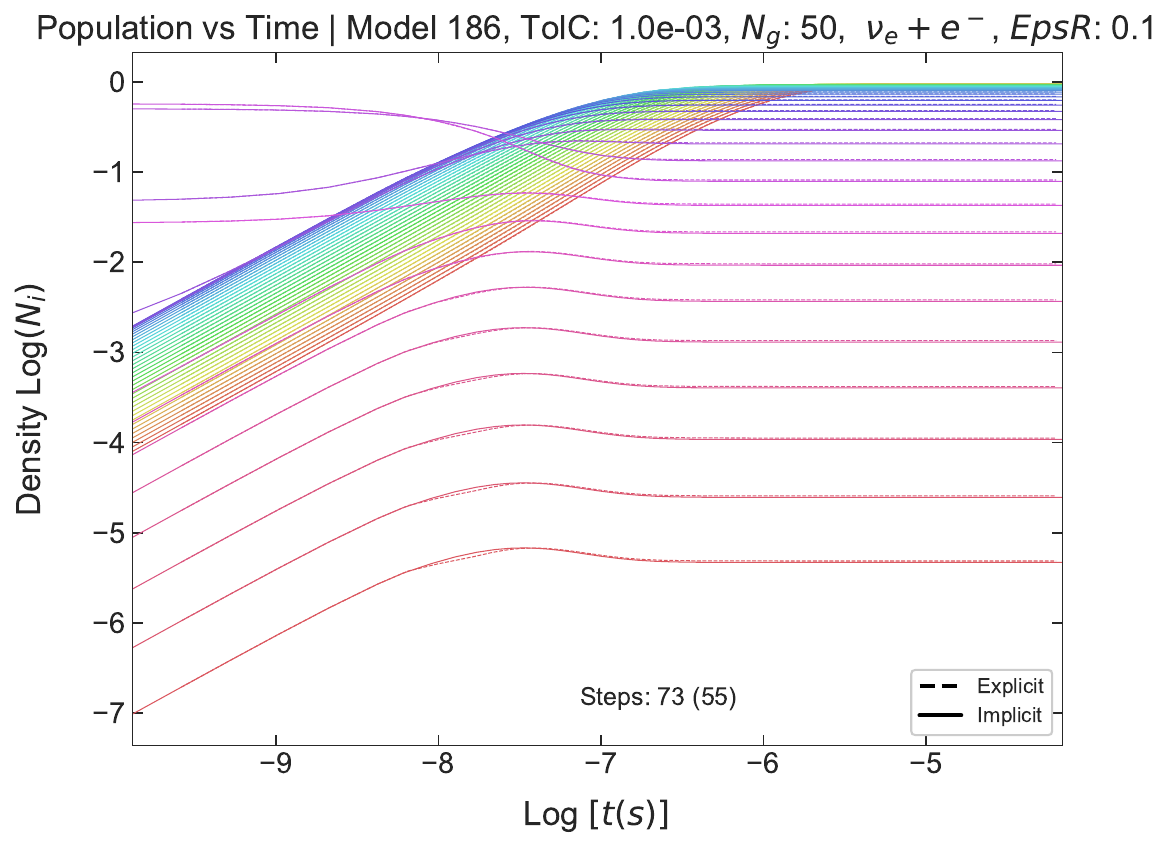}
    \caption{Population vs time for Model 186 - Fast Case.}
    \label{fig:model186popfast}
\end{figure}

\begin{figure}[H]
    \centering
    \includegraphics[width=5in]{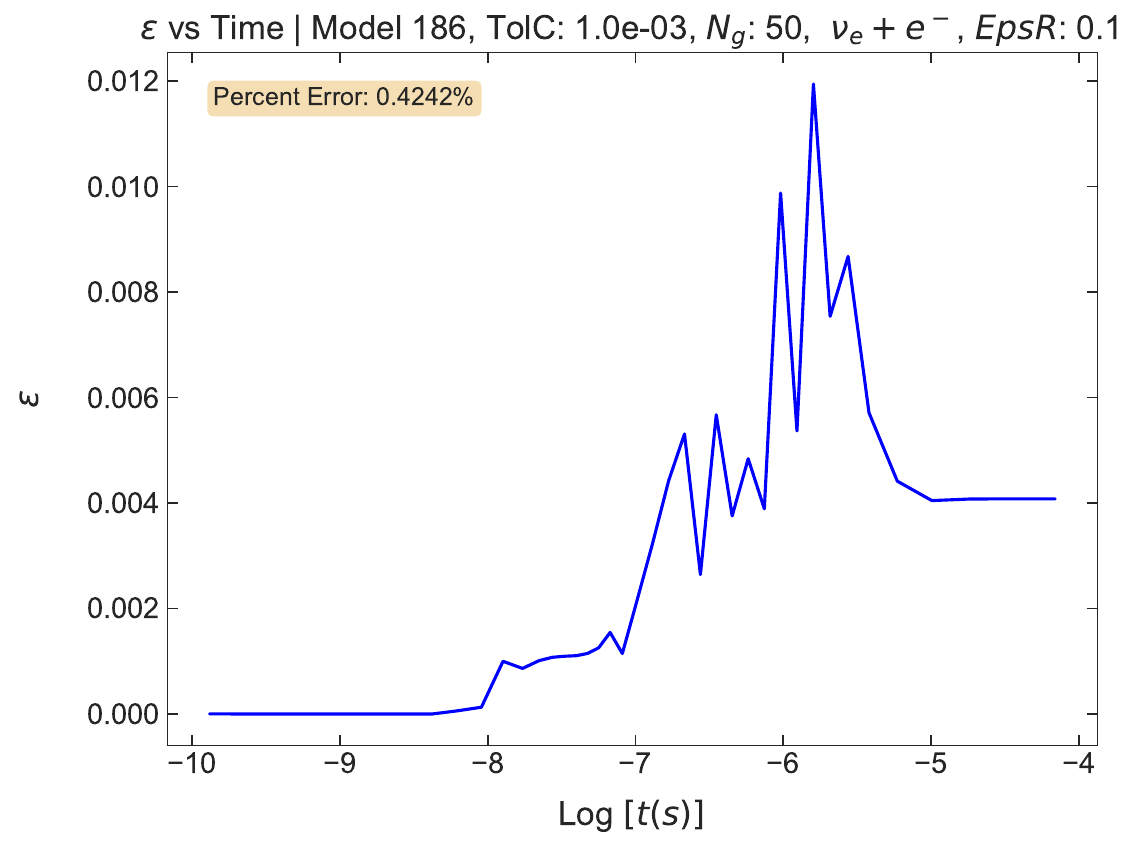}
    \caption{RMS vs time for Model 186 - Fast Case.}
    \label{fig:model186rmsfast}
\end{figure}

\begin{figure}[H]
    \centering
    \includegraphics[width=5in]{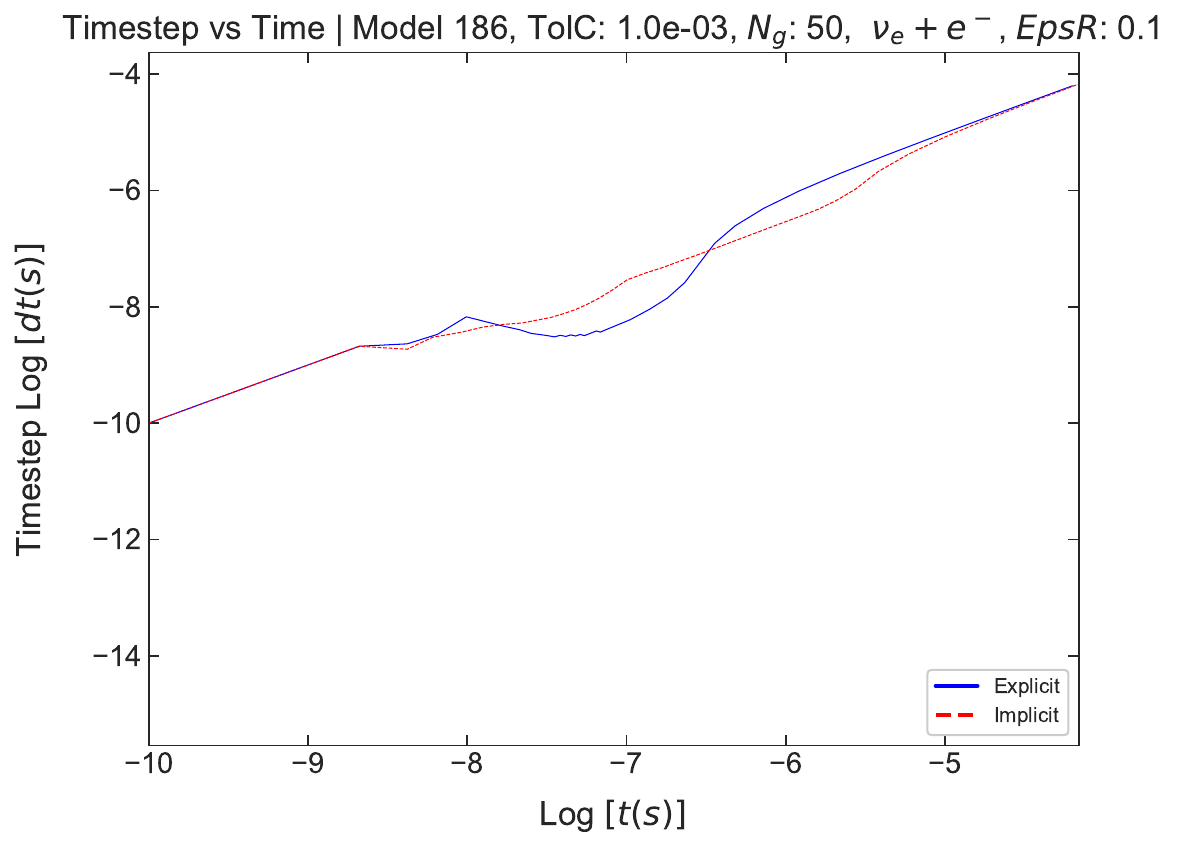}
    \caption{Timestep vs time for Model 186 - Fast Case.}
    \label{fig:model186timestepfast}
\end{figure}

\begin{table}[H]
\centering
\begin{tabular}{lccc}
\hline
Case & Steps Explicit & Steps Implicit & Error (\%) \\
\hline
Accurate & 209 & 55 & 0.0261 \\
Intermediate & 175 & 55 & 0.0771 \\
Fast & 73 & 55 & 0.4242 \\
\hline
\end{tabular}
\caption{Summary of steps and error for Model 186.}
\label{tab:model186summary}
\end{table}

\subsection{Model 213 Analysis}
\label{subsec:model213}

Similarly, we can provide an analysis of Model 213, for comparison.
\subsubsection{Accurate Case}
\label{subsubsec:model213accurate}

\begin{figure}[H]
    \centering
    \includegraphics[width=5in]{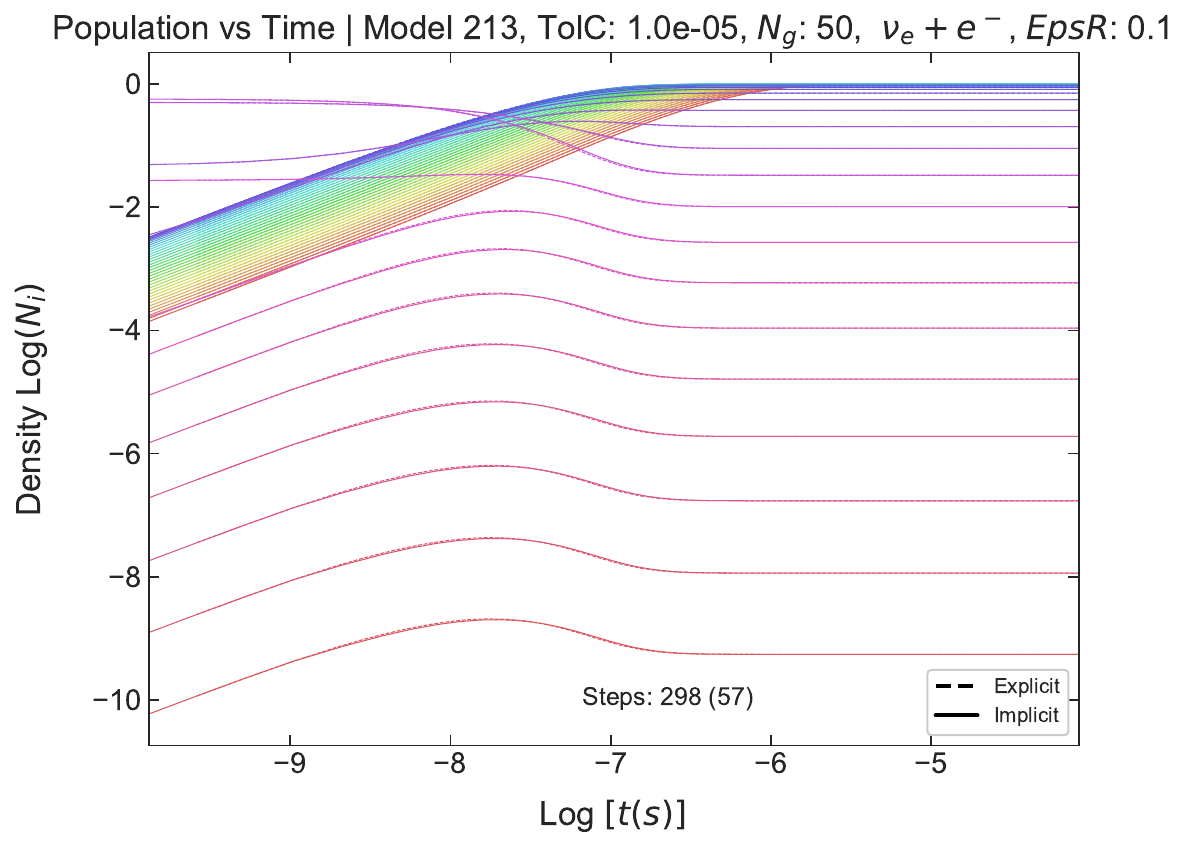}
    \caption{Population vs time for Model 213 - Accurate Case.}
    \label{fig:model213popacc}
\end{figure}

\begin{figure}[H]
    \centering
    \includegraphics[width=5in]{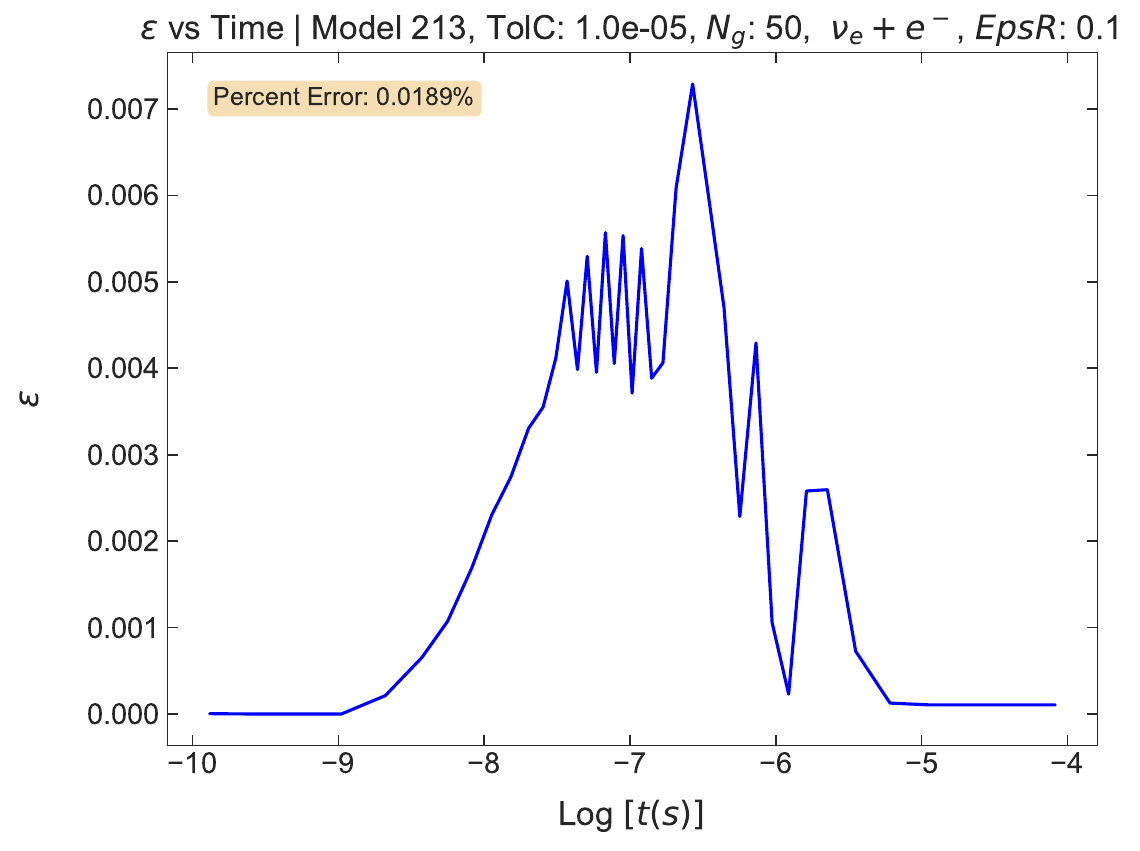}
    \caption{RMS vs time for Model 213 - Accurate Case.}
    \label{fig:model213rmsacc}
\end{figure}

\begin{figure}[H]
    \centering
    \includegraphics[width=5in]{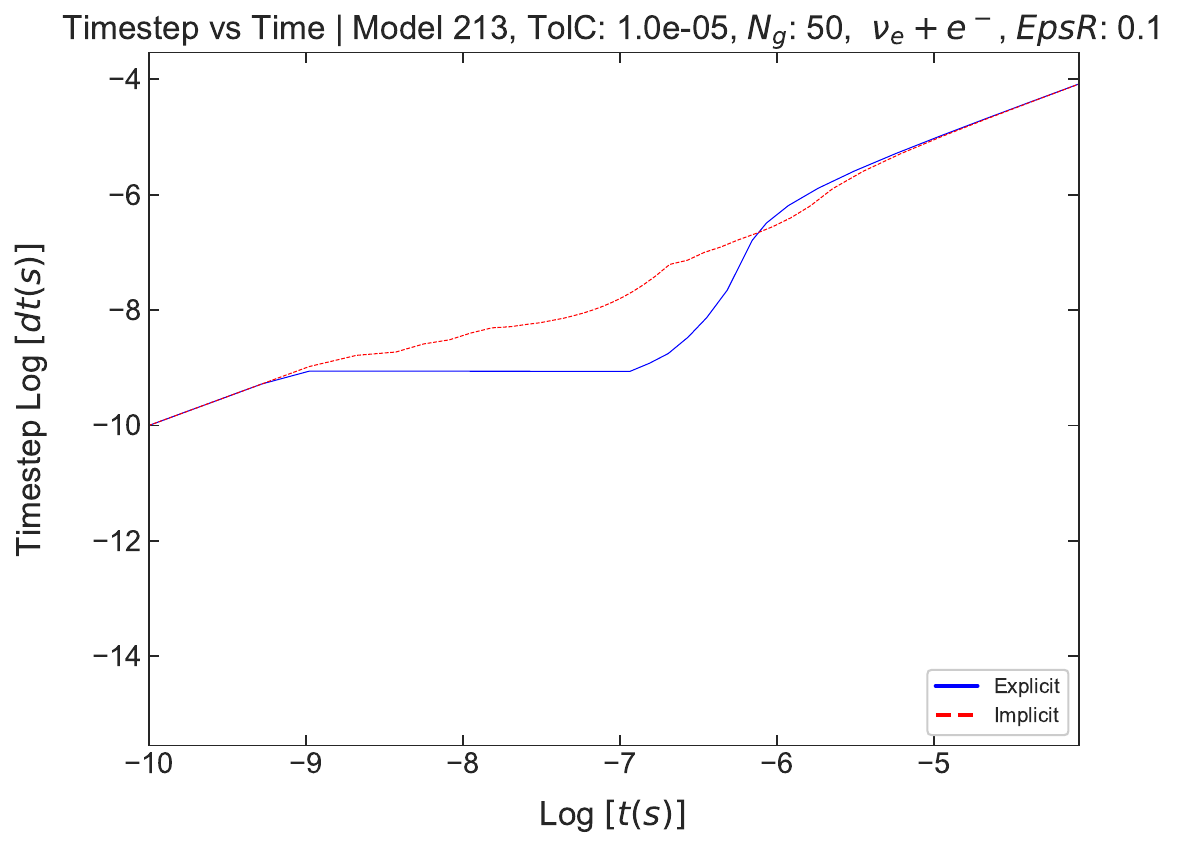}
    \caption{Time step vs time for Model 213 - Accurate Case.}
    \label{fig:model213timestepacc}
\end{figure}

\subsubsection{Intermediate Case}
\label{subsubsec:model213intermediate}

\begin{figure}[H]
    \centering
    \includegraphics[width=5in]{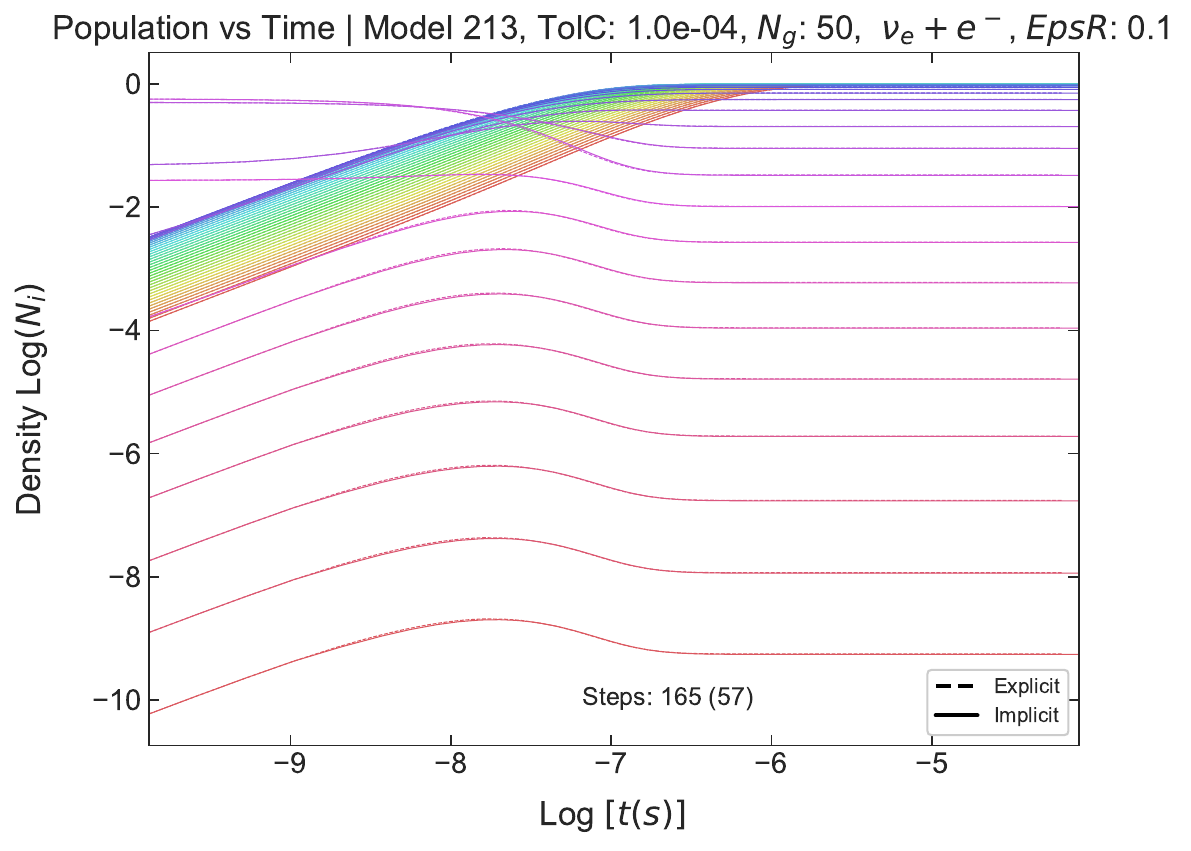}
    \caption{Population vs time for Model 213 - Intermediate Case.}
    \label{fig:model213popint}
\end{figure}

\begin{figure}[H]
    \centering
    \includegraphics[width=5in]{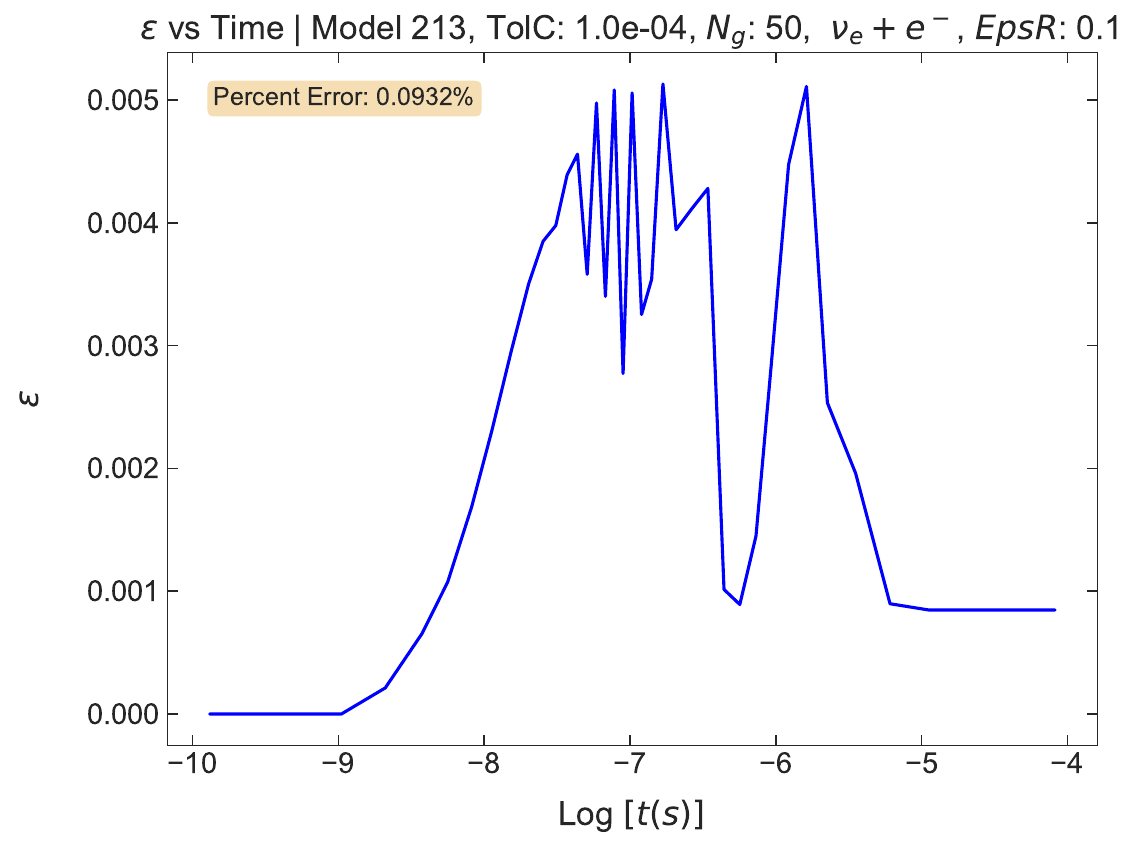}
    \caption{RMS vs time for Model 213 - Intermediate Case.}
    \label{fig:model213rmsint}
\end{figure}

\begin{figure}[H]
    \centering
    \includegraphics[width=5in]{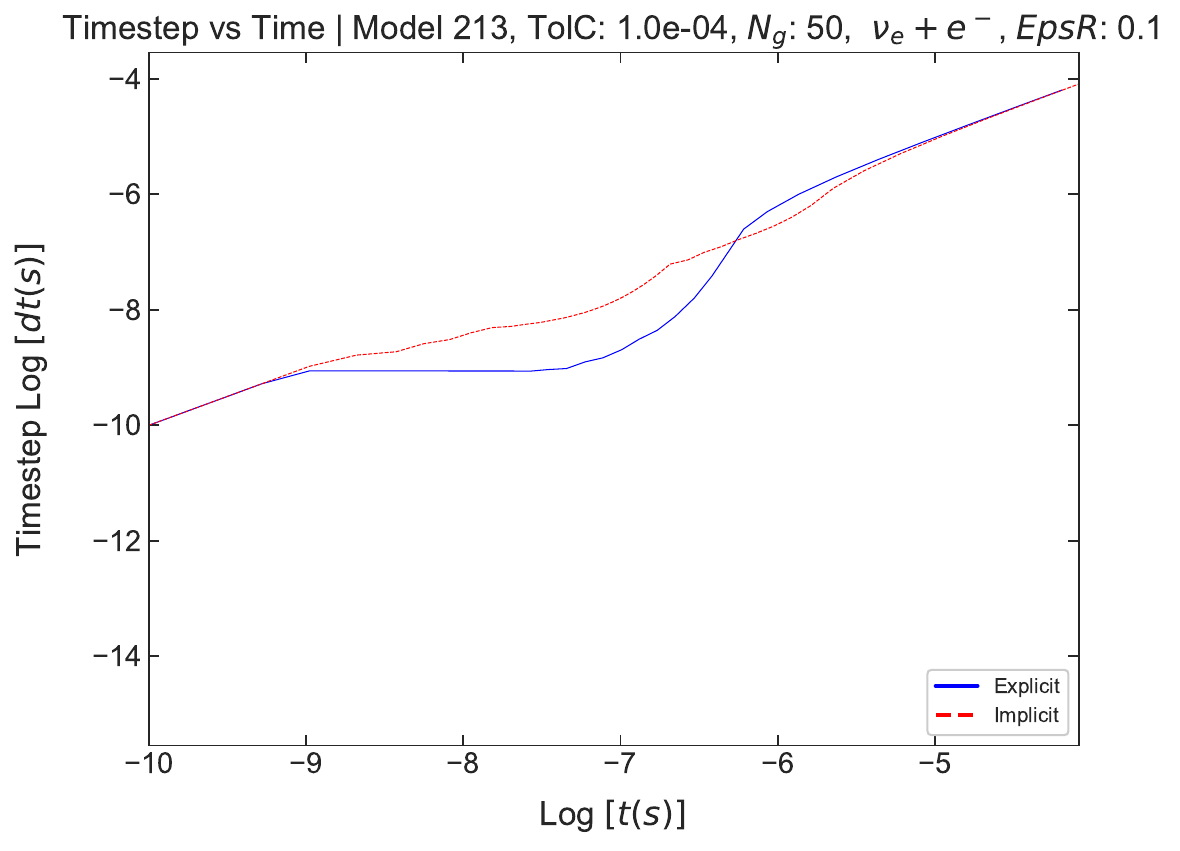}
    \caption{Time step vs time for Model 213 - Intermediate Case.}
    \label{fig:model213timestepint}
\end{figure}

\subsubsection{Fast Case}
\label{subsubsec:model213fast}

\begin{figure}[H]
    \centering
    \includegraphics[width=5in]{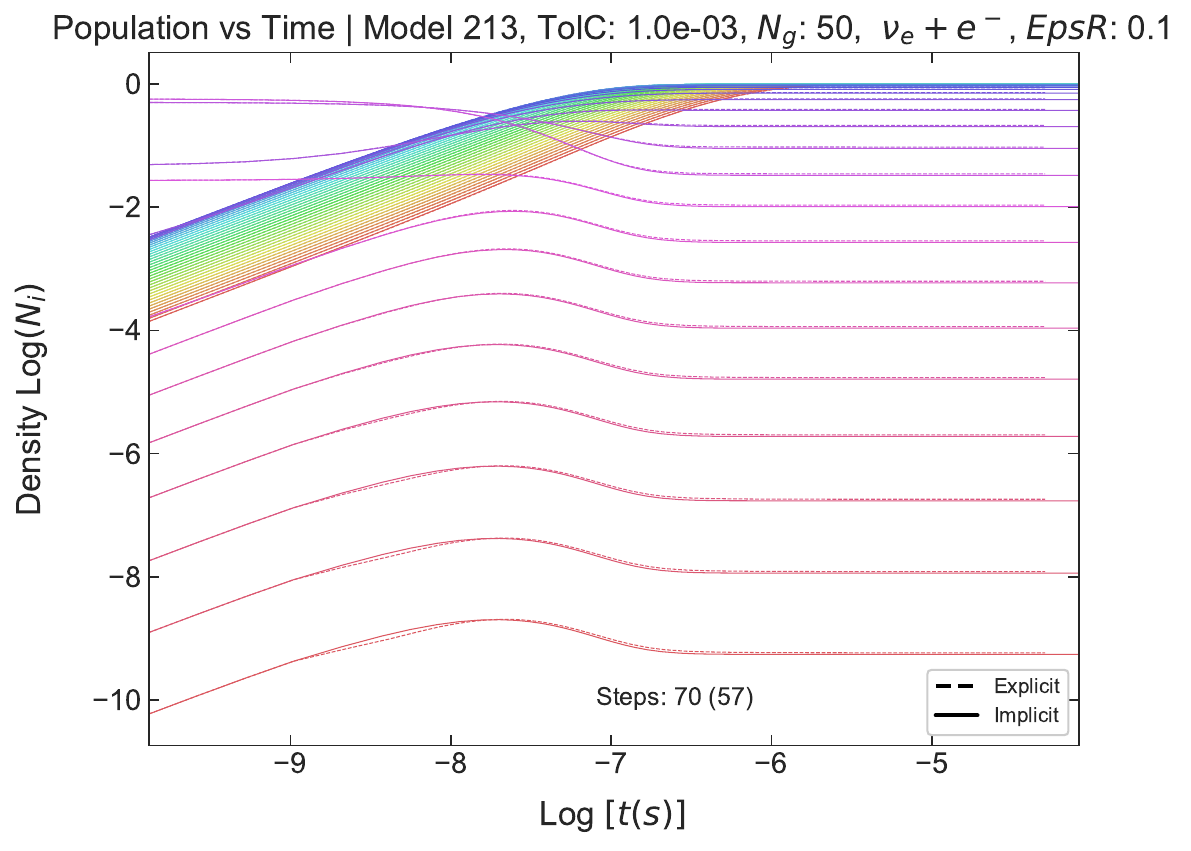}
    \caption{Population vs time for Model 213 - Fast Case.}
    \label{fig:model213popfast}
\end{figure}

\begin{figure}[H]
    \centering
    \includegraphics[width=5in]{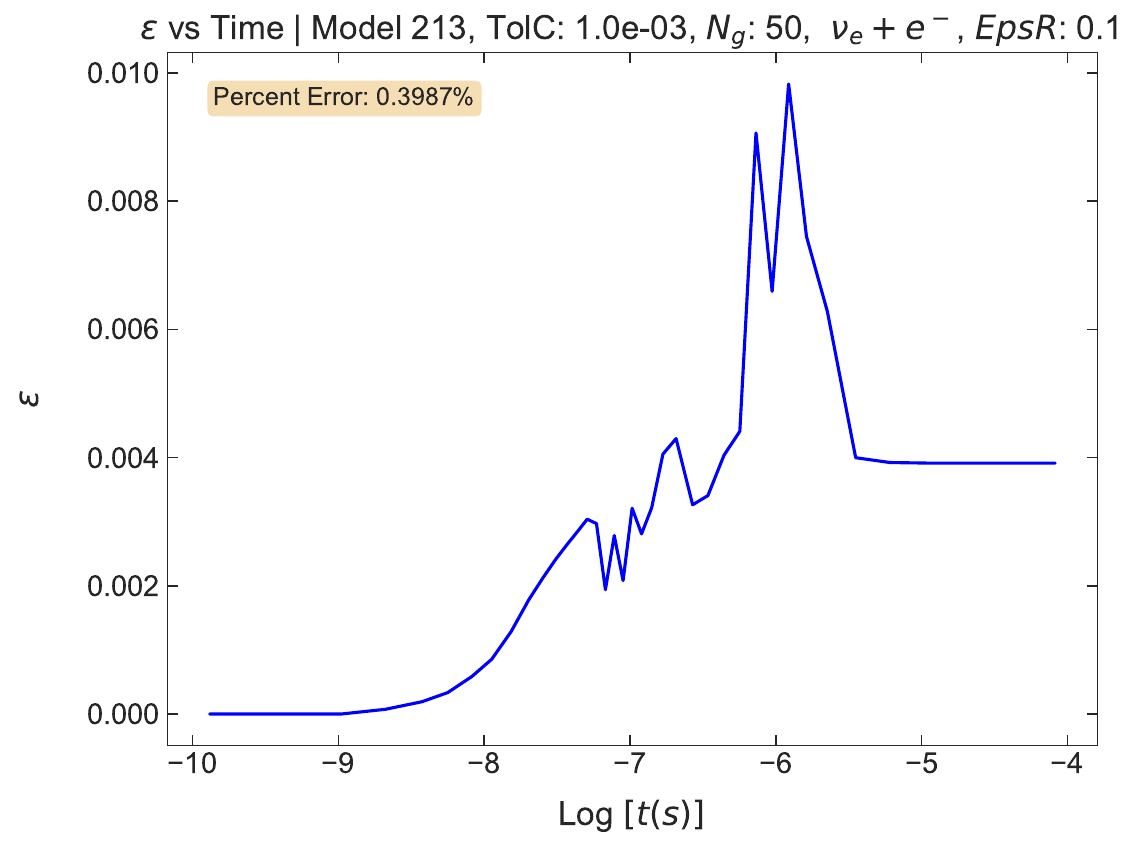}
    \caption{RMS vs time for Model 213 - Fast Case.}
    \label{fig:model213rmsfast}
\end{figure}

\begin{figure}[H]
    \centering
    \includegraphics[width=5in]{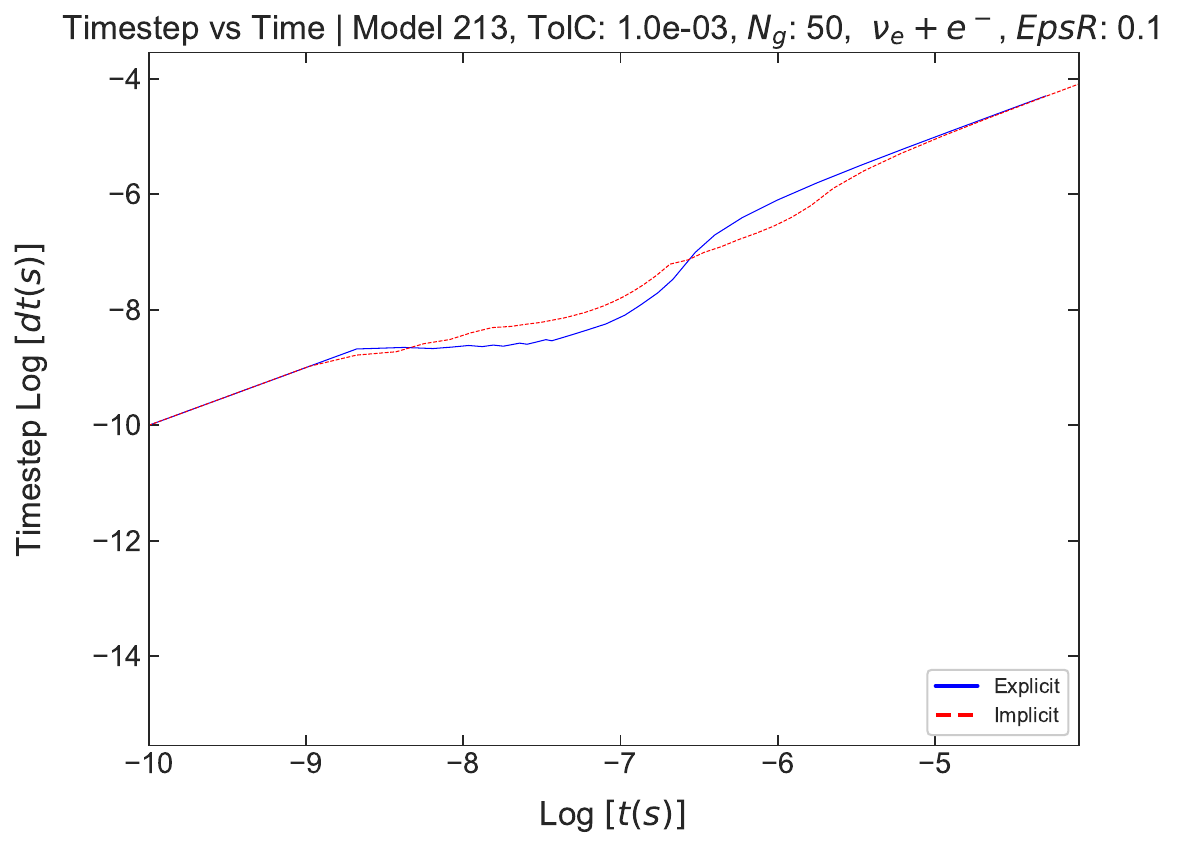}
    \caption{Time step vs time for Model 213 - Fast Case.}
    \label{fig:model213timestepfast}
\end{figure}

\begin{table}[H]
\centering
\begin{tabular}{lccc}
\hline
Case & Steps Explicit & Steps Implicit & Error (\%) \\
\hline
Accurate & 298 & 57 & 0.0189 \\
Intermediate & 165 & 57 & 0.0932 \\
Fast & 70 & 57 & 0.3987 \\
\hline
\end{tabular}
\caption{Summary of steps and error for Model 213.}
\label{tab:model213summary}
\end{table}
\subsection{Speed vs Accuracy}

We can see in sections \ref{subsec:model81}, \ref{subsec:model186}, and  \ref{subsec:model213} that we demonstrate that as we increase the speed of our explicit algorithms by decreasing our steps through increasing our tolerance conditions (TolC), we obtain a trade off of speed vs accuracy. 
The TolC conditions are tabulated in table \ref{tab:tolc_conditions}.

\begin{table}[H]
\centering
\begin{tabular}{lccc}
\hline
Model / Case & Accurate & Intermediate & Fast \\
\hline
Model 81 & \(10^{-8}\) & \(10^{-6}\) & \(10^{-4}\) \\
Model 186 & \(10^{-5}\) & \(10^{-4}\) & \(10^{-3}\) \\
Model 213 & \(10^{-5}\) & \(10^{-4}\) & \(10^{-3}\) \\
\hline
\end{tabular}
\caption{Tolerance Conditions (TolC) for Each Model and Case.}
\label{tab:tolc_conditions}
\end{table}
We can necessitate a trade off of the accuracy by comparing the steps and error in tables \ref{tab:model81summary}, \ref{tab:model186summary}, \ref{tab:model213summary}.
For speed, we can adopt Eq. \ref{wallClockRatio} for a 50-Bin Network, as we see in \ref{tab:speedup} the speedup factor is approximately equal. 

In the case of Model 213, the backward Euler implicit calculation took 57 integration steps. Referring to our earlier calculations in figures \ref{fig:model213popacc}, \ref{fig:model213popint}, \ref{fig:model213popfast}   the number of explicit asymptotic (Asy) integration steps taken for the Accurate, Intermediate, and Fast cases were 298, 165, and 70, respectively.

For the Accurate case of Model 213, we calculate the ratio of wall clock times for asymptotic integration relative to implicit integration using Eq. \ref{wallClockRatio}.
\[
\frac{\Delta t_{\mathrm{EA}}}{\Delta t_{\mathrm{BE}}} \simeq 0.25 \times \frac{298}{57} \simeq 1.31.
\]
This indicates that the explicit asymptotic integration is approximately \((1.31)^{-1} \simeq 0.77\) times as fast as the backward Euler calculation. This result implies that even though the explicit method requires approximately \(5.2\) times as many integration steps as the implicit method, it computes each step about \(4\) times faster on the same system due to much easier matrix vector multiplication instead of costly matrix inversions.

For the intermediate calculation in Model 213, we proceed similarly:
\[
\frac{\Delta t_{\mathrm{EA}}}{\Delta t_{\mathrm{BE}}} \simeq 0.25 \times \frac{165}{57} \simeq 0.72,
\]
resulting in the calculation being about \((0.72)^{-1} \simeq 1.38\) times faster than the implicit \(\mathrm{BE}\) calculation. Here, the explicit method requires roughly \(2.9\) times as many steps as the implicit method.

Lastly, for the fast model 213 example, we find:
\[
\frac{\Delta t_{\mathrm{EA}}}{\Delta t_{\mathrm{BE}}} \simeq 0.25 \times \frac{70}{57} \simeq 0.31,
\]
indicating that the fast asymptotic calculation is estimated to be about \((0.31)^{-1} \simeq 3.26\) times faster than the implicit \(\mathrm{BE}\) calculation. In this scenario, the explicit method only requires about \(1.2\) times more steps than the implicit method.

So, while the explicit asymptotic integration often requires more steps, its faster computation per step on equivalent machines leads to significant reductions in overall computational time compared to the backward Euler method. This analysis shows how leveraging computational efficiency can compensate for an increased number of steps in certain dynamic models. 

These calculations lead to the following table \ref{tab:speed_factors}, which summarizes the wall clock times for the three models referenced across different cases:

\begin{table}[H]
\centering
\begin{tabular}{lccc}
\hline
\textbf{Case} & \textbf{Model 81} & \textbf{Model 186} & \textbf{Model 213} \\
\hline
Accurate & 0.63 & 1.05 & 0.77 \\
Intermediate & 1.19 & 1.26 & 1.38 \\
Fast & 3.41 & 3.01 & 3.26 \\
\hline
\end{tabular}
\caption{Summary of wall clock times for Models 81, 186, 213}
\label{tab:speed_factors}
\end{table}
The explicit asymptotic calculation is similar in speed to the implicit reference calculation for the accurate case on a single CPU, but by speeding up the integration by choosing specific tolerance conditions from Eq. \ref{TolC}, we can take fewer explicit steps to increase an estimate 1-3 times faster for our calculations for the intermediate and fast cases, while maintaining an acceptable error.

       \chapter{Neutrino Flavors} \label{ch:flavors}

\section{Introduction to Neutrino Flavors and Electron Scattering}

In the results presented in Chapter \ref{ch:results}, \citep{PhysRevD.109.103019}, as well as \cite{lac2020} only include simple neutrino electron scattering $\nu_e + e^-$. Here we introduce different flavors to the problem. 

It is important to consider that neutrinos exist in three flavors: electron neutrinos ($\nu_e$), muon neutrinos ($\nu_\mu$), and tau neutrinos ($\nu_\tau$), each associated with their respective charged leptons (electron $e^-$, muon $\mu^-$, and tau $\tau^-$). Additionally, for each neutrino, there exists a corresponding antineutrino defined as the electron antineutrino ($\bar{\nu}_e$), muon antineutrino ($\bar{\nu}_\mu$), and tau antineutrino ($\bar{\nu}_\tau$). We can introduce these various flavors into our problem to visualize how considering neutrino flavors incorporates additional realistic physics into the simulation at a fundamental level.

In this thesis, we use the notation $\nu_{\mu,\tau} + \mu^-,\tau^-$ and $\bar{\nu}_{\mu,\tau} + \mu^-,\tau^-$ to  represent the scattering processes involving muon and tau neutrinos (and their antineutrinos) with their corresponding charged leptons. We did not consider muon or tau chemical potentials due to their negligible presence in the scenarios we are modeling. Our primary concern was to examine the impact of electron scattering processes, which are more relevant in core-collapse supernova environments.

\subsection{$\bar{\nu}_e + e^-$ Scattering}

Following $\nu_e + e^-$ in chapter \ref{ch:results}, we can introduce AntiNeutrino Electron Scattering:

\begin{figure}[H]
    \centering
    \includegraphics[width=5in]{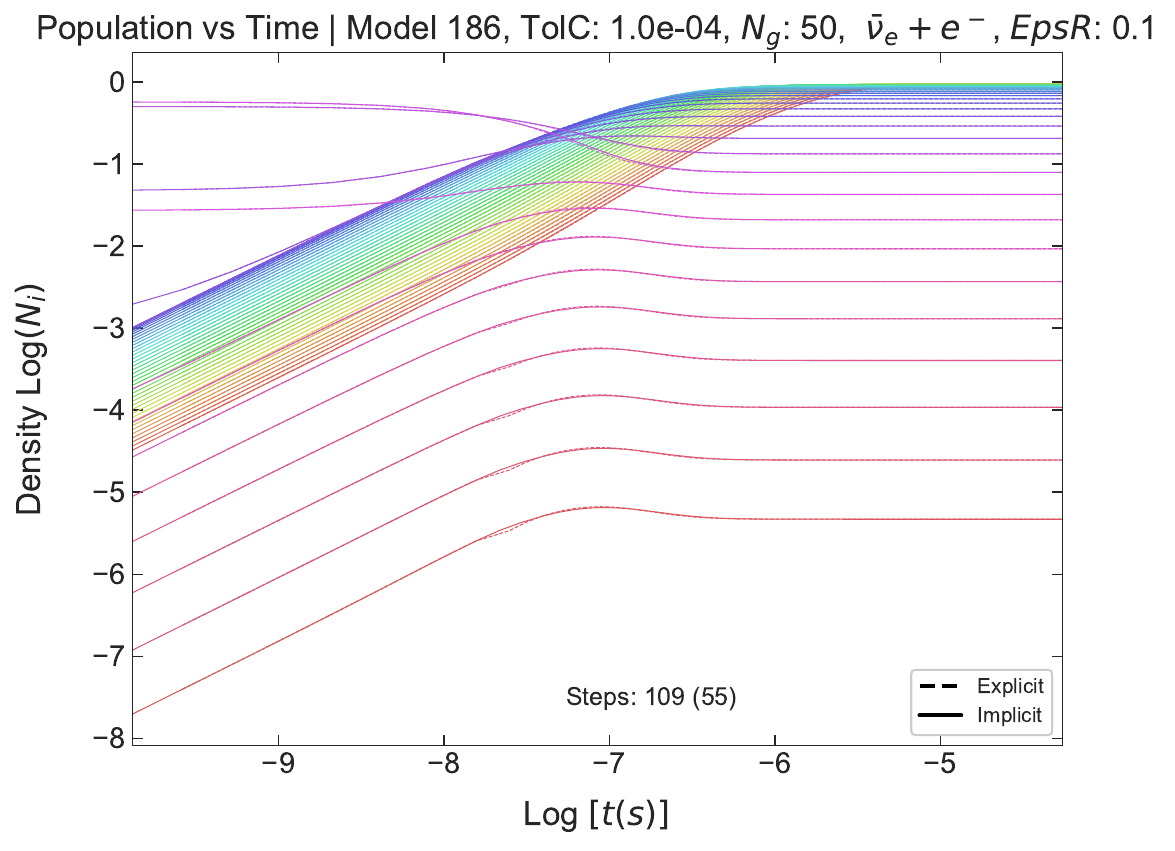}
    \caption{Population vs time for Model 186 - TolC = $10^{-4}$}
    \label{fig:AntiNeutrinoPopvsTime}
\end{figure}

\begin{figure}[H]
    \centering
    \includegraphics[width=5in]{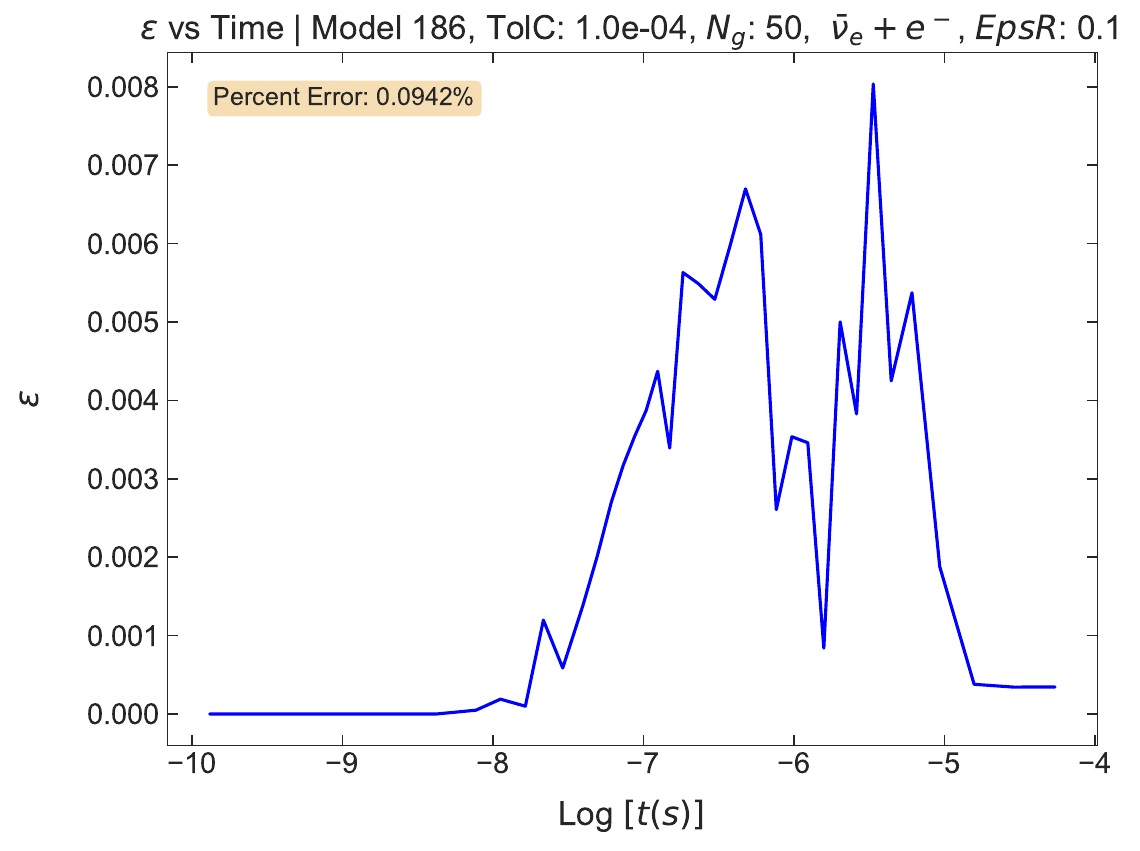}
    \caption{RMS vs time for Model 186 - TolC = $10^{-4}$}
    \label{fig:AntiNeutrinoepsilonvsTime}
\end{figure}

\begin{figure}[H]
    \centering
    \includegraphics[width=5in]{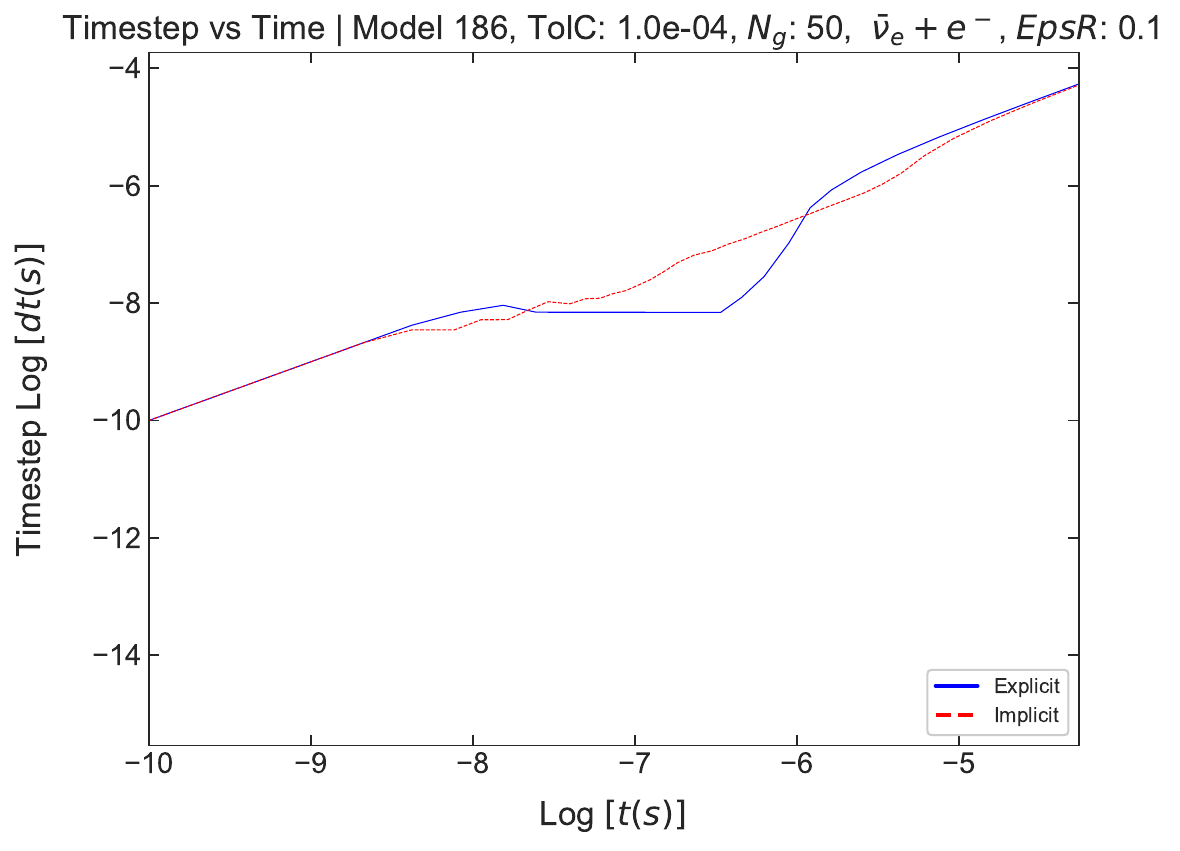}
    \caption{Time step vs time for Model 186 - TolC = $10^{-4}$}
    \label{fig:AntiNeutrinoTimestepvsTime}
\end{figure}

\begin{figure}[H]
    \centering
    \includegraphics[width=5in]{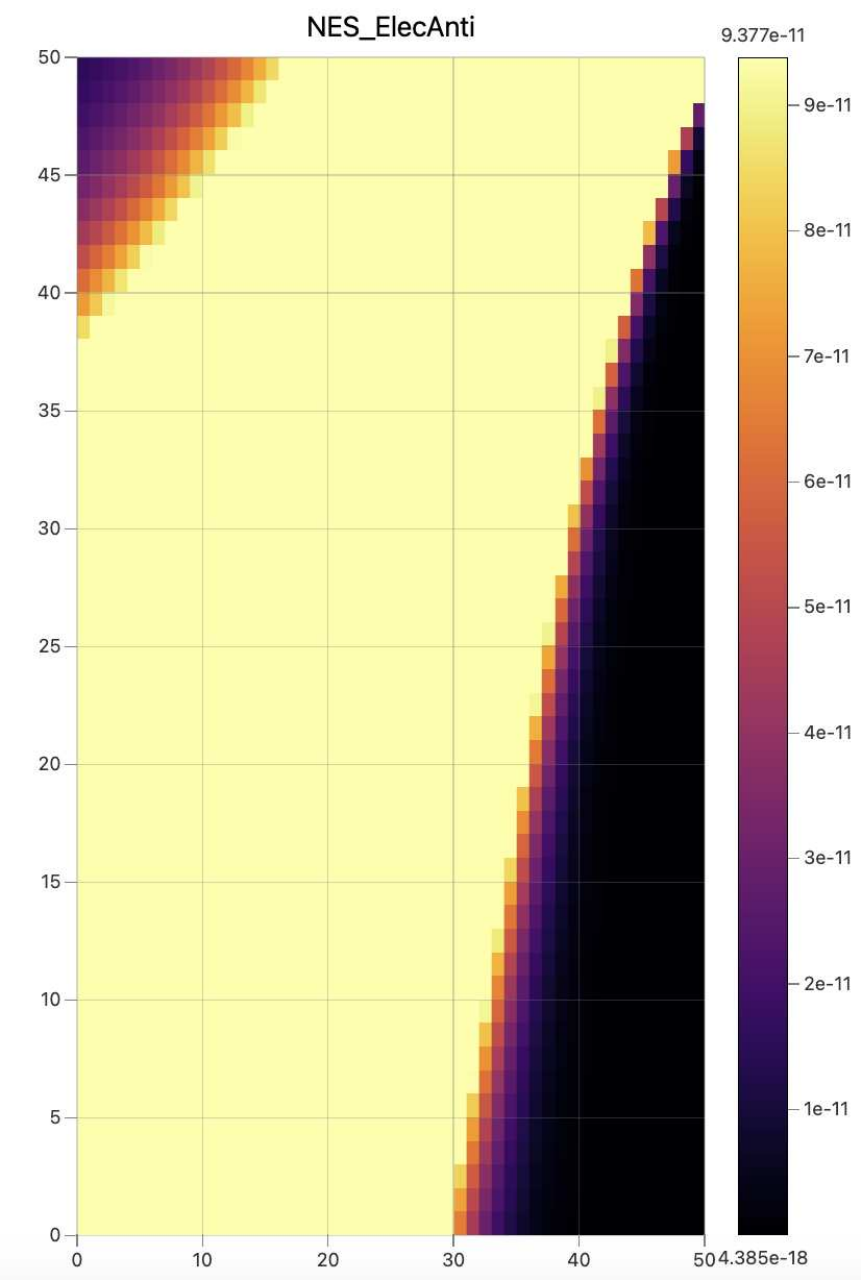}
    \caption{Heatmap of Scattering Rates for $\bar{\nu}_e + e^-$}
    \label{fig:AntiNeutrinoHeatMap}
\end{figure}

We can see here that the anti-neutrinos have a negligible effect, in comparison with Fig. \ref{fig:model186rmsint}. 

\subsection{$\bar{\nu}_{\mu,\tau} + \mu^-,\tau^-$ Scattering}

\begin{figure}[H]
    \centering
    \includegraphics[width=5in]{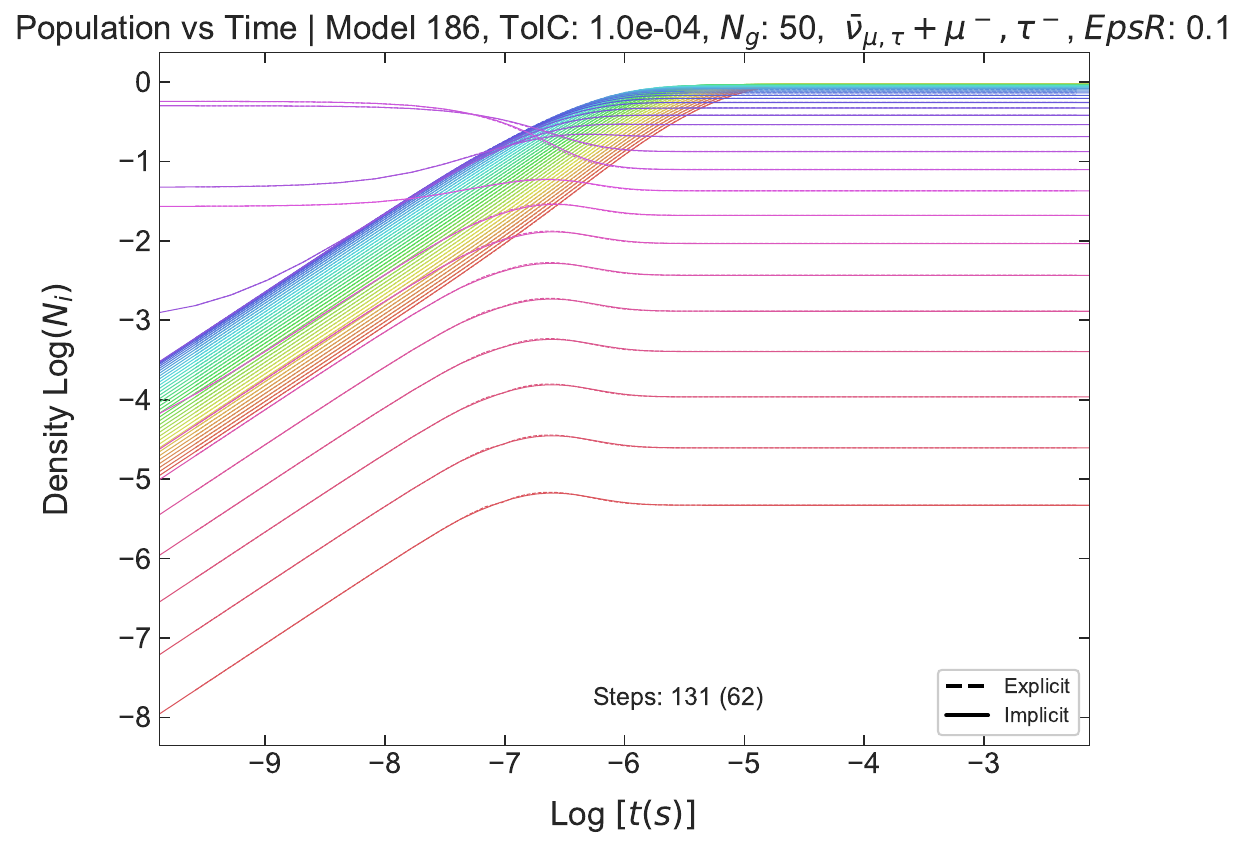}
    \caption{Population vs time for Model 186 - TolC = $10^{-4}$}
    \label{fig:mutauNeutrinoPopvsTime}
\end{figure}

Similarly we can show results for $\bar{\nu}_{\mu,\tau} + \mu^-,\tau^-$ scattering in fig. \ref{fig:mutauNeutrinoPopvsTime}.

\begin{figure}[H]
    \centering
    \includegraphics[width=5in]{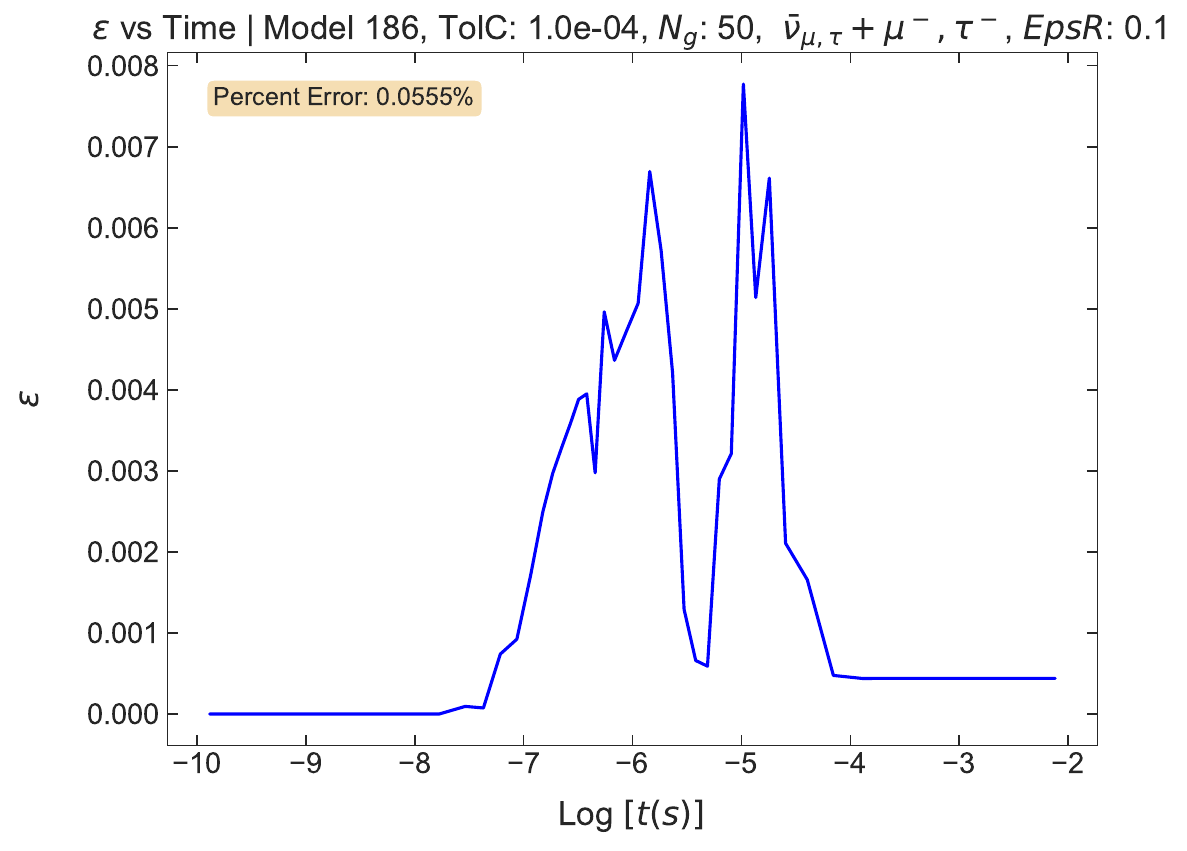}
    \caption{RMS vs time for Model 186 - TolC = $10^{-4}$}
    \label{fig:mutauNeutrinoepsilonvsTime}
\end{figure}

\begin{figure}[H]
    \centering
    \includegraphics[width=5in]{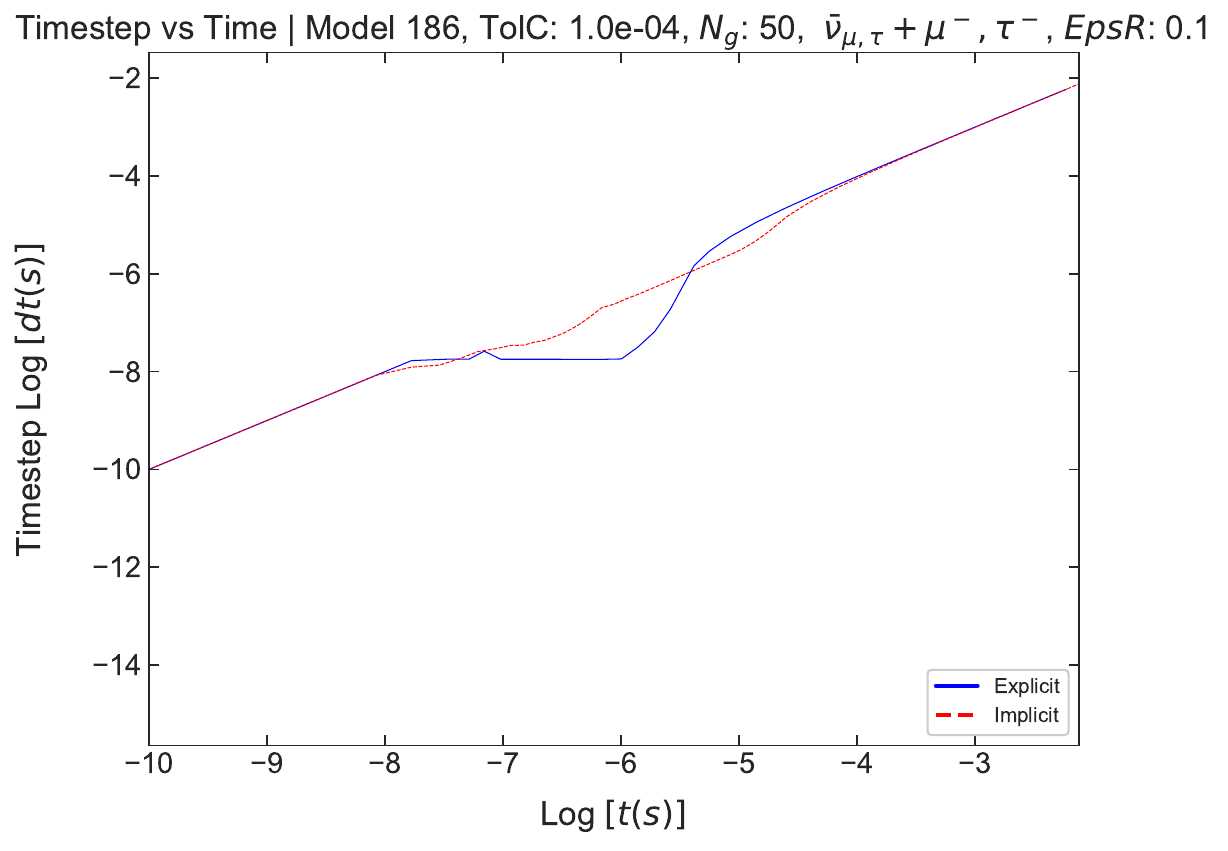}
    \caption{Time step vs time for Model 186 - TolC = $10^{-4}$}
    \label{fig:mutauNeutrinoTimestepvsTime}
\end{figure}

\begin{figure}[H]
    \centering
    \includegraphics[width=5in]{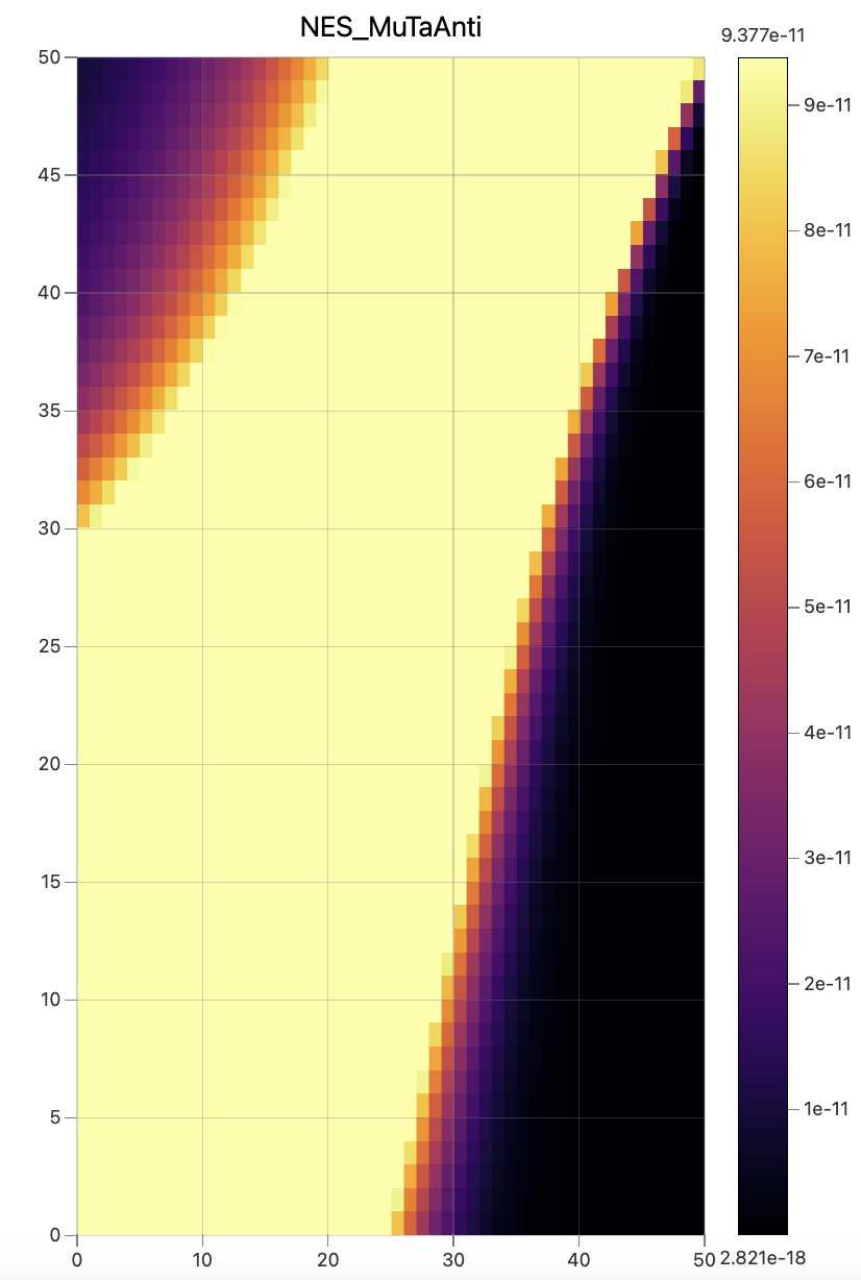}
    \caption{Heatmap of Scattering Rates for $\bar{\nu}_{\mu,\tau} + \mu^-,\tau^-$}
    \label{fig:MuTaAntiHeatMap}
\end{figure}

We can see here the error is comparable with Fig. \ref{fig:model186rmsint}, however the solutions take longer to each equilibrium in reference to fig. \ref{fig:model186popint}. 

\subsection{$\nu_{\mu,\tau} + \mu^-,\tau^-$ Scattering}

Results for $\nu_{\mu,\tau} + \mu^-,\tau^-$  are as follows: 

\begin{figure}[H]
    \centering
    \includegraphics[width=5in]{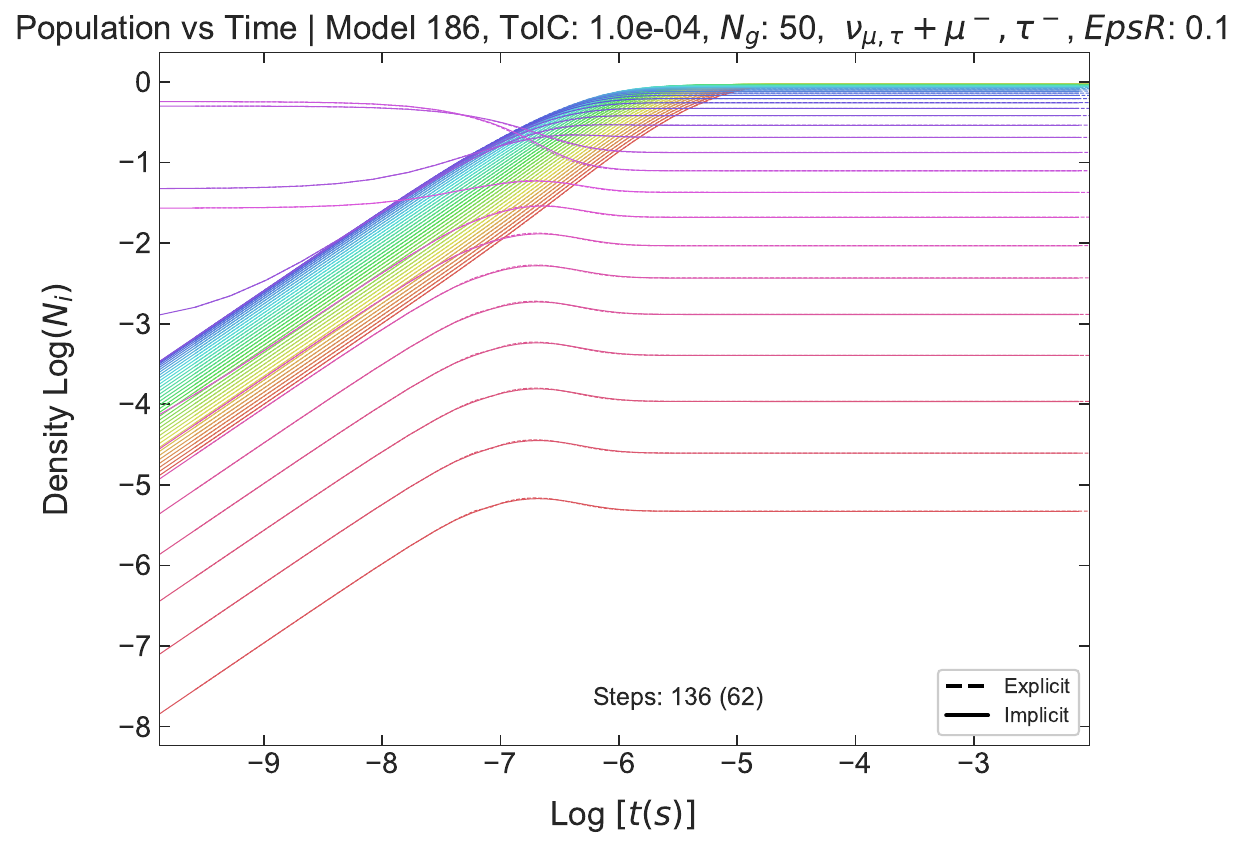}
    \caption{Population vs time for Model 186 - TolC = $10^{-4}$}
    \label{fig:mutaumutauNeutrinoPopvsTime}
\end{figure}

\begin{figure}[H]
    \centering
    \includegraphics[width=5in]{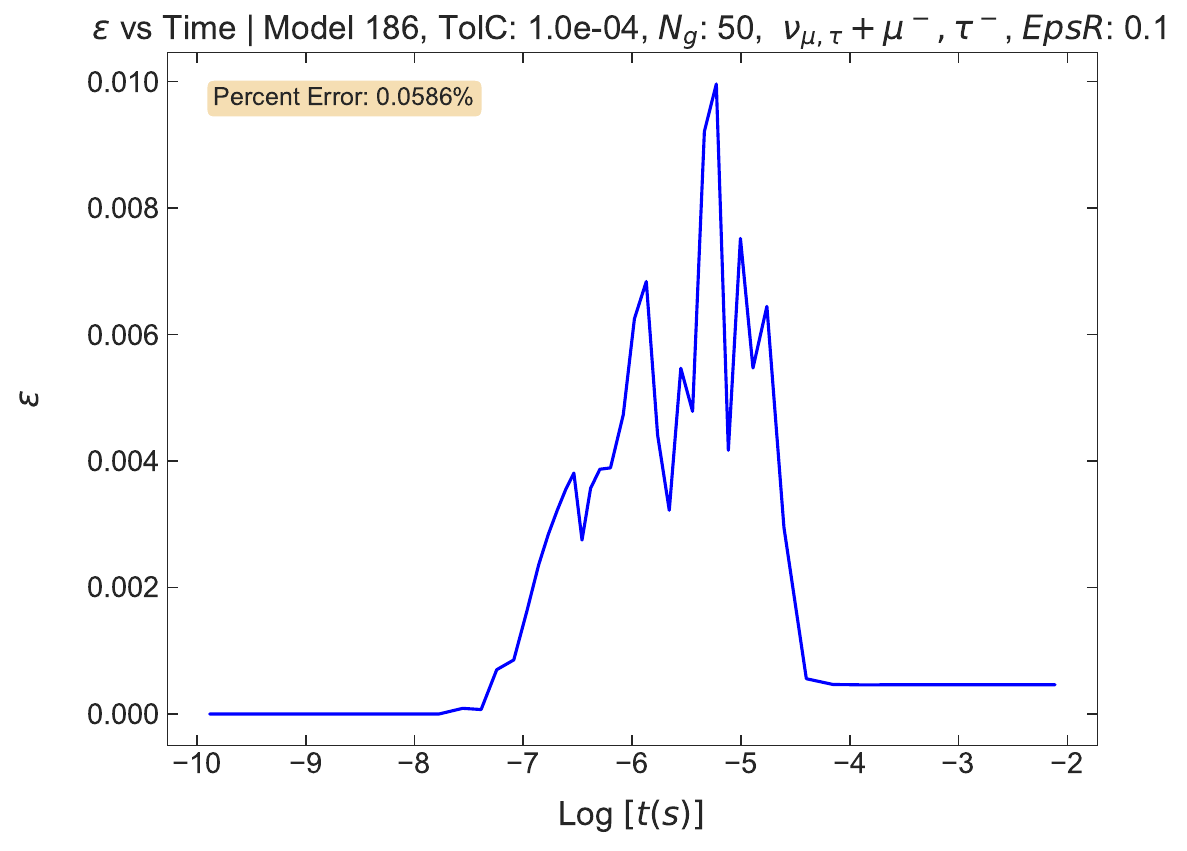}
    \caption{RMS vs time for Model 186 - TolC = $10^{-4}$}
    \label{fig:mutaumutauNeutrinoepsilonvsTime}
\end{figure}

\begin{figure}[H]
    \centering
    \includegraphics[width=5in]{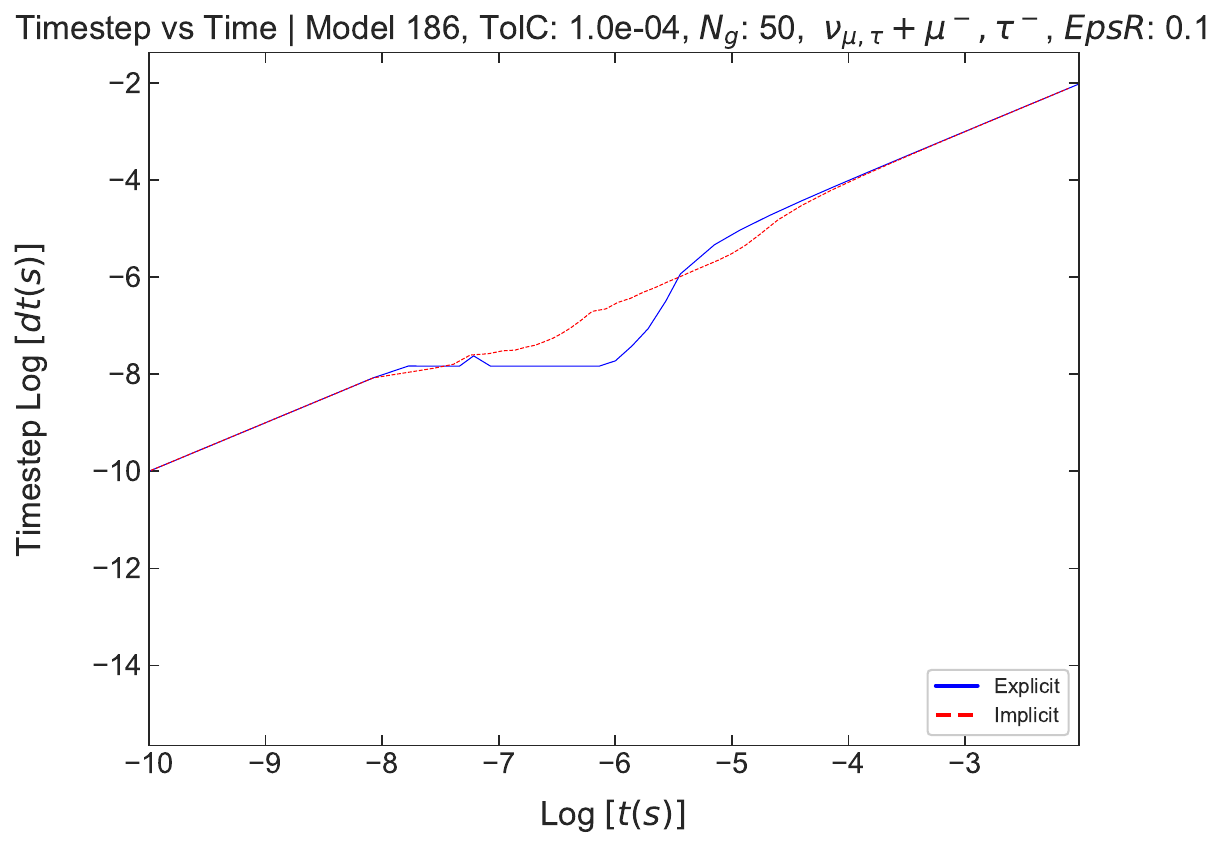}
    \caption{Time step vs time for Model 186 - TolC = $10^{-4}$}
    \label{fig:mutaumutauNeutrinoTimestepvsTime}
\end{figure}

\begin{figure}[H]
    \centering
    \includegraphics[width=5in]{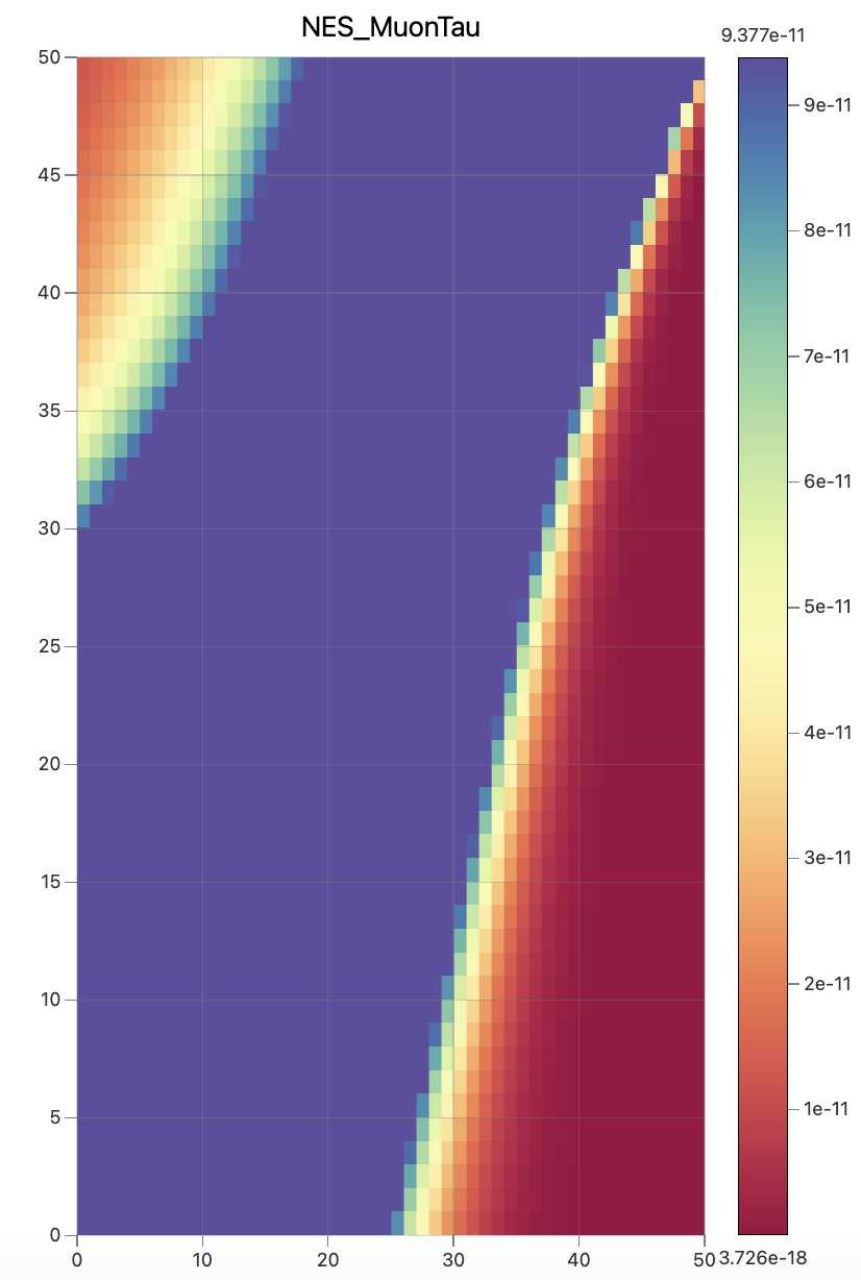}
    \caption{Heatmap of Scattering Rates for $\nu_{\mu,\tau} + \mu^-,\tau^-$}
    \label{fig:MuonTauHeatMap}
\end{figure}

Similarly, the integration takes longer and the error is comparable to the results from fig. \ref{fig:mutauNeutrinoepsilonvsTime}. 

The results showcase that the neutrino flavors may introduce a small but negligible correction for the present  solutions.
This introduces a realistic ingredient in our simulations, with the possibility of capturing the rich interplay of muon and tau neutrinos with their associated leptons in more complex cases.

The mixing of different neutrino flavors introduced into our models exerted some influence on the dynamics of a system. Still, the level of error is reasonable which is crucial to maintain the integrity of the model under different conditions. However, if one wants to model the neutrino behavior in astrophysical environment with all necessary and sufficient details, a complete neutrino transport regime is necessary. 

Oscillations of flavor make the calculation and transport of neutrinos complex because they take place when neutrinos change their flavor or state while propagating. Present models often make simplifications in these processes, which in turn may omit interactions that are crucial and may be responsible for a range of large-scale phenomena. Such as supernova explosion mechanisms or rates of cooling of neutron stars. While the full neutrino transport coupling is not included here, this introduces foundational work which FENN will build upon for a more realistic model. 

    \chapter{Scaling with Network Size} \label{ch:NetworkScaling}

\section{Analysis of Scaled Networks}
\label{sec:scaled_networks}

In chapter \ref{ch:results} we presented the accurate, intermediate, and fast results for the standard scenario of 40 energy bins. After integration with WeakLib, we can now make the same analysis for increased network sizes at fixed resolution, giving an opportunity to enhance the complexity of our simulations even more. Note that this implies in the changing the size of our collision matrix in Eq. \ref{CollisionMatrix}, letting us handle more detailed interactions and higher energy resolutions.

This section gives a detailed comparative analysis of the network scaling process with respect to size for Model 186, and elaborates on the effect of increasing network size on the system dynamics. Making our comparative study with larger networks, it is possible to appreciate scaling with system size in terms of both complexity and computational demand. We analyze and compare results using network sizes of 40, 50, 80, 100, 130, 160, and 180, energy bins, comparing  Population vs. Time, Epsilon vs. Time, and Timestep vs. Time.

The following are similar figures from chapter \ref{ch:results}, with the mentioned energy sizes.
\subsection{Network Size 40}
\label{subsec:network_40}

\begin{figure}[H]
    \centering
    \includegraphics[width=5in]{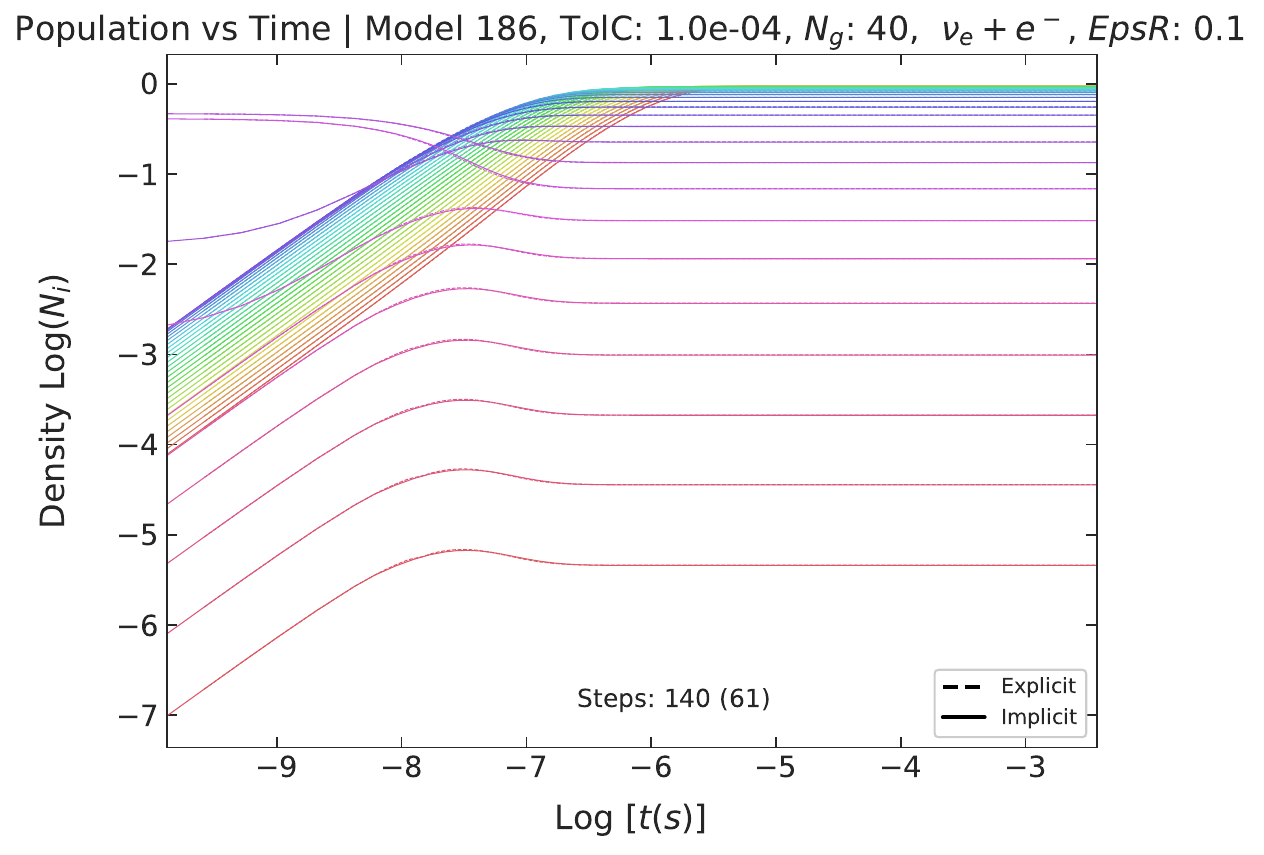}
    \caption{Population vs. time for Network Size 40.}
    \label{fig:pop40}
\end{figure}

\begin{figure}[H]
    \centering
    \includegraphics[width=5in]{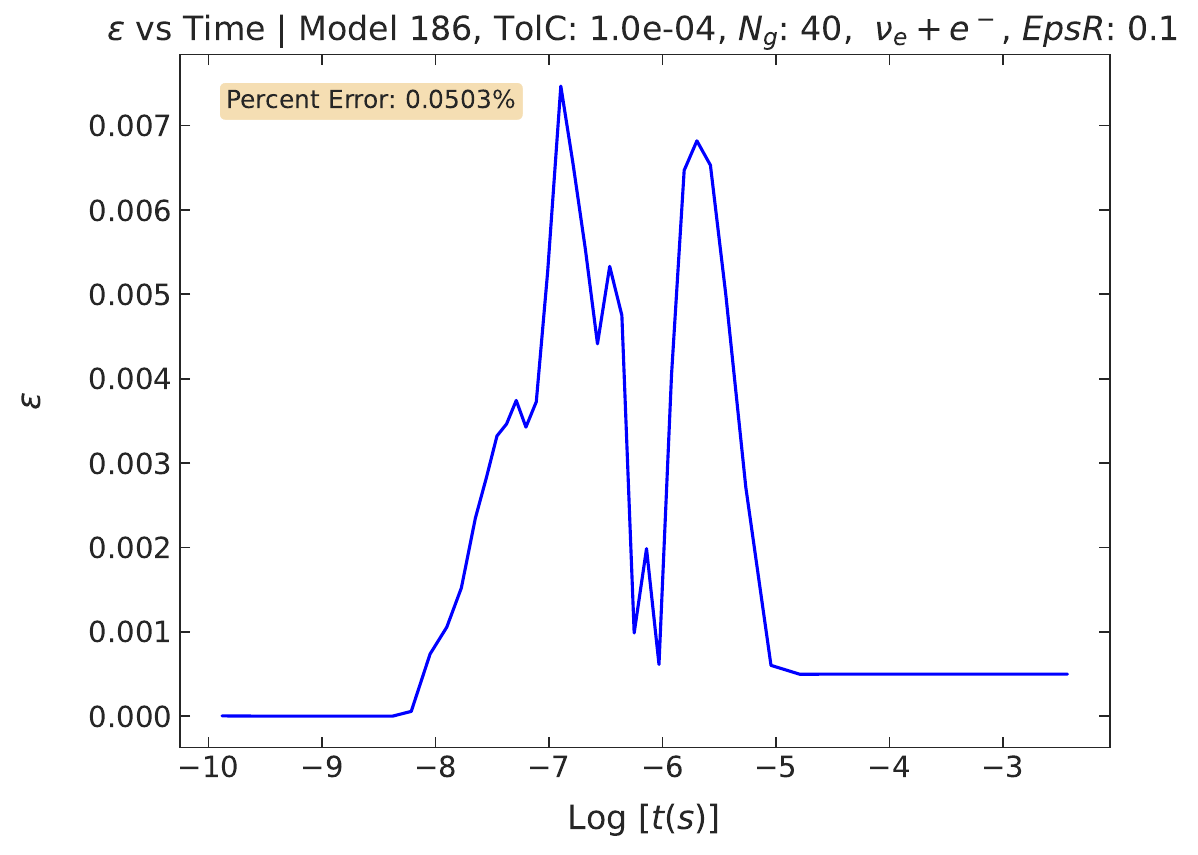}
    \caption{Epsilon vs. time for Network Size 40.}
    \label{fig:epsilon40}
\end{figure}

\begin{figure}[H]
    \centering
    \includegraphics[width=5in]{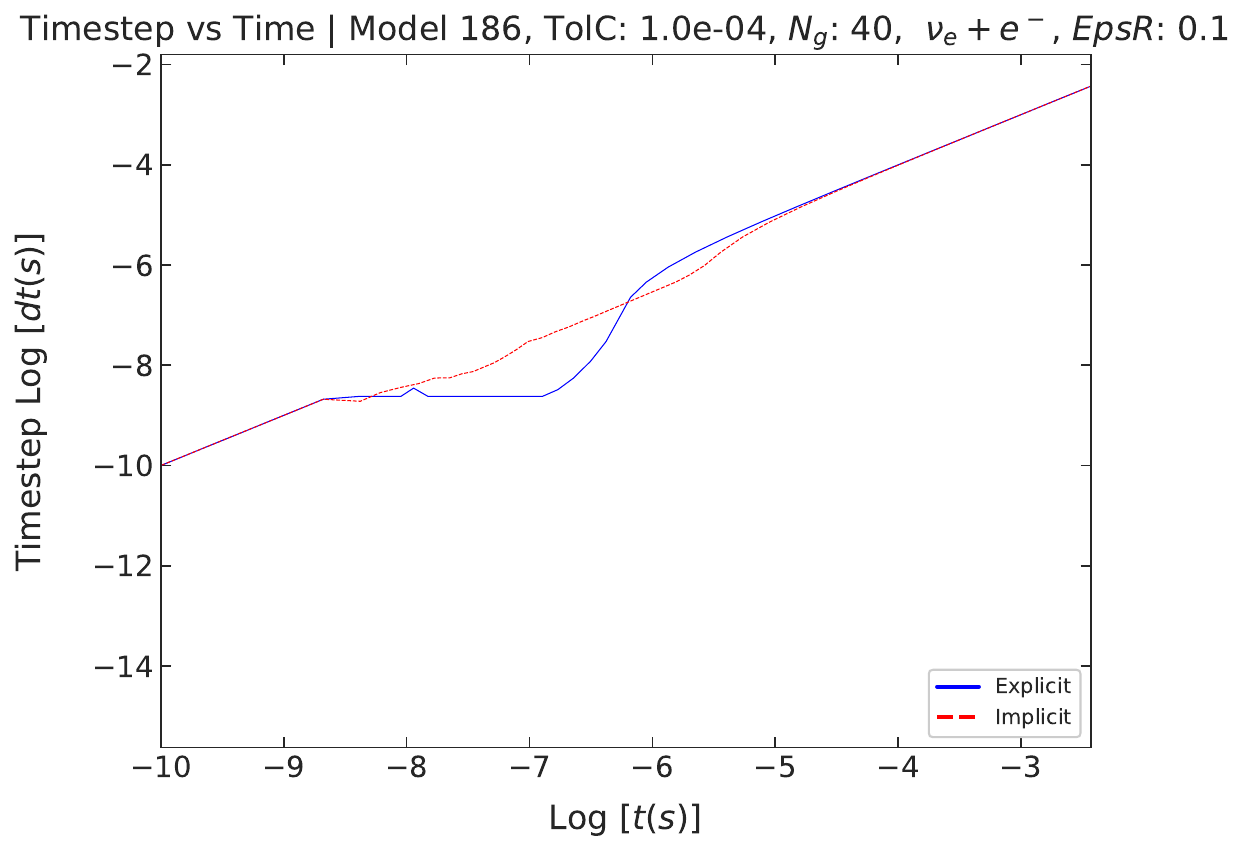}
    \caption{Timestep vs. time for Network Size 40.}
    \label{fig:timestep40}
\end{figure}

\subsection{Network Size 50}
\label{subsec:network_50}

\begin{figure}[H]
    \centering
    \includegraphics[width=5in]{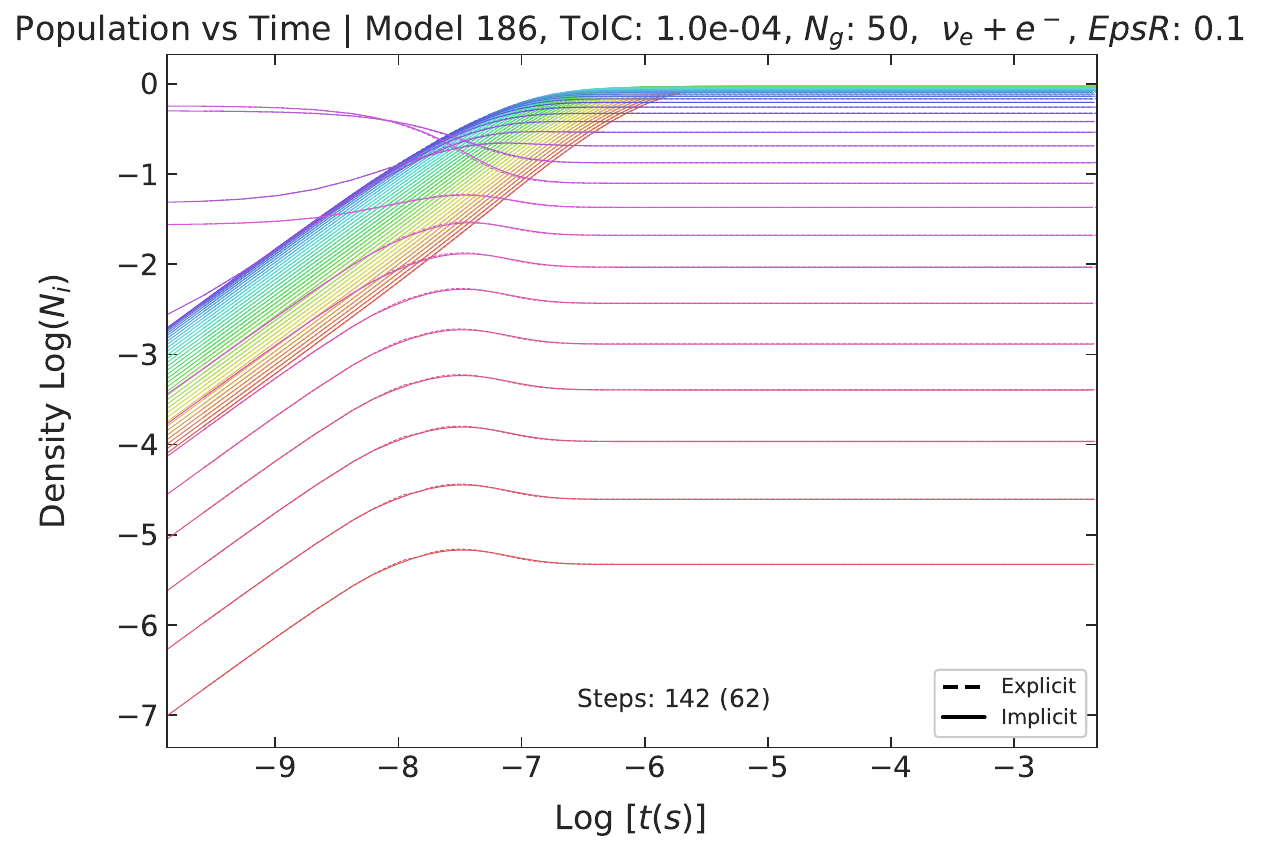}
    \caption{Population vs. time for Network Size 50.}
    \label{fig:pop50}
\end{figure}

\begin{figure}[H]
    \centering
    \includegraphics[width=5in]{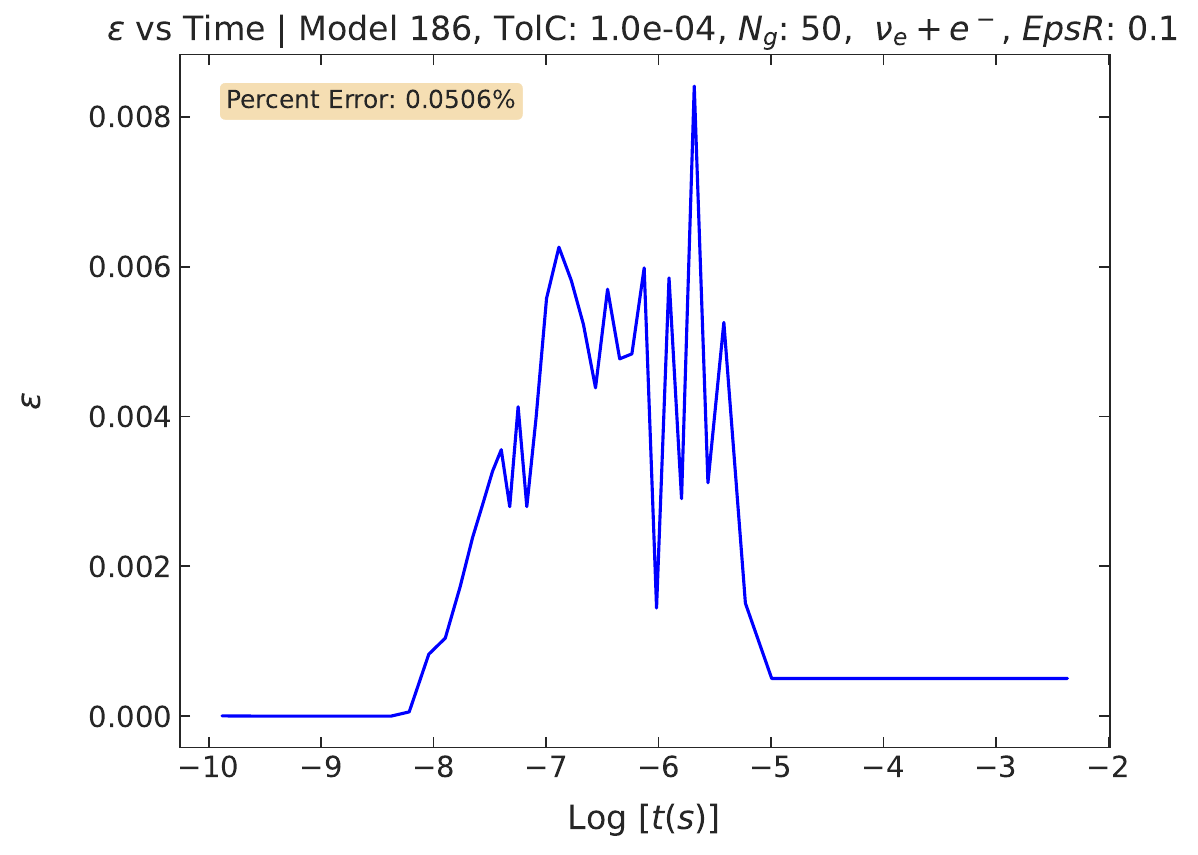}
    \caption{Epsilon vs. time for Network Size 50.}
    \label{fig:epsilon50}
\end{figure}

\begin{figure}[H]
    \centering
    \includegraphics[width=5in]{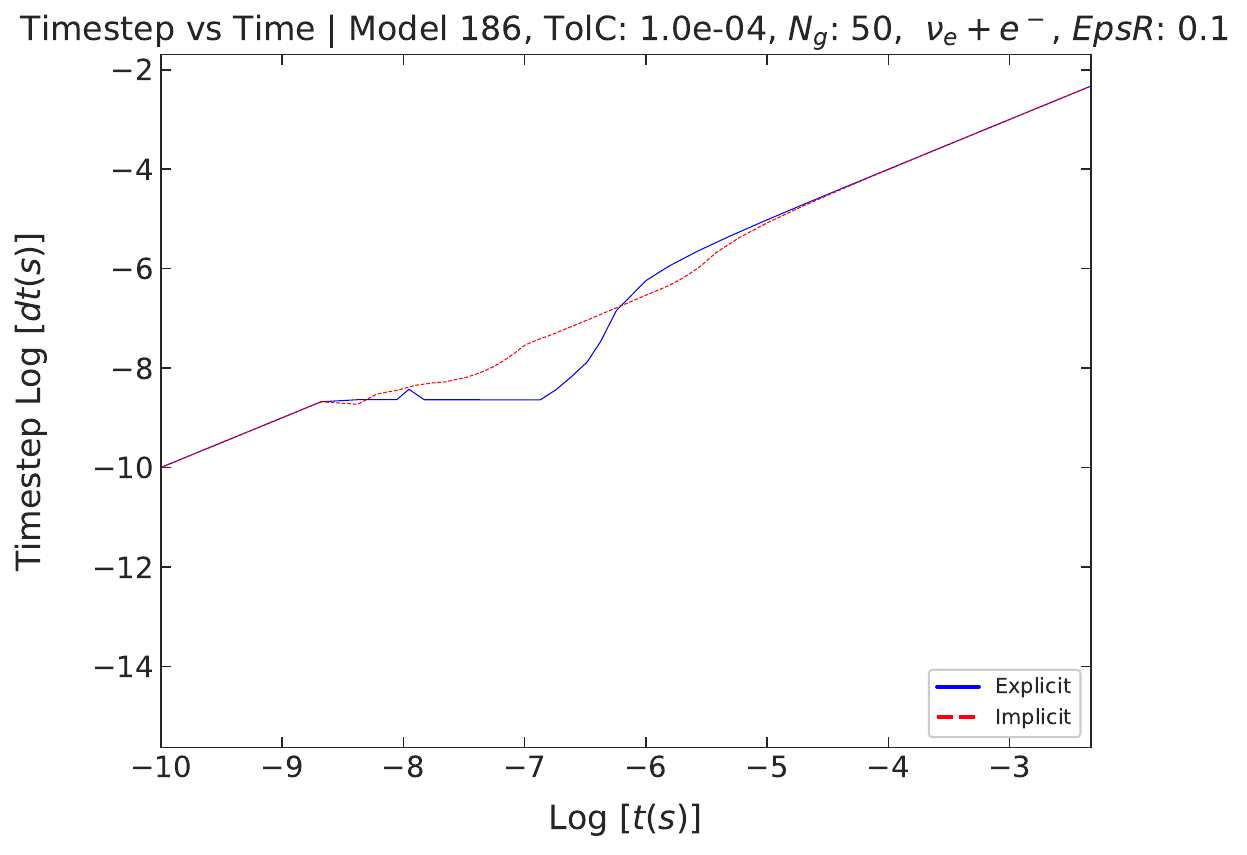}
    \caption{Timestep vs. time for Network Size 50.}
    \label{fig:timestep50}
\end{figure}

\subsection{Network Size 80}
\label{subsec:network_80}

\begin{figure}[H]
    \centering
    \includegraphics[width=5in]{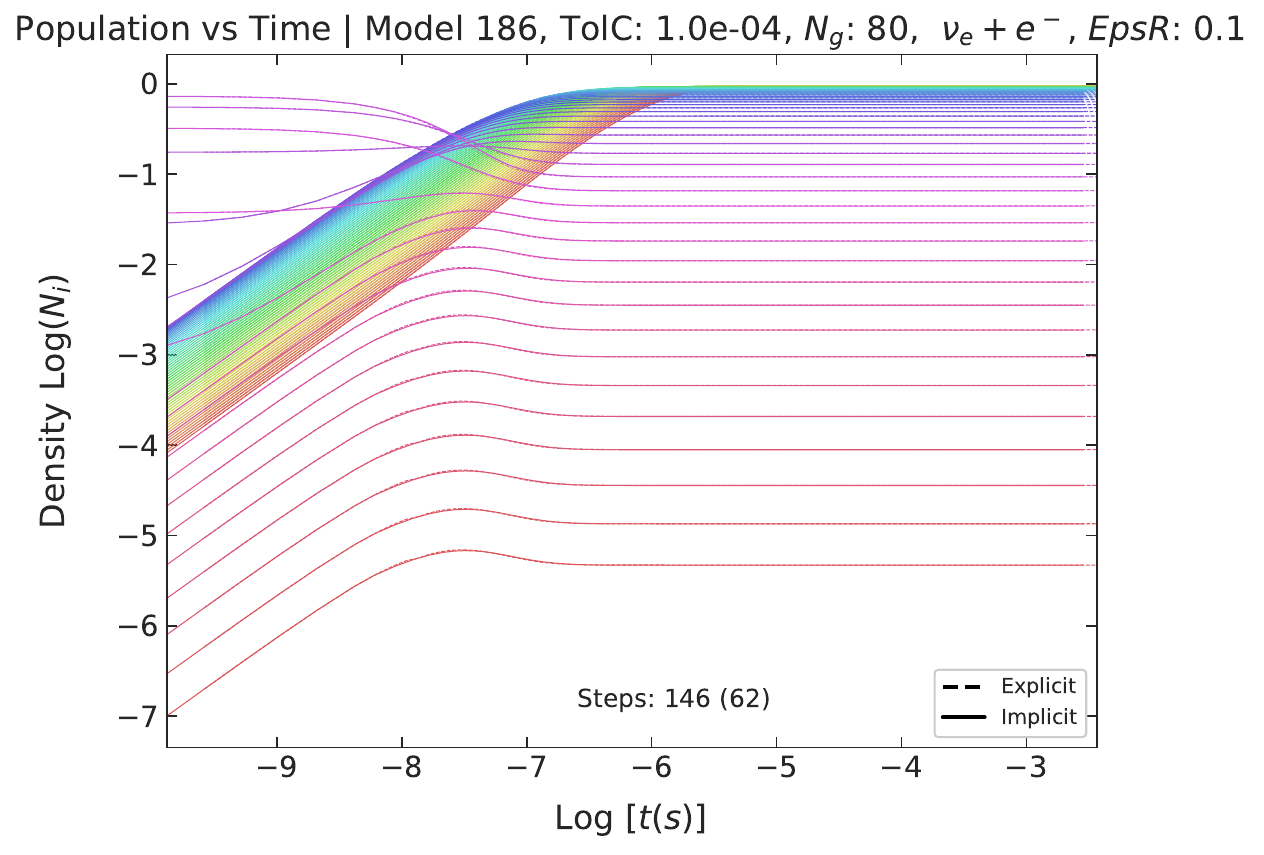}
    \caption{Population vs. time for Network Size 80.}
    \label{fig:pop80}
\end{figure}

\begin{figure}[H]
    \centering
    \includegraphics[width=5in]{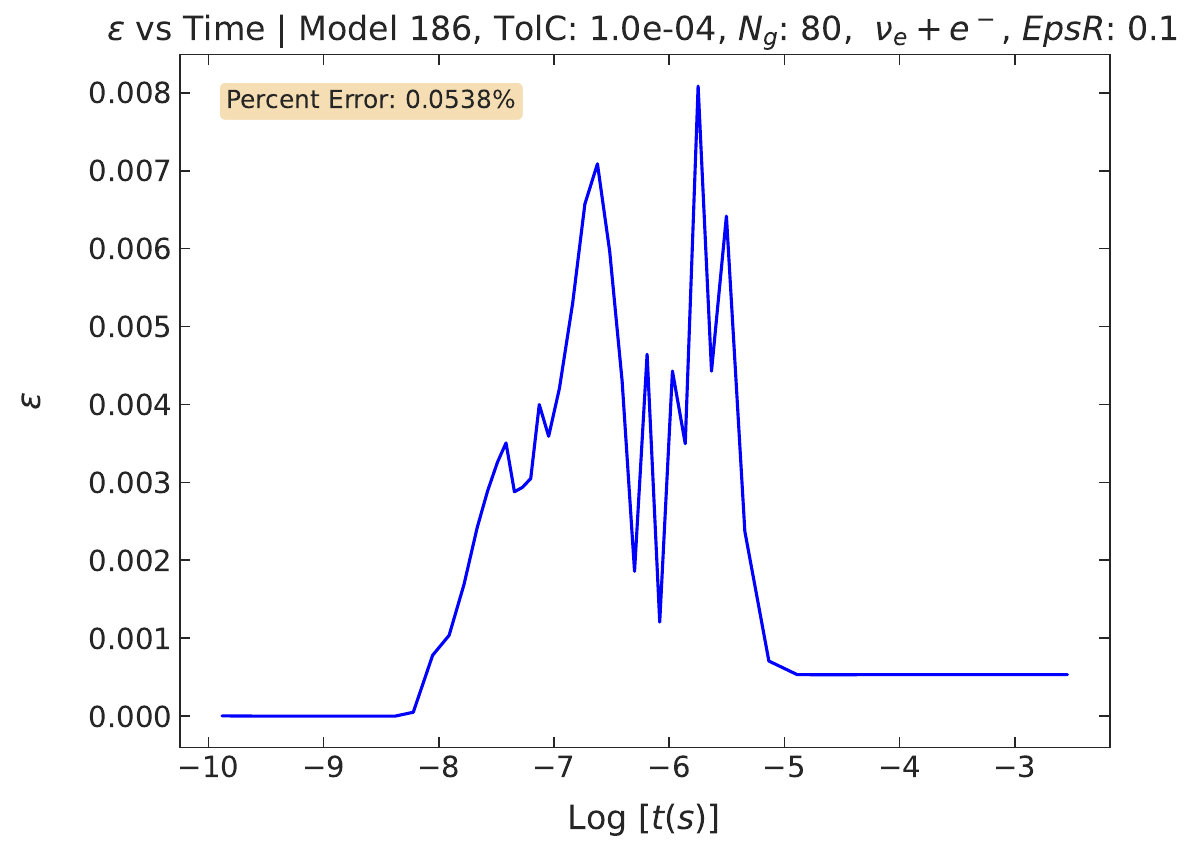}
    \caption{Epsilon vs. time for Network Size 80.}
    \label{fig:epsilon80}
\end{figure}

\begin{figure}[H]
    \centering
    \includegraphics[width=5in]{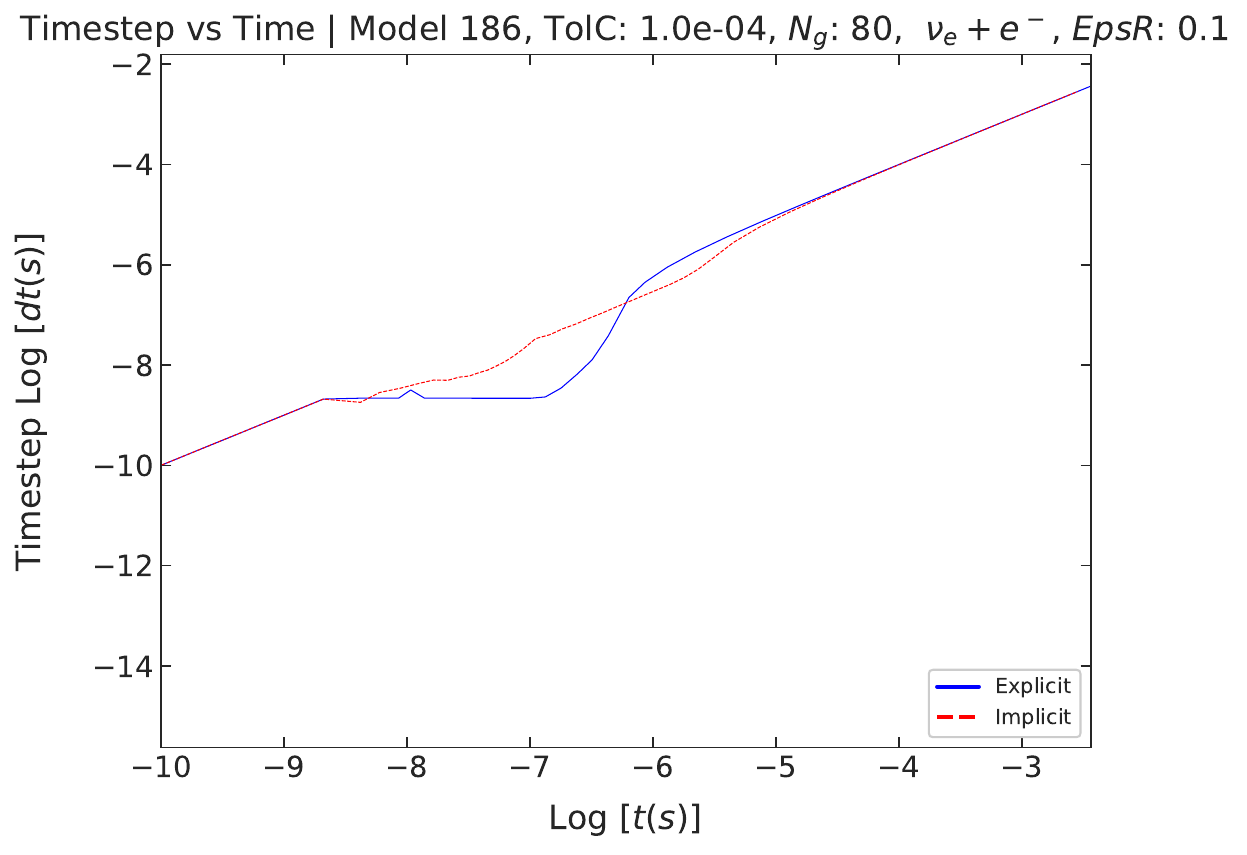}
    \caption{Timestep vs. time for Network Size 80.}
    \label{fig:timestep80}
\end{figure}

\subsection{Network Size 100}
\label{subsec:network_100}

\begin{figure}[H]
    \centering
    \includegraphics[width=5in]{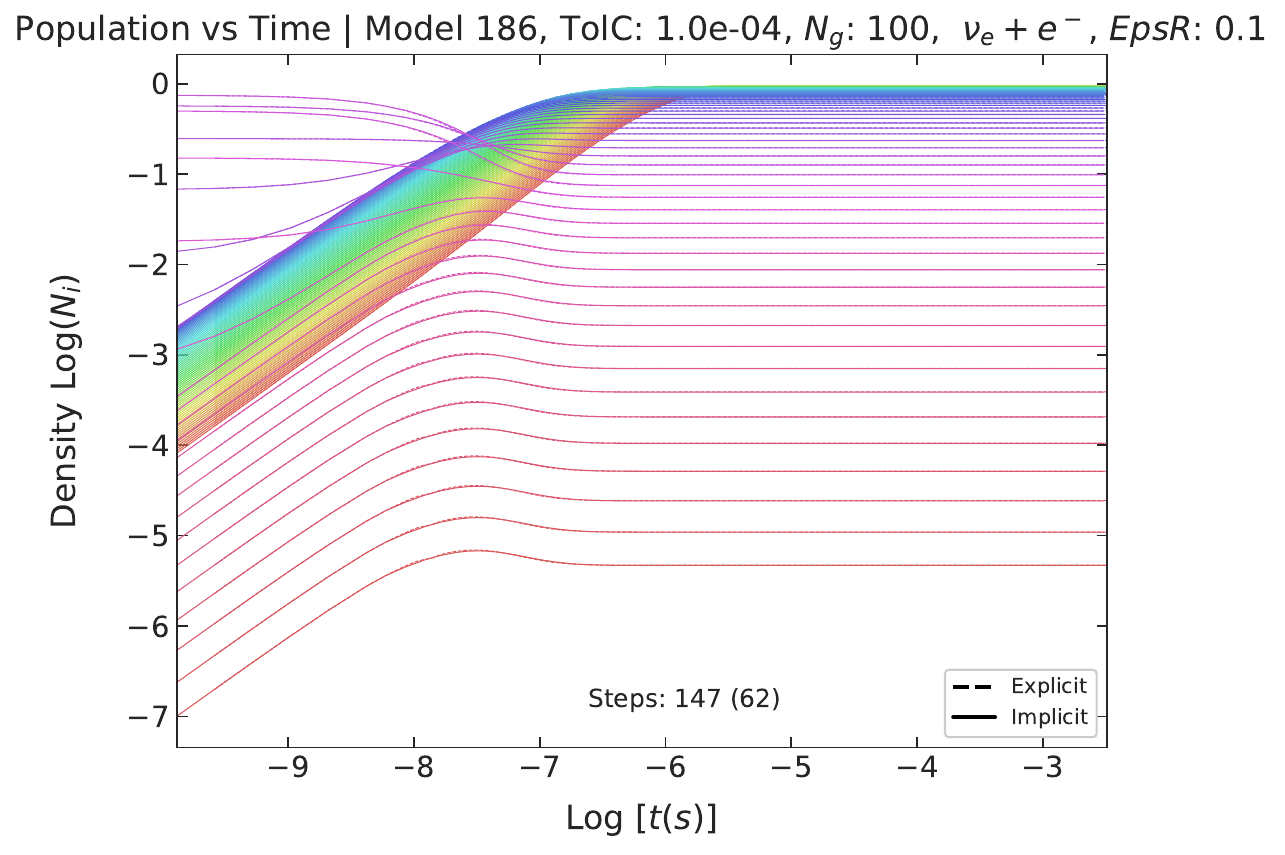}
    \caption{Population vs. time for Network Size 100.}
    \label{fig:pop100}
\end{figure}

\begin{figure}[H]
    \centering
    \includegraphics[width=5in]{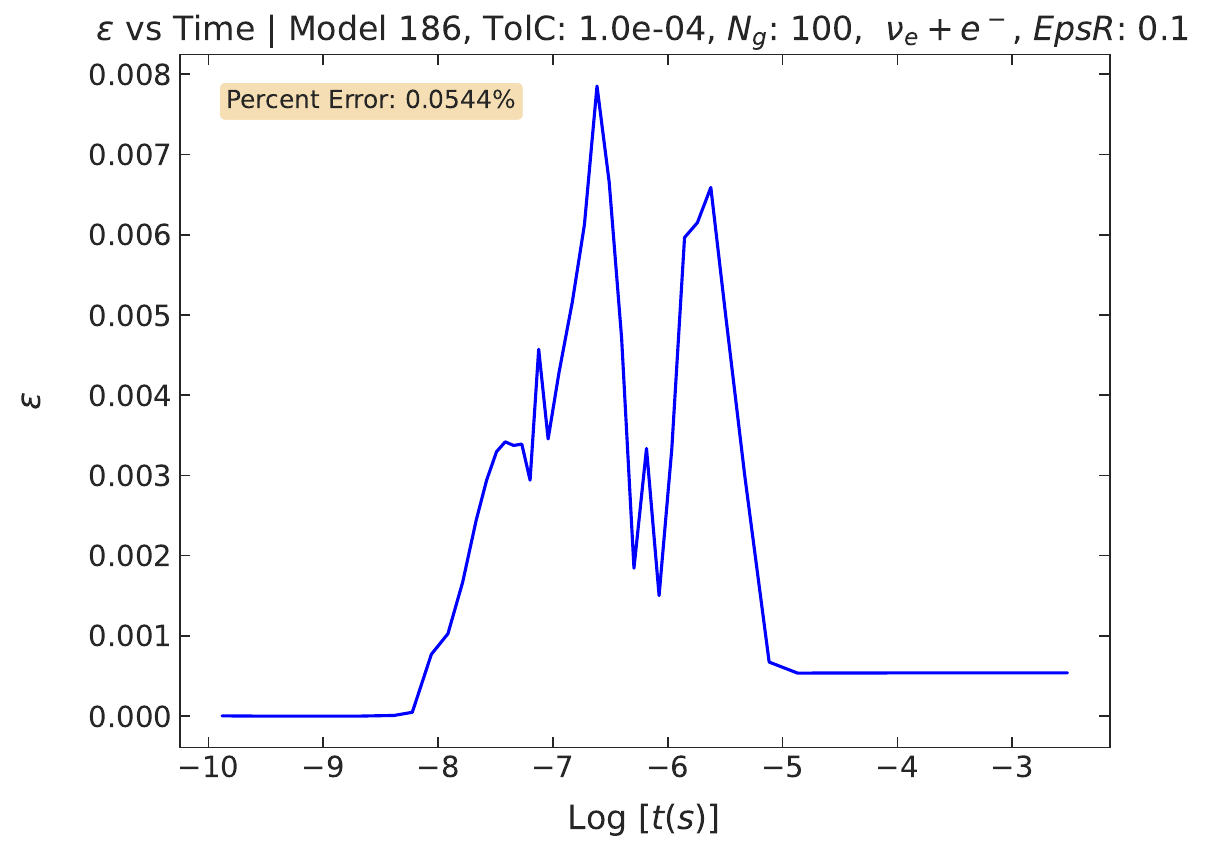}
    \caption{Epsilon vs. time for Network Size 100.}
    \label{fig:epsilon100}
\end{figure}

\begin{figure}[H]
    \centering
    \includegraphics[width=5in]{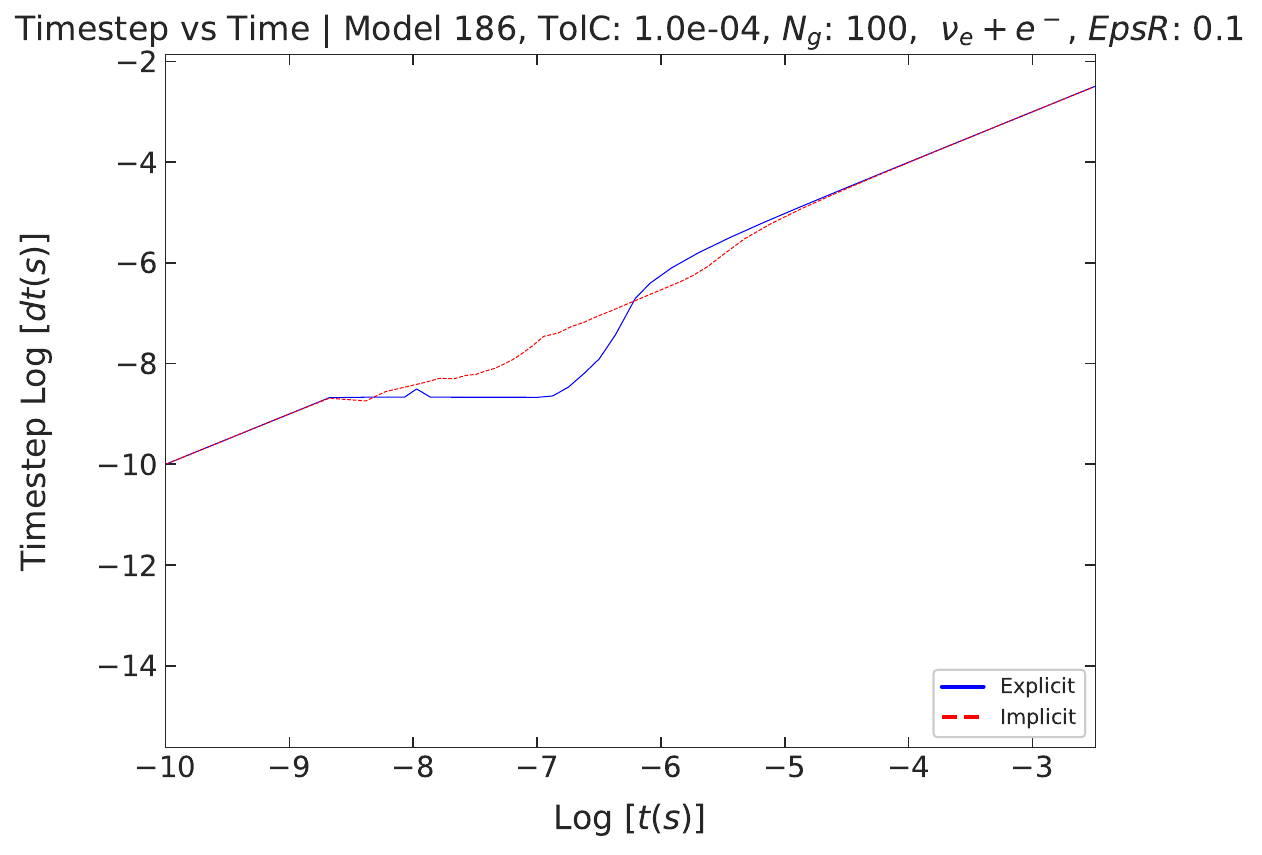}
    \caption{Timestep vs. time for Network Size 100.}
    \label{fig:timestep100}
\end{figure}

\subsection{Network Size 130}
\label{subsec:network_130}

\begin{figure}[H]
    \centering
    \includegraphics[width=5in]{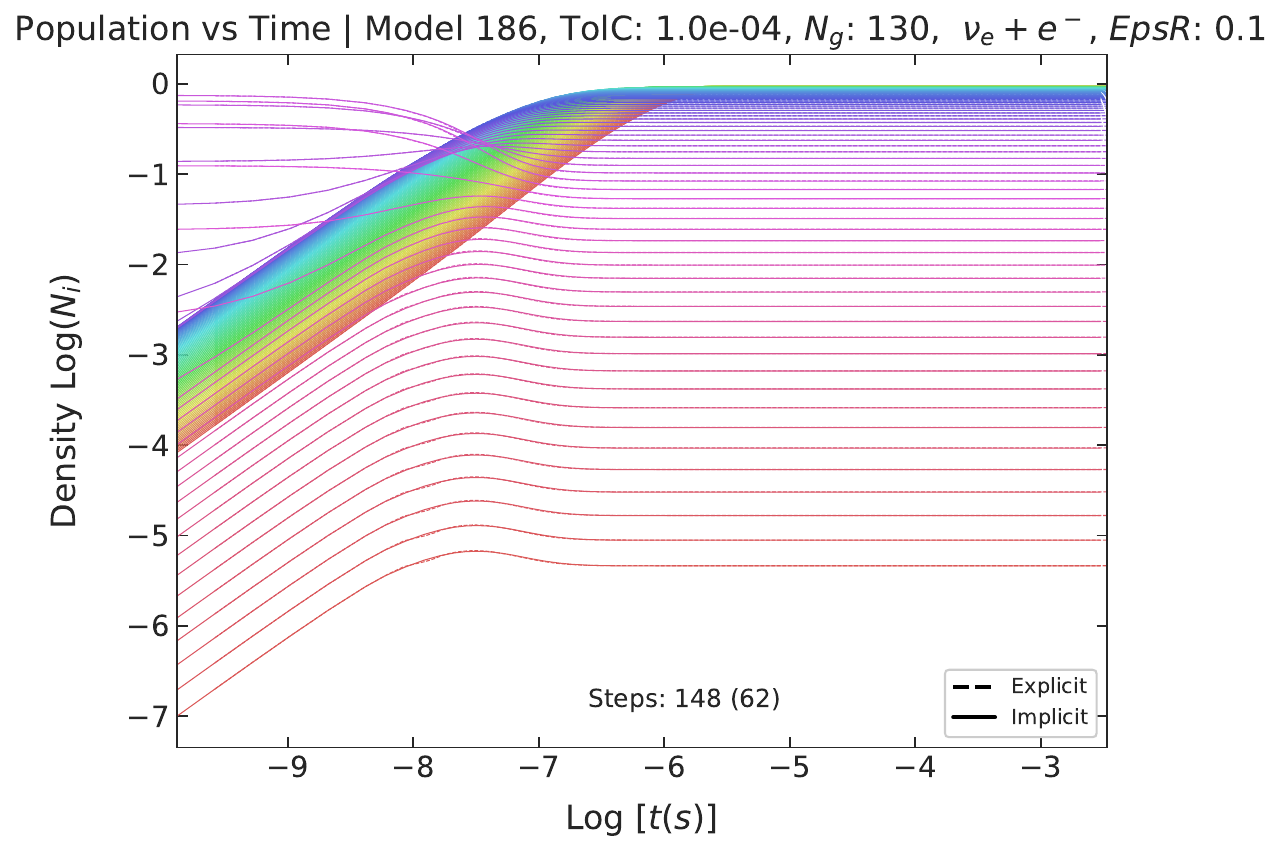}
    \caption{Population vs. time for Network Size 160.}
    \label{fig:pop160}
\end{figure}

\begin{figure}[H]
    \centering
    \includegraphics[width=5in]{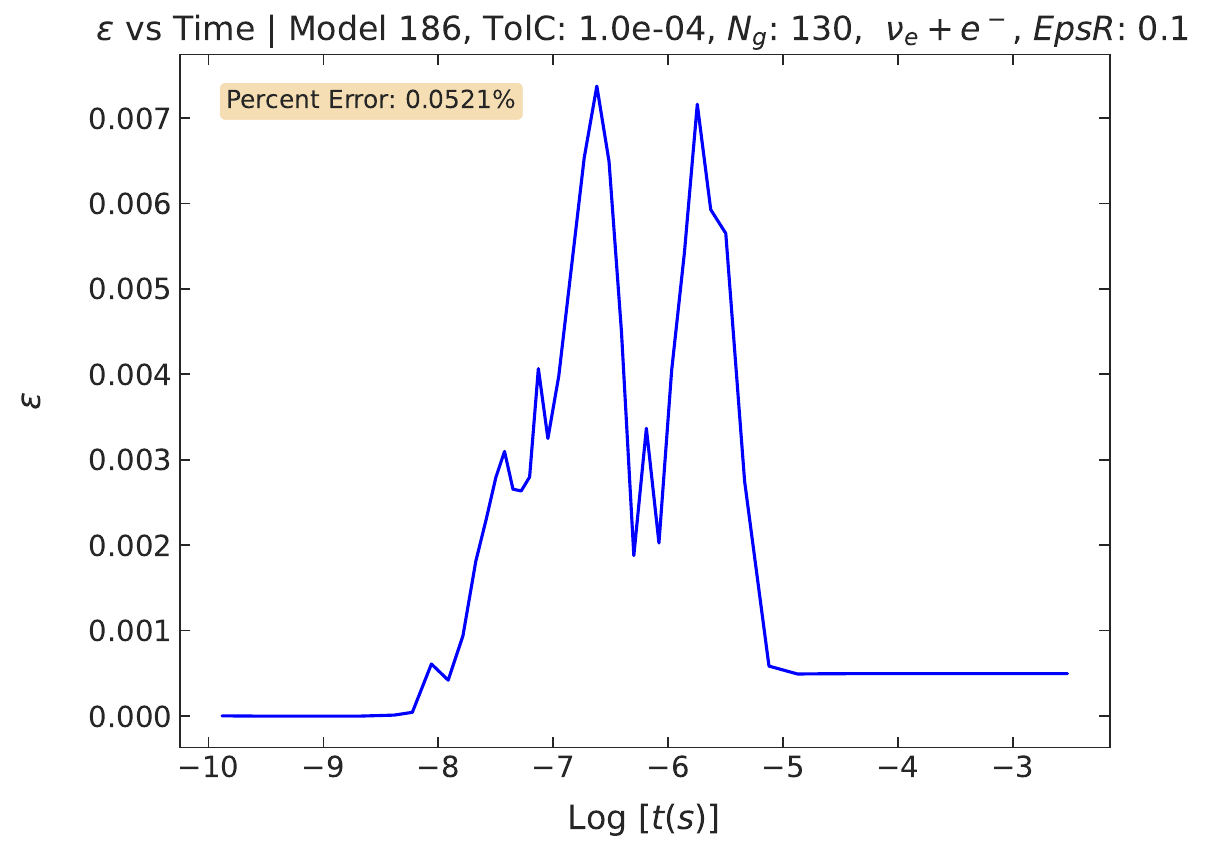}
    \caption{Epsilon vs. time for Network Size 160.}
    \label{fig:epsilon160}
\end{figure}

\begin{figure}[H]
    \centering
    \includegraphics[width=5in]{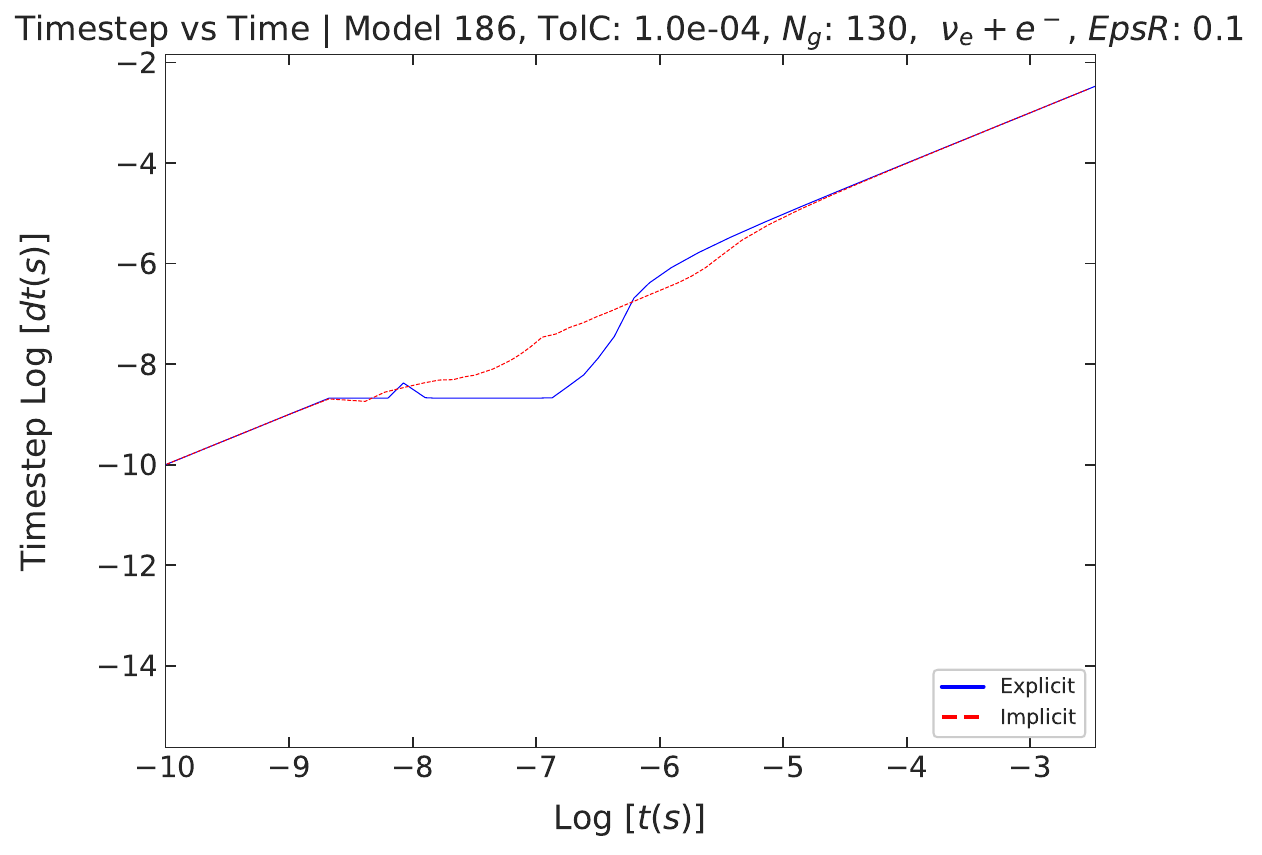}
    \caption{Timestep vs. time for Network Size 160.}
    \label{fig:timestep160}
\end{figure}

\subsection{Network Size 160}
\label{subsec:network_160}

\begin{figure}[H]
    \centering
    \includegraphics[width=5in]{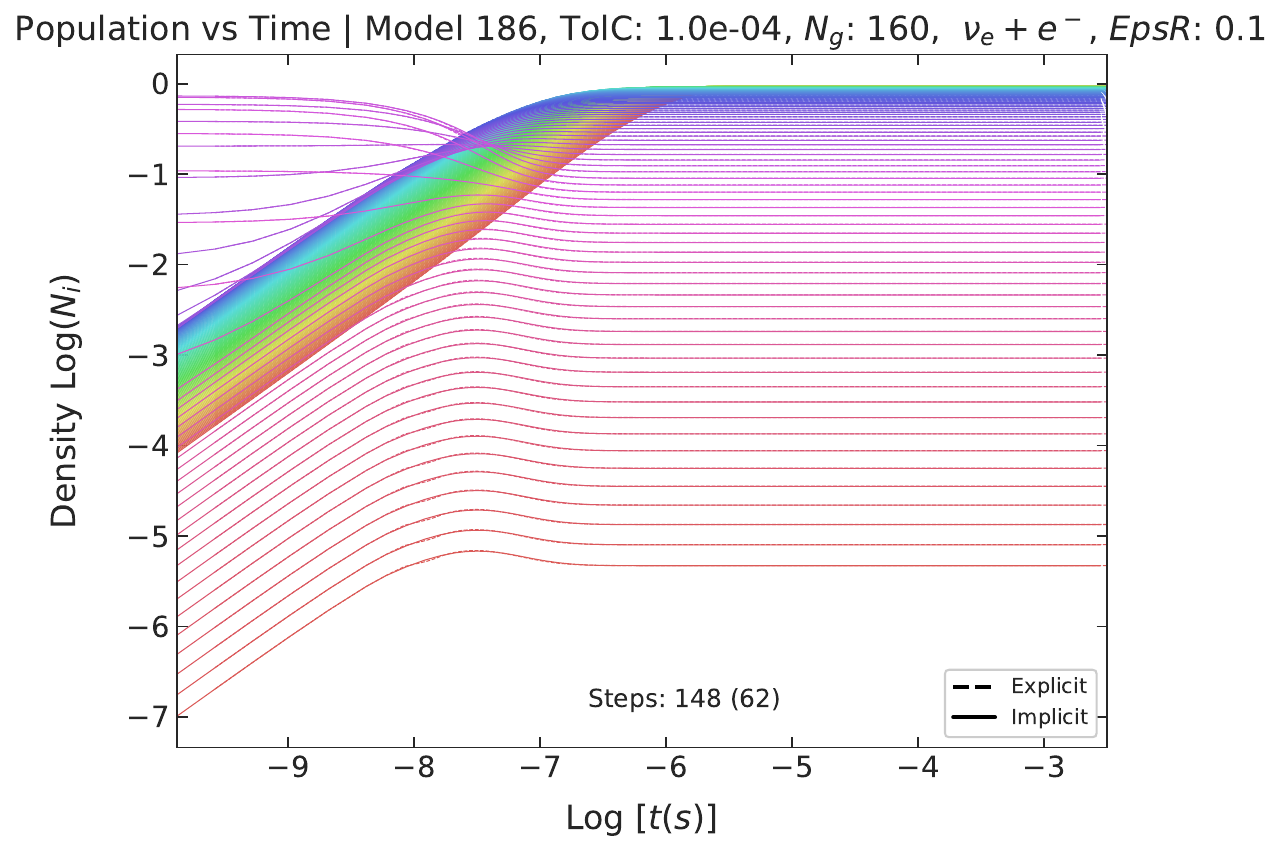}
    \caption{Population vs. time for Network Size 160.}
    \label{fig:pop160}
\end{figure}

\begin{figure}[H]
    \centering
    \includegraphics[width=5in]{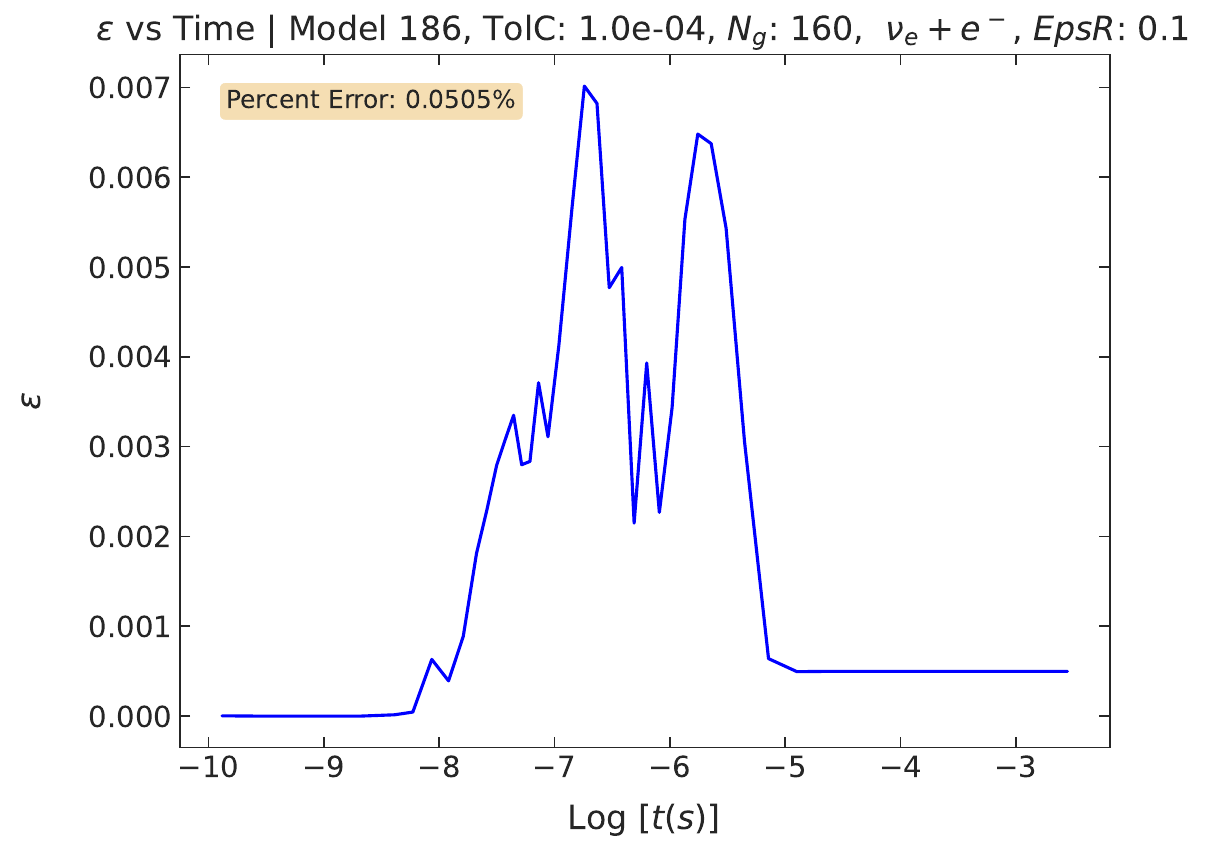}
    \caption{Epsilon vs. time for Network Size 160.}
    \label{fig:epsilon160}
\end{figure}

\begin{figure}[H]
    \centering
    \includegraphics[width=5in]{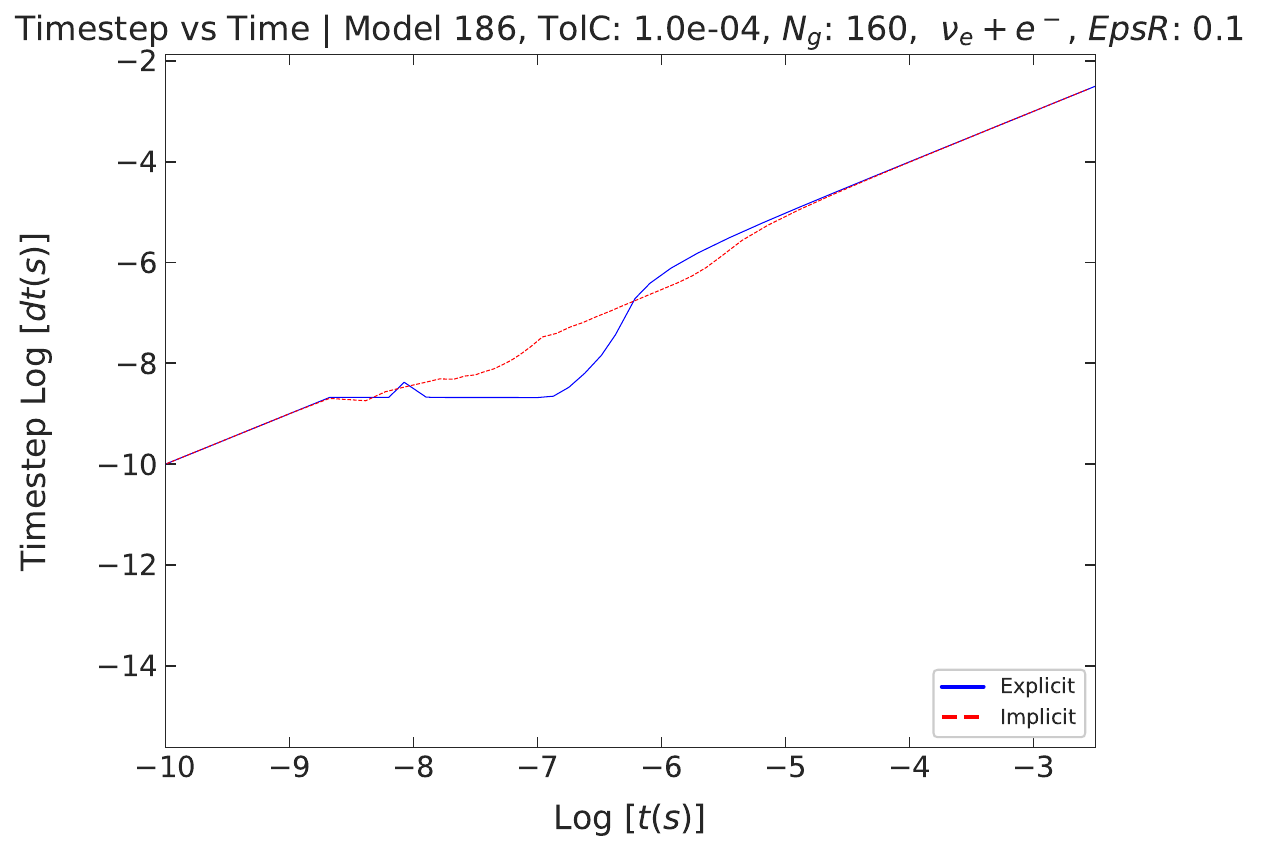}
    \caption{Timestep vs. time for Network Size 160.}
    \label{fig:timestep160}
\end{figure}

\subsection{Network Size 180}
\label{subsec:network_180}

\begin{figure}[H]
    \centering
    \includegraphics[width=5in]{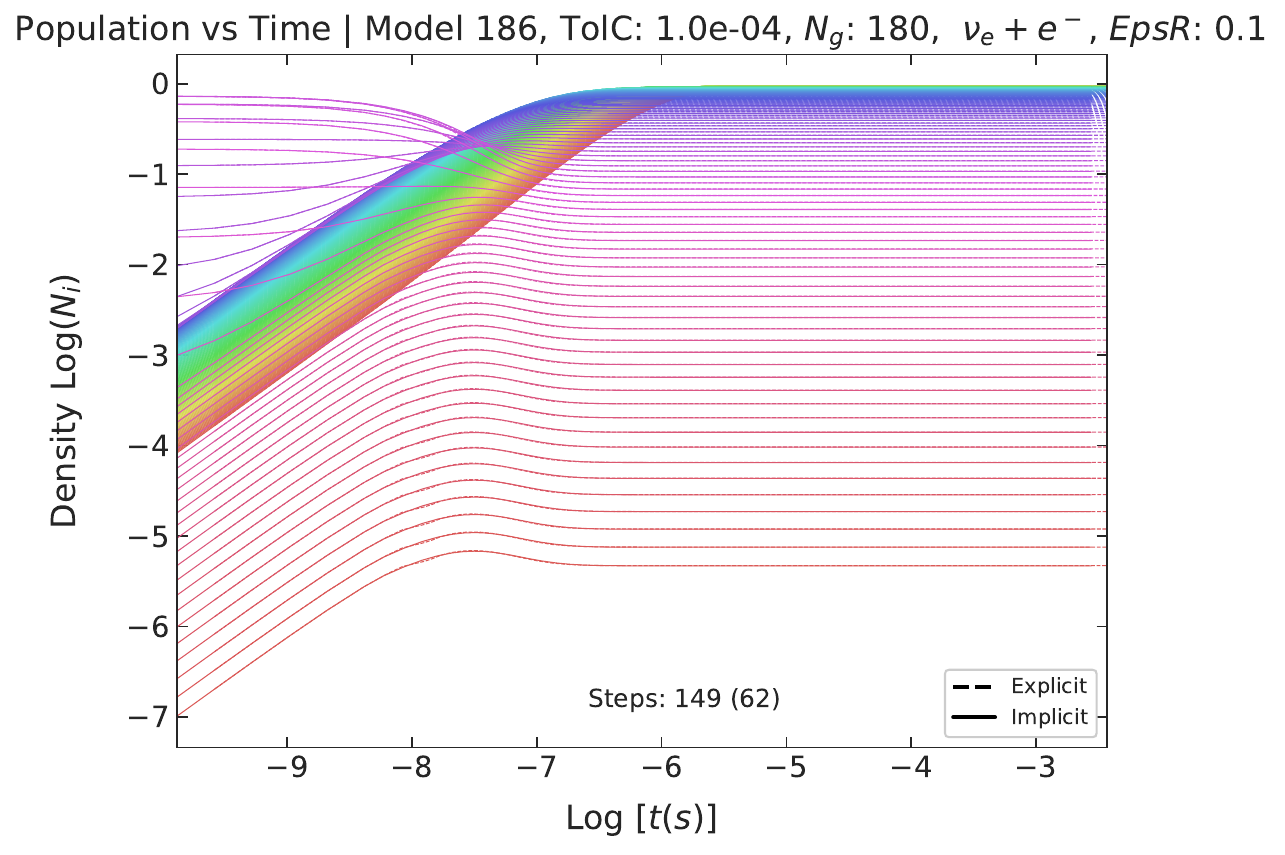}
    \caption{Population vs. time for Network Size 180.}
    \label{fig:pop180}
\end{figure}

\begin{figure}[H]
    \centering
    \includegraphics[width=5in]{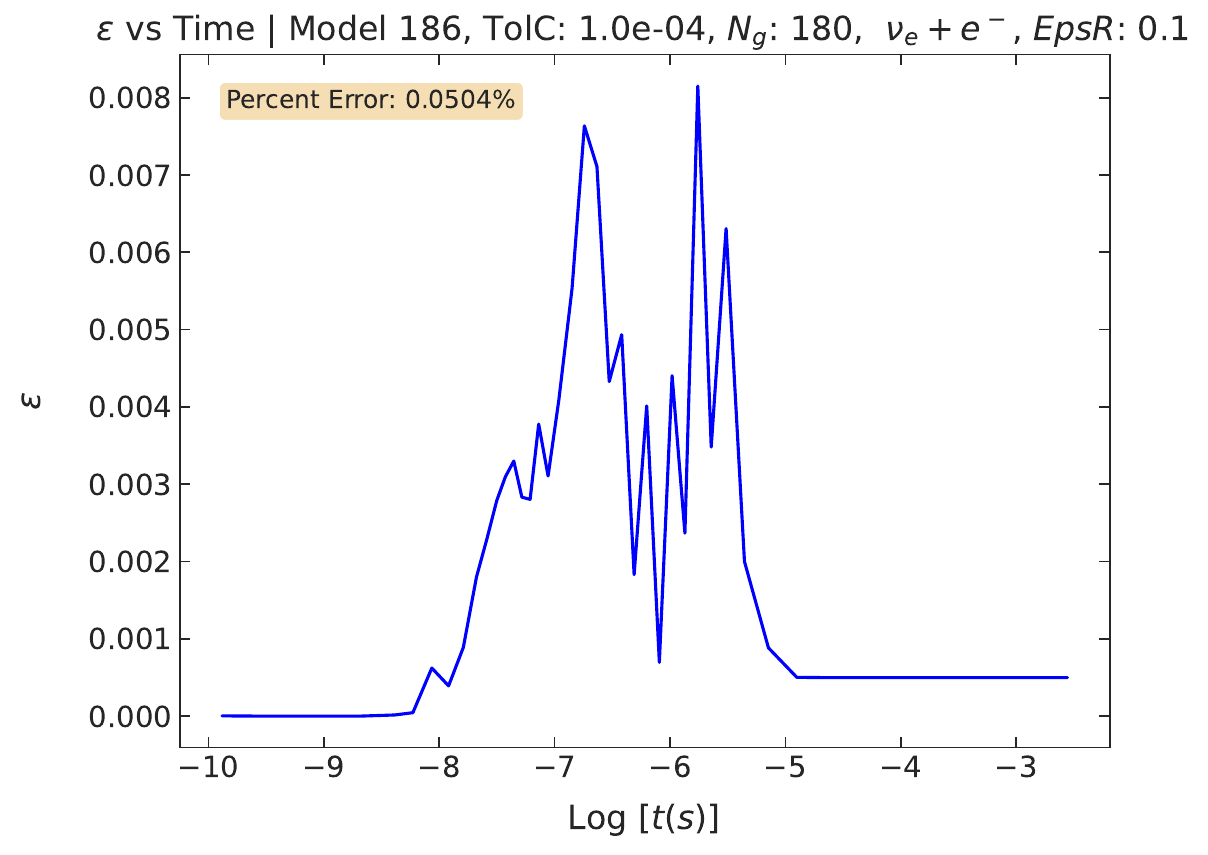}
    \caption{Epsilon vs. time for Network Size 180.}
    \label{fig:epsilon180}
\end{figure}

Additionally, we can display one of our cases (180 Size Network) in a heatmap to illustrate the differences in the scattering kernels.
\begin{figure}[H]
    \centering
    \includegraphics[width=5in]{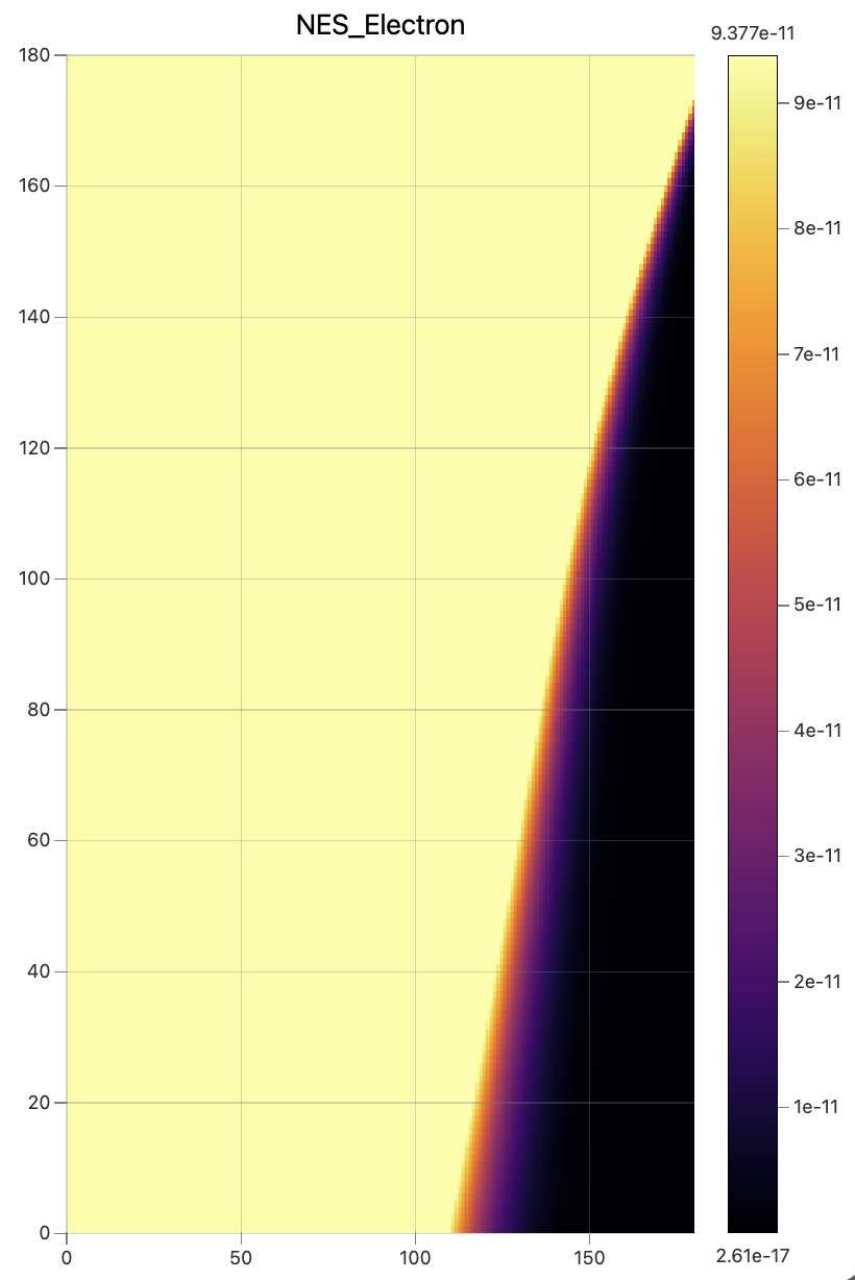}
    \caption{Heat map of Scattering Rates for a 180 Size Network}
    \label{fig:ScaledHeatmap}
\end{figure}

The error from these figures is tabulated into table \ref{tab:steps_and_errors}.

\begin{table}[H]
\centering
\begin{tabular}{ccccc}
\hline
\textbf{Network Size} & \textbf{Explicit Steps} & \textbf{Backward Euler Steps} & \textbf{Error (\%)} \\
\hline
40 & 140 & 61 & 0.0503 \\
50 & 142 & 62 & 0.0506 \\
80 & 146 & 62 & 0.0538 \\
100 & 147 & 62 & 0.0544 \\
130 & 148 & 62 & 0.0521 \\
160 & 148 & 62 & 0.0505 \\
180 & 149 & 62 & 0.0504 \\
\hline
\end{tabular}
\caption{Comparison of Explicit Asymptotic and Backward Euler Steps with Error across Network Sizes.}
\label{tab:steps_and_errors}
\end{table}

It is observed that the number of steps required for both algorithms increases marginally as the network size increases. Importantly, the error remains approximately equal across the various network sizes, which indicates that the precision of the algorithms is not compromised as the networks scale in size while maintaining competitive time stepping.

We can compare the wall clock time for the explicit methods for these various sizes to the wall clock time for the backward Euler method.

\begin{table}[H]
\centering
\begin{tabular}{ccc}
\hline
\textbf{Network Size} & \textbf{Time (Explicit Asymptotic)} & \textbf{Time (LU Solver)} \\
\hline
40 & 211 ms & 800 ms \\
50 & 319 ms & 1220 ms \\
60 & 450 ms & 1780 ms \\
70 & 611 ms & 2484 ms \\
80 & 784 ms & 3308 ms \\
90 & 984 ms & 4298 ms \\
100 & 1208 ms & 5751 ms \\
130 & 2049 ms & 1028 ms \\
160 & 3078 ms & 16997 ms \\
180 & 3873 ms & 22931 ms \\
\hline
\end{tabular}
\caption{Comparison of simulation times for explicit asymptotic and LU solver methods across network sizes.}
\label{tab:explicit_vs_lu}
\end{table}
Using table \ref{tab:explicit_vs_lu} we can derive our speed-up factors:

\begin{table}[ht]
\centering
\begin{tabular}{cccc}
\hline
Network Size & Time (Backward Euler) & Time (Explicit Asymptotic) & Speed Up Factor \\ \hline
40 & 800 ms & 211 ms & $\approx$ 4 \\
50 & 1220 ms & 319 ms & $\approx$ 4 \\
60 & 1789 ms & 450 ms & $\approx$ 4 \\
70 & 2484 ms & 611 ms & $\approx$ 4 \\
80 & 3308 ms & 784 ms & $\approx$ 4.2 \\
90 & 4298 ms & 984 ms & $\approx$ 4.33 \\
100 & 5451 ms & 1208 ms & $\approx$ 4.5 \\
130 & 10280 ms & 2049 ms & $\approx$ 5 \\
160 & 16997 ms & 3078 ms & $\approx$ 5.5 \\
180 & 22931 ms & 3873 ms & $\approx$ 6 \\ \hline
\end{tabular}
\caption{Comparison of computational time and speed up factor for different network sizes}
\label{tab:speedup}
\end{table}
While we used the LU solver in the figures presented in this chapter, we can also show that LU is the most consistent compared to other solvers used in backward Euler. It is important to note we use the LU solver in the backward Euler method for consistency. 

\begin{table}[H]
\centering
\small 
\begin{tabular}{cccc}
\hline
\textbf{Network Size} & \textbf{Time (QR)} & \textbf{Time (Cholesky)} & \textbf{Time (LU)} \\
\hline
40 & 874 ms & 2205 ms & 800 ms \\
50 & 1328 ms & 3463 ms & 1220 ms \\
60 & 1895 ms & 5431 ms & 1789 ms \\
70 & 2647 ms & 7998 ms & 2484 ms \\
80 & 3461 ms & 11207 ms & 3308 ms \\
90 & 4501 ms & 15173 ms & 4298 ms \\
100 & 5714 ms & 20044 ms & 5451 ms \\
130 & 9643 ms & 41092 ms & 10280 ms \\
160 & 15604 ms & 73308 ms & 16997 ms \\
180 & 20646 ms & 101942 ms & 22931 ms \\
\hline
\end{tabular}
\caption{Comparison of Simulation Times for Different Implicit Solvers across Network Sizes.}
\label{tab:implicit_solvers}
\end{table}
These results from tables \ref{tab:implicit_solvers} and \ref{tab:speedup} can be illustrated in the following figures:

\begin{figure}[H]
\centering
\includegraphics[width=5in]{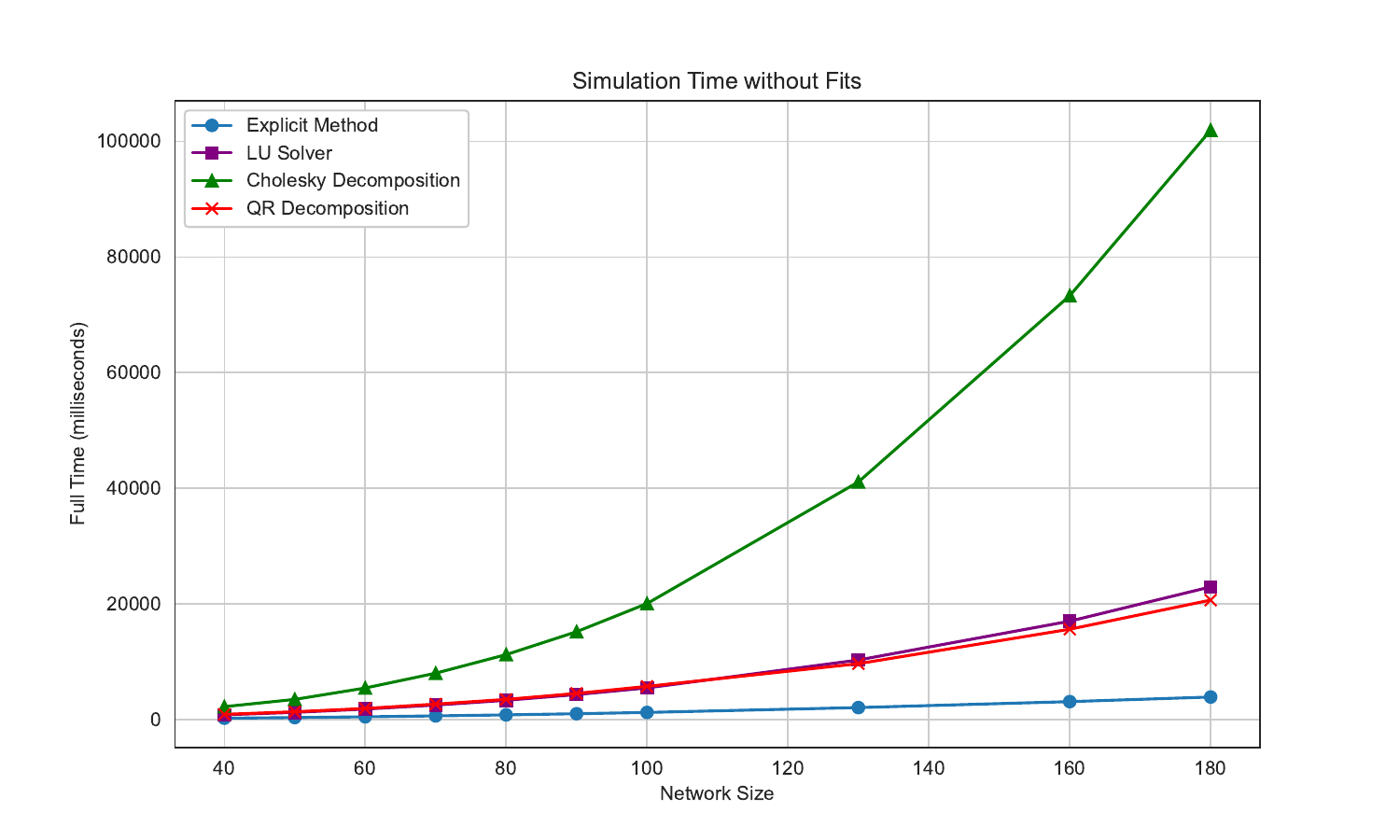}
\caption{Full wall clock times. Includes times for different implicit solvers, along with polynomial fits to show the trends for each method. }
\label{fig:no_fits}
\end{figure}

\begin{figure}[H]
\centering
\includegraphics[width=5in]{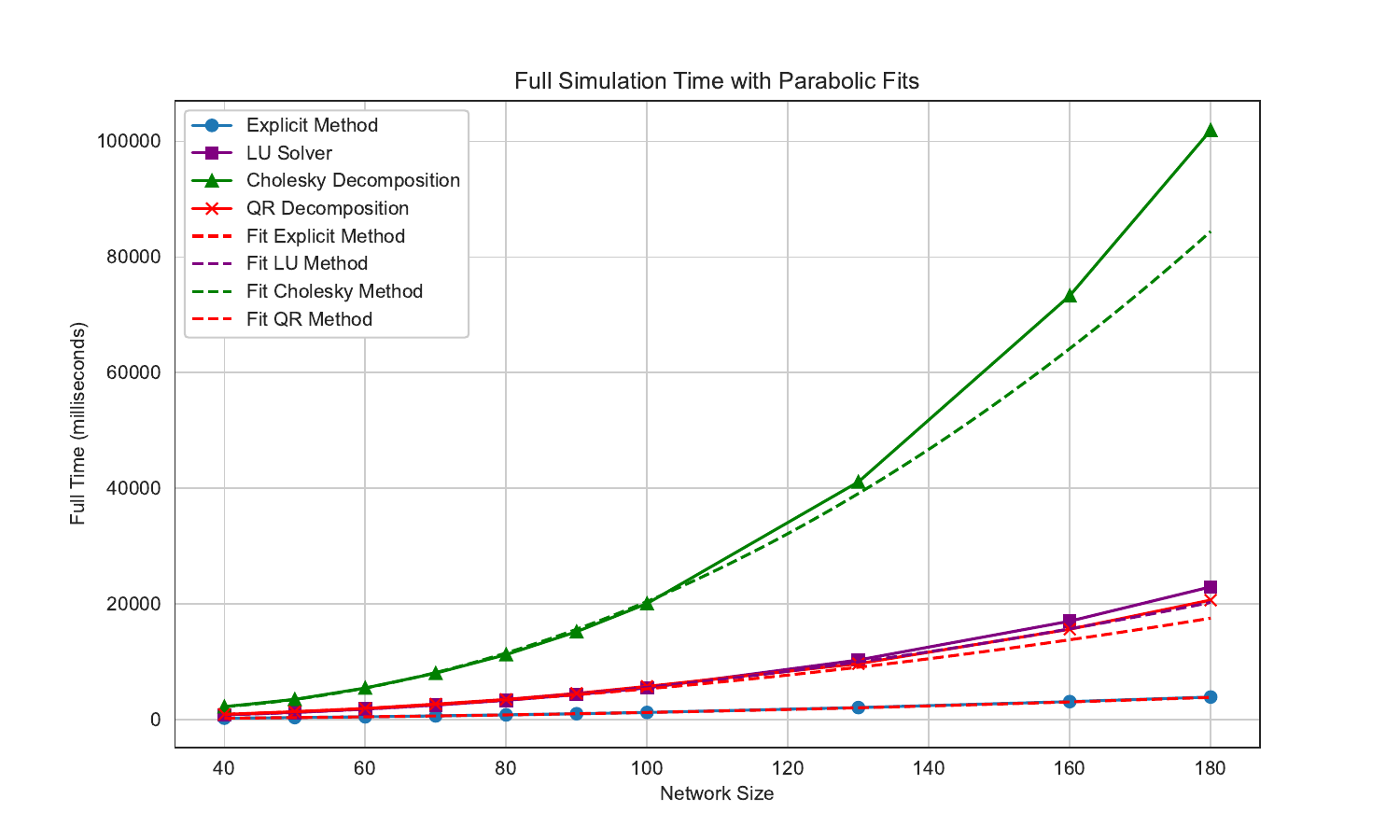}
\caption{Full wall clock times from \ref{fig:no_fits} with polynomial Fits for the first 3 points.}
\label{fig:parabolic_fits}
\end{figure}

\begin{figure}[H]
\centering
\includegraphics[width=5in]{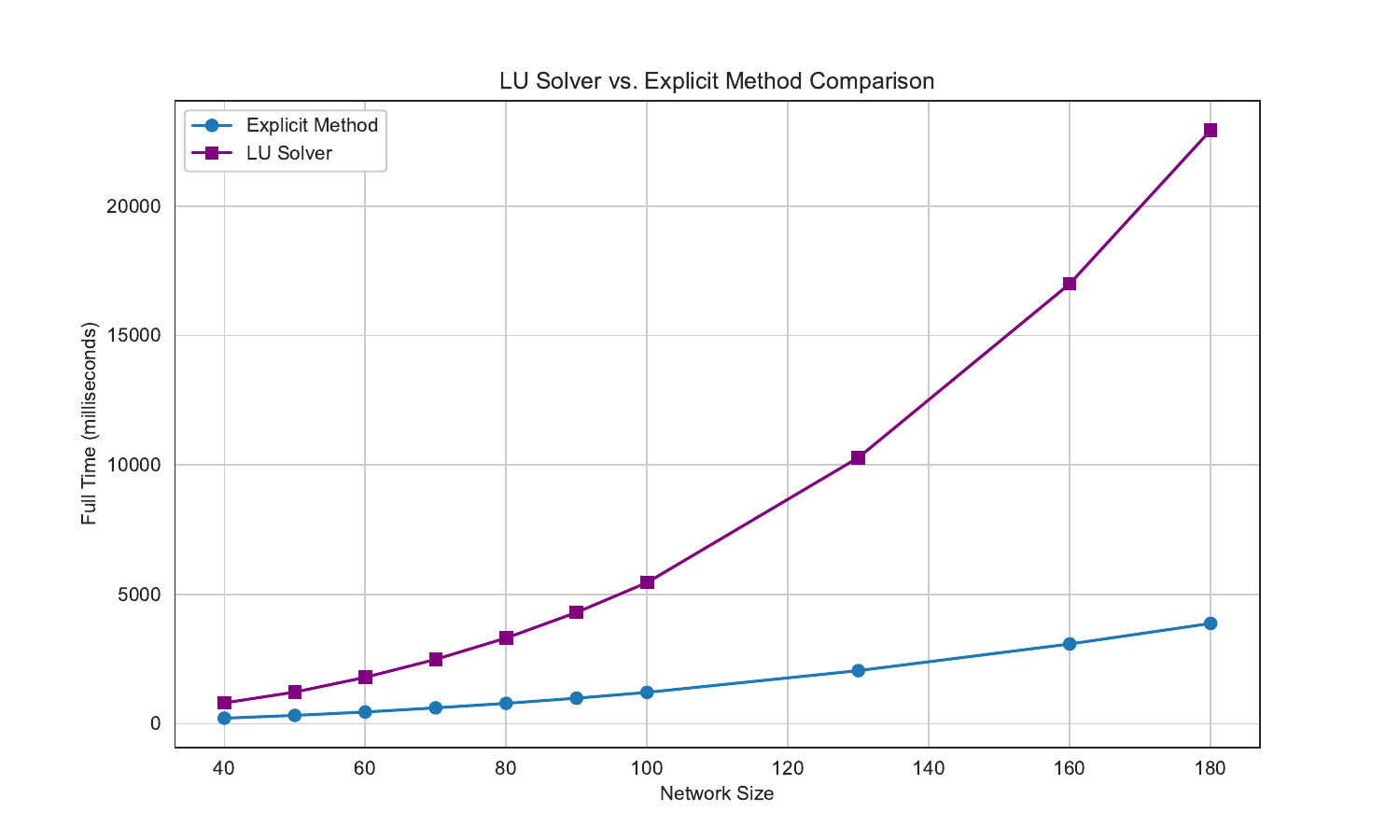}
\caption{LU solver vs. explicit method comparison. LU solver is used in all calculations from \ref{ch:NetworkScaling}.}
\label{fig:lu_vs_explicit}
\end{figure}

\begin{figure}[H]
\centering
\includegraphics[width=5in]{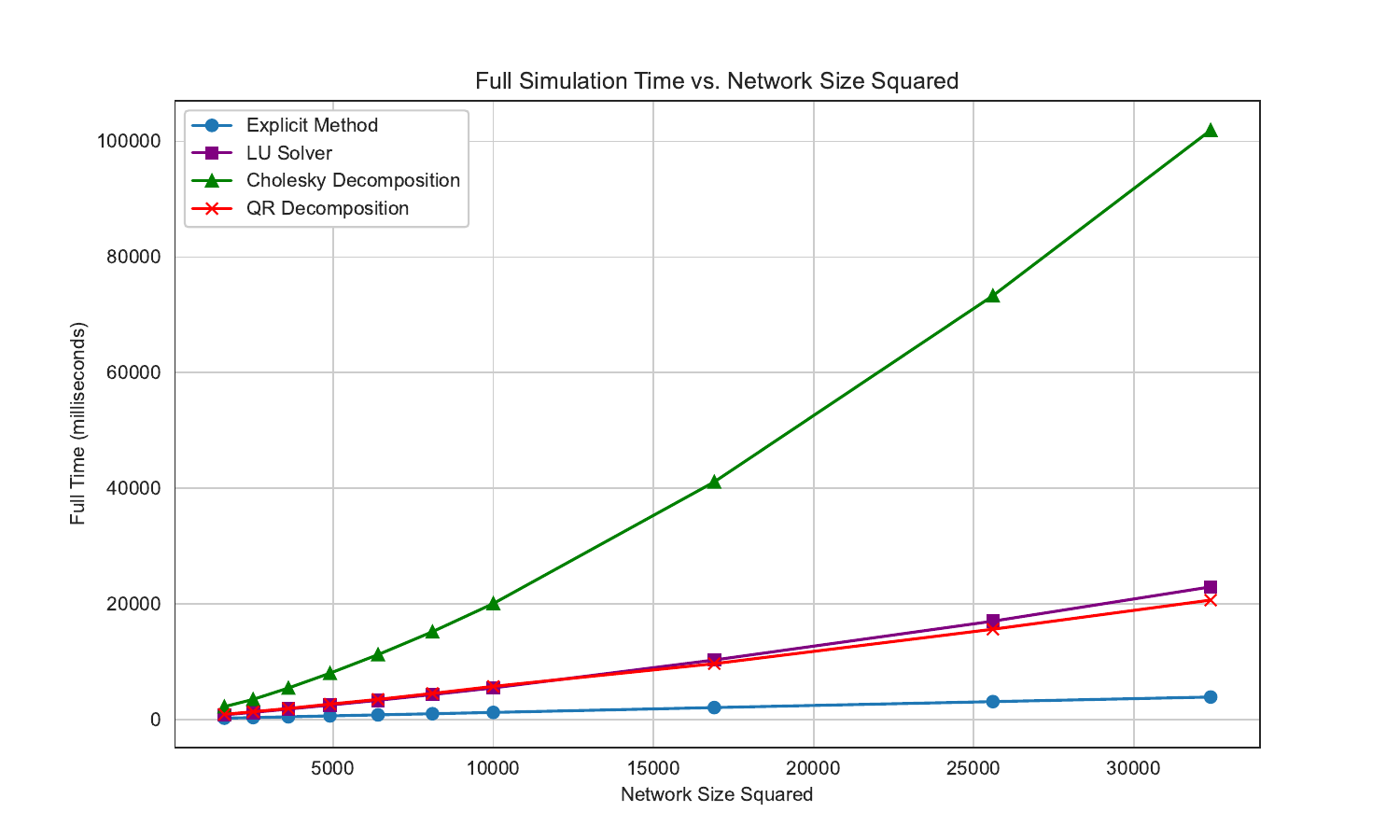}
\caption{Full simulation time vs. network Size squared. This can represent the computational cost and general trends for Network Sizes.}
\label{fig:network_size_squared}
\end{figure}

For a larger network, while the number of steps does increase for computation, but the error stays very constant. This gives the strongest indication of the effectiveness of the methods used numerically, precisely the Explicit Asymptotic and Backward Euler methods, as depicted in table \ref{tab:steps_and_errors}. 

On the other hand, from tables \ref{tab:explicit_vs_lu} and \ref{tab:implicit_solvers}, computation times for the LU solver increase quadratically with network size, sacrificing speed and reliability for larger networks. On the other hand, Cholesky solvers are computationally intensive, and they would be the most preferred in cases where numerical stability is of core importance, especially with the expanding size of the network.

The QR solver proves a viable alternative and outperforms the Cholesky method in smaller networks due to the balance in computation time and accuracy. It is seen that the computational times with respect to these solvers in Figures \ref{fig:no_fits}, \ref{fig:parabolic_fits}, and \ref{fig:network_size_squared} grow polynomially with respect to network size, especially for Cholesky and QR solvers. This trend underlines the significant aspect of choice of the appropriate solver for the network size and the computational load to be expected.

These results overall indicate the scaling calculations of the explicit algebraic algorithms, with network size have a much stronger advantage than those of the implicit methods, in both cases having a quadratic dependence on size. The scaling of the implicit method has a much steeper increase in computational costs, as indicated by a stronger curvature of its scaling parabola. This difference means that explicit methods are much more efficient for large networks. The likely reasons for such a higher curvature, or the increased increase in computational demand for implicit methods, presumably lies in the enhanced complexity and an increased number of matrix inversions at each step, which act non-linearly on computation time. Such a result underscores the potential of algebraic explicit schemes in treating large-scale simulations, and it points in a straightforward way toward more computationally viable and scalable solutions for complex models of neutrino interaction.
    \chapter{Conclusion} \label{ch:Conclusion}

In our model of neutrino-electron scattering ($\nu_e + e^-$), we have shown that the explicit asymptotic approximation yields controlled errors and operates at speeds 1-3 times faster than traditional implicit methods (backward Euler) for networks with 40 neutrino energy bins. We have also shown that this efficiency increases for larger network sizes, demonstrating that these methods tend to scale better than implicit methods with network size. Given that explicit methods are generally more efficient per time step as they avoid the need for matrix inversions—the explicit asymptotic approach to neutrino transport presents a potential for enhanced speed and efficiency in many large-scale simulations. 

We have given an introduction to scattering of various neutrino flavors using FENN from ($\nu_e$), electron anti-neutrinos ($\bar{\nu}_e$), and muon/tau neutrinos ($\nu_{\mu,\tau}$) and their anti-particles ($\bar{\nu}_{\mu,\tau}$). In a full neutrino transport scheme these methods would be more systematically considered, and this lays groundwork for flexibility to include the full neutrino transport regimes in future work.

Additionally, we have verified that for larger networks these methods may scale better than implicit methods, because of the absence of matrix inversions. Therefore, we advocate for the use of algebraically stabilized explicit methods, capable of integrating realistic neutrino (and thermonuclear) networks within multidimensional hydrodynamics on current high-performance platforms.

The C++ codebase FENN: ``Fast Explicit Neutrino Networks'' will  be released to the astrophysics community as documented open-source code for the rapid solution of large kinetic networks coupled to fluid dynamics, and to the broader scientific community as an open-source template for disciplines benefiting from the scalability and efficiency of algebraically stabilized explicit integration, especially when deployed on modern GPUs.

First introduced for Neutrino Networks by \citep{2024AAS...24326034C}, these methods have shown potential performance in terms of effectively introducing far greater efficiency on GPU architectures than that exhibited by more traditional implicit approaches. The explicit methods show better performance when scaling the the neutrino networks, especially when ported to GPUs, because of efficient parallel processing. Most importantly, the ability to perform well for networks sizes that are larger than the standard 40 energy bins is paramount since it can be used to handle the large and complex nature of simulations with more efficiency. On the other hand, explicit methods architecture reduces dependence between computational elements and hence makes better use of the parallelism of the GPU. 

Further work will be needed to optimize and port the FENN codebase to GPU environments. This move is expected to bring significant gains in performance, specifically with the processing speed and ability of handling larger datasets more effectively. 

Additionally, alternative approximations such as Quasi Steady State \citep{guidQSS} and Partial Equilibrium \citep{guidPE} can be applied to certain time periods of the integration to best approximate the full integration. All the calculations done in this thesis run a full Explicit Asymptotic integration over the entire network, however we can better approximate this by using combinations of all three methods. In future work, we hope to address this as well by incorporating the already established methods for thermonuclear networks into the neutrino networks addressed by FENN.

    \makeBibliographyPage 
    \bibliographystyle{apalike} 
    \bibliography{RaghavChari_Thesis/references/PaxHeader/my_dissertation} 

\begin{thebibliography}{}

\bibitem[Brey, 2022]{brey2022}
Brey, N. (2022).
\newblock Analysis of controlled approximations for explicit integration of stiff thermonuclear networks.
\newblock MS thesis, University of Tennessee; \url{https://trace.tennessee.edu/utk_gradthes/9259}.

\bibitem[Bruenn, 1985]{bru85}
Bruenn, S.~W. (1985).
\newblock {\em App.\ J. Supp.}, 58:771.

\bibitem[{Chari} et~al., 2024]{2024AAS...24313505C}
{Chari}, R., {Cole}, A., {Guidry}, M., {Brey}, N., {Endeve}, E., and {Crowley}, R. (2024).
\newblock {Advancing Astrophysical Models through FENN: Algebraically Stabilized Explicit Integration for Neutrino Electron Scattering in Stellar Explosions and Mergers}.
\newblock In {\em American Astronomical Society Meeting Abstracts}, volume~56 of {\em American Astronomical Society Meeting Abstracts}, page 135.05.

\bibitem[Chupryna, 2008]{chup2008}
Chupryna, V. (2008).
\newblock Explicit methods in the nuclear burning problem for supernova ia models.
\newblock doctoral thesis, University of Tennessee; \url{https://trace.tennessee.edu/utk_graddiss/484}.

\bibitem[{Cole} et~al., 2024]{2024AAS...24326034C}
{Cole}, A., {Chari}, R., {Brey}, N., {Crowley}, R., {Guidry}, M., and {Endeve}, E. (2024).
\newblock {Controlled and Parallelizable Approximation for Evolution Populations and Neutrino Distributions in Stellar Explosions and Mergers.}
\newblock In {\em American Astronomical Society Meeting Abstracts}, volume~56 of {\em American Astronomical Society Meeting Abstracts}, page 260.34.

\bibitem[Feger, 2011]{fege2011}
Feger, E. (2011).
\newblock Evaluating explicit methods for solving astrophysical nuclear reaction networks.
\newblock doctoral thesis, University of Tennessee; \url{https://trace.tennessee.edu/utk_graddiss/1048}.

\bibitem[Gear, 1971]{gear71}
Gear, C.~W. (1971).
\newblock {\em Numerical Initial Value Problems in Ordinary Differential Equations}.
\newblock Prentice Hall, Englewood Cliffs, N. J.

\bibitem[Guidry et~al., 2023]{guid2023a}
Guidry, M., Brey, N., Billings, J., and Hix, R. (2023).
\newblock A controlled approximation for solving large kinetic networks coupled to fluid dynamics.
\newblock Manuscipt in preparation.

\bibitem[Guidry, 2012]{guidJCP}
Guidry, M.~W. (2012).
\newblock {\em J. Comp.\ Phys.}, 231:5266.

\bibitem[Guidry, 2016]{guid2016}
Guidry, M.~W. (2016).
\newblock Efficient gpu acceleration for integrating large thermonuclear networks in astrophysics.
\newblock In {\em EPJ Web of Conferences {\bf 109}}, page 06003. EDP Sciences.

\bibitem[Guidry et~al., 2013a]{guidPE}
Guidry, M.~W., Billings, J.~J., and Hix, W.~R. (2013a).
\newblock {\em Comput.\ Sci.\ Disc.}, 6:015003.

\bibitem[Guidry et~al., 2013b]{guidAsy}
Guidry, M.~W., Budiardja, R., Feger, E., Billings, J.~J., Hix, W.~R., Messer, O. E.~B., Roche, K.~J., McMahon, E., and He, M. (2013b).
\newblock {\em Comput.\ Sci.\ Disc.}, 6:015001.

\bibitem[Guidry and Harris, 2013]{guidQSS}
Guidry, M.~W. and Harris, J.~A. (2013).
\newblock {\em Comput.\ Sci.\ Disc.}, 6:015002.

\bibitem[Haidar et~al., 2016]{haid2016}
Haidar, A., Brock, B., Tomov, S., Guidry, M., Billings, J.~J., Shyles, D., and Dongarra, J. (2016).
\newblock Performance analysis and acceleration of explicit integration for large kinetic networks using batched gpu computations.
\newblock IEEE High Performance Extreme Computing Conference, HPEC.

\bibitem[Haidar et~al., 2015]{haid2015}
Haidar, A., Dong, T., Tomov, S., Luszczek, P., and Dongarra, J. (2015).
\newblock Framework for batched and gpu-resident factorization algorithms to block householder transformations.
\newblock ISC High Performance, Springer Frankfurt.

\bibitem[Hix and Thielemann, 1999]{hix1999}
Hix, W.~R. and Thielemann, F.~K. (1999).
\newblock {\em J. Comp. Appl.\ Math.}, 109:321.

\bibitem[Lackey-Stewart, 2020]{lac2020}
Lackey-Stewart, A. (2020).
\newblock An explicit asymptotic approach applied to neutrino-electron scattering in the neutrino transport problem.
\newblock MS thesis, University of Tennessee; \url{https://trace.tennessee.edu/utk_gradthes/5845}.

\bibitem[Lackey-Stewart et~al., 2024]{PhysRevD.109.103019}
Lackey-Stewart, A., Chari, R., Cole, A., Brey, N., Gregory, K., Crowley, R., Guidry, M., and Endeve, E. (2024).
\newblock Fast explicit solutions for neutrino-electron scattering: Explicit asymptotic methods.
\newblock {\em Phys. Rev. D}, 109:103019.

\bibitem[Lambert, 1991]{lam91}
Lambert, J. (1991).
\newblock {\em Numerical Methods for Ordinary Differential Equations}.
\newblock Wiley, New York.

\bibitem[{Mezzacappa} and {Bruenn}, 1993]{1993ApJ...410..740M}
{Mezzacappa}, A. and {Bruenn}, S.~W. (1993).
\newblock {Stellar Core Collapse: A Boltzmann Treatment of Neutrino-Electron Scattering}.
\newblock {\em APJ}, 410:740.

\bibitem[Oran and Boris, 2005]{oran05}
Oran, E. and Boris, J. (2005).
\newblock {\em Numerical Simulation of Reactive Flow}.
\newblock Cambridge University Press, Cambridge.

\bibitem[Press et~al., 1992]{press92}
Press, W., Teukolsky, S., Vettering, W., and Flannery, B. (1992).
\newblock {\em Numerical Recipes in Fortran}.
\newblock Cambridge University Press, Cambridge.

\bibitem[Smit and Cernohorsky, 1996]{smit96}
Smit, J.~M. and Cernohorsky, J. (1996).
\newblock {\em A \& A}, 311:347.

\bibitem[Yueh and Buchler, 1976]{Yueh1976}
Yueh, W.~R. and Buchler, J.~R. (1976).
\newblock Scattering functions for neutrino transport.
\newblock {\em Astrophysics and Space Science}, 39(2):429--435.

\end{thebibliography}
    \makeAppendixPage   
    \appendix 
    \chapter{Supernova Conditions} \label{ch:appendix}
\addcontentsline{lot}{table}{\protect\numberline{}Supernova Conditions}

\begin{longtable}{ccccc}
\toprule
Model & $r$ & $\rho$ & $T$ & $Y_e$ \\
\midrule
\endfirsthead

\multicolumn{5}{c}%
{{\bfseries \tablename\ \thetable{} -- continued from previous page}} \\
\toprule
Model & $r$ & $\rho$ & $T$ & $Y_e$ \\
\midrule
\endhead

\midrule
\multicolumn{5}{r}{{Continued on next page}} \\
\midrule
\endfoot

\bottomrule
\endlastfoot
1 & 2.28E+07 & 1.00E+08 & 6.91E+09 & 0.4980 \\
2 & 2.25E+07 & 1.02E+08 & 6.92E+09 & 0.4980 \\
3 & 2.22E+07 & 1.04E+08 & 6.94E+09 & 0.4980 \\
4 & 2.20E+07 & 1.06E+08 & 6.96E+09 & 0.4979 \\
5 & 2.17E+07 & 1.07E+08 & 6.98E+09 & 0.4979 \\
6 & 2.14E+07 & 1.09E+08 & 6.99E+09 & 0.4979 \\
7 & 2.12E+07 & 1.11E+08 & 7.01E+09 & 0.4979 \\
8 & 2.10E+07 & 1.13E+08 & 7.03E+09 & 0.4978 \\
9 & 2.07E+07 & 1.14E+08 & 7.04E+09 & 0.4978 \\
10 & 2.05E+07 & 1.16E+08 & 7.06E+09 & 0.4978 \\
11 & 2.02E+07 & 1.18E+08 & 7.07E+09 & 0.4978 \\
12 & 2.00E+07 & 1.20E+08 & 7.09E+09 & 0.4977 \\
13 & 1.98E+07 & 1.22E+08 & 7.11E+09 & 0.4977 \\
14 & 1.95E+07 & 1.24E+08 & 7.13E+09 & 0.4977 \\
15 & 1.93E+07 & 1.26E+08 & 7.15E+09 & 0.4977 \\
16 & 1.90E+07 & 1.29E+08 & 7.17E+09 & 0.4977 \\
17 & 1.88E+07 & 1.30E+08 & 7.20E+09 & 0.4977 \\
18 & 1.86E+07 & 1.33E+08 & 7.21E+09 & 0.4977 \\
19 & 1.84E+07 & 1.35E+08 & 7.28E+09 & 0.4976 \\
20 & 1.82E+07 & 1.37E+08 & 7.21E+09 & 0.4976 \\
21 & 1.80E+07 & 1.50E+08 & 7.41E+09 & 0.4976 \\
22 & 1.77E+07 & 2.59E+08 & 8.60E+09 & 0.4976 \\
23 & 1.75E+07 & 6.22E+08 & 1.16E+10 & 0.4976 \\
24 & 1.73E+07 & 1.03E+09 & 1.46E+10 & 0.4975 \\
25 & 1.71E+07 & 1.34E+09 & 1.52E+10 & 0.4968 \\
26 & 1.69E+07 & 1.54E+09 & 1.54E+10 & 0.4961 \\
27 & 1.67E+07 & 1.62E+09 & 1.56E+10 & 0.4951 \\
28 & 1.65E+07 & 1.70E+09 & 1.60E+10 & 0.4937 \\
29 & 1.63E+07 & 1.78E+09 & 1.62E+10 & 0.4918 \\
30 & 1.61E+07 & 1.92E+09 & 1.64E+10 & 0.4902 \\
31 & 1.59E+07 & 2.06E+09 & 1.65E+10 & 0.4882 \\
32 & 1.57E+07 & 2.17E+09 & 1.68E+10 & 0.4852 \\
33 & 1.55E+07 & 2.28E+09 & 1.72E+10 & 0.4819 \\
34 & 1.53E+07 & 2.42E+09 & 1.74E+10 & 0.4794 \\
35 & 1.51E+07 & 2.61E+09 & 1.75E+10 & 0.4757 \\
36 & 1.50E+07 & 2.75E+09 & 1.78E+10 & 0.4709 \\
37 & 1.48E+07 & 2.85E+09 & 1.82E+10 & 0.4659 \\
38 & 1.46E+07 & 2.99E+09 & 1.86E+10 & 0.4612 \\
39 & 1.44E+07 & 3.20E+09 & 1.87E+10 & 0.4557 \\
40 & 1.42E+07 & 3.41E+09 & 1.90E+10 & 0.4492 \\
41 & 1.40E+07 & 3.51E+09 & 1.96E+10 & 0.4422 \\
42 & 1.39E+07 & 3.65E+09 & 2.00E+10 & 0.4347 \\
43 & 1.37E+07 & 3.85E+09 & 2.04E+10 & 0.4274 \\
44 & 1.35E+07 & 4.04E+09 & 2.08E+10 & 0.4193 \\
45 & 1.34E+07 & 4.19E+09 & 2.14E+10 & 0.4104 \\
46 & 1.32E+07 & 4.29E+09 & 2.23E+10 & 0.4012 \\
47 & 1.30E+07 & 4.42E+09 & 2.31E+10 & 0.3927 \\
48 & 1.29E+07 & 4.62E+09 & 2.37E+10 & 0.3837 \\
49 & 1.27E+07 & 4.77E+09 & 2.45E+10 & 0.3748 \\
50 & 1.25E+07 & 4.89E+09 & 2.54E+10 & 0.3661 \\
51 & 1.24E+07 & 5.02E+09 & 2.63E+10 & 0.3581 \\
52 & 1.22E+07 & 5.20E+09 & 2.70E+10 & 0.3502 \\
53 & 1.21E+07 & 5.40E+09 & 2.76E+10 & 0.3427 \\
54 & 1.19E+07 & 5.60E+09 & 2.83E+10 & 0.3358 \\
55 & 1.18E+07 & 5.81E+09 & 2.89E+10 & 0.3293 \\
56 & 1.16E+07 & 6.02E+09 & 2.96E+10 & 0.3233 \\
57 & 1.15E+07 & 6.23E+09 & 3.02E+10 & 0.3178 \\
58 & 1.13E+07 & 6.44E+09 & 3.09E+10 & 0.3135 \\
59 & 1.12E+07 & 6.66E+09 & 3.16E+10 & 0.3095 \\
60 & 1.10E+07 & 6.94E+09 & 3.21E+10 & 0.3055 \\
61 & 1.09E+07 & 7.25E+09 & 3.26E+10 & 0.3018 \\
62 & 1.07E+07 & 7.56E+09 & 3.31E+10 & 0.2986 \\
63 & 1.06E+07 & 7.86E+09 & 3.37E+10 & 0.2962 \\
64 & 1.05E+07 & 8.20E+09 & 3.42E+10 & 0.2939 \\
65 & 1.03E+07 & 8.57E+09 & 3.47E+10 & 0.2915 \\
66 & 1.02E+07 & 8.99E+09 & 3.51E+10 & 0.2888 \\
67 & 1.00E+07 & 9.43E+09 & 3.56E+10 & 0.2865 \\
68 & 9.92E+06 & 9.89E+09 & 3.61E+10 & 0.2845 \\
69 & 9.79E+06 & 1.04E+10 & 3.66E+10 & 0.2829 \\
70 & 9.66E+06 & 1.08E+10 & 3.71E+10 & 0.2811 \\
71 & 9.53E+06 & 1.14E+10 & 3.76E+10 & 0.2790 \\
72 & 9.40E+06 & 1.20E+10 & 3.80E+10 & 0.2766 \\
73 & 9.28E+06 & 1.27E+10 & 3.84E+10 & 0.2739 \\
74 & 9.15E+06 & 1.34E+10 & 3.88E+10 & 0.2709 \\
75 & 9.03E+06 & 1.43E+10 & 3.92E+10 & 0.2677 \\
76 & 8.90E+06 & 1.52E+10 & 3.96E+10 & 0.2643 \\
77 & 8.78E+06 & 1.61E+10 & 4.00E+10 & 0.2606 \\
78 & 8.66E+06 & 1.72E+10 & 4.03E+10 & 0.2568 \\
79 & 8.54E+06 & 1.84E+10 & 4.07E+10 & 0.2530 \\
80 & 8.43E+06 & 1.96E+10 & 4.10E+10 & 0.2490 \\
81 & 8.31E+06 & 2.10E+10 & 4.14E+10 & 0.2451 \\
82 & 8.20E+06 & 2.25E+10 & 4.18E+10 & 0.2409 \\
83 & 8.08E+06 & 2.41E+10 & 4.22E+10 & 0.2367 \\
84 & 7.97E+06 & 2.59E+10 & 4.25E+10 & 0.2327 \\
85 & 7.86E+06 & 2.78E+10 & 4.29E+10 & 0.2286 \\
86 & 7.75E+06 & 2.99E+10 & 4.33E+10 & 0.2247 \\
87 & 7.64E+06 & 3.22E+10 & 4.37E+10 & 0.2206 \\
88 & 7.53E+06 & 3.46E+10 & 4.41E+10 & 0.2167 \\
89 & 7.42E+06 & 3.73E+10 & 4.45E+10 & 0.2127 \\
90 & 7.32E+06 & 4.03E+10 & 4.50E+10 & 0.2089 \\
91 & 7.21E+06 & 4.34E+10 & 4.54E+10 & 0.2053 \\
92 & 7.11E+06 & 4.69E+10 & 4.59E+10 & 0.2017 \\
93 & 7.01E+06 & 5.06E+10 & 4.64E+10 & 0.1982 \\
94 & 6.91E+06 & 5.47E+10 & 4.70E+10 & 0.1947 \\
95 & 6.81E+06 & 5.92E+10 & 4.75E+10 & 0.1912 \\
96 & 6.71E+06 & 6.40E+10 & 4.81E+10 & 0.1881 \\
97 & 6.61E+06 & 6.93E+10 & 4.87E+10 & 0.1851 \\
98 & 6.51E+06 & 7.50E+10 & 4.93E+10 & 0.1820 \\
99 & 6.42E+06 & 8.12E+10 & 5.00E+10 & 0.1789 \\
100 & 6.32E+06 & 8.80E+10 & 5.06E+10 & 0.1762 \\
101 & 6.23E+06 & 9.53E+10 & 5.14E+10 & 0.1737 \\
102 & 6.13E+06 & 1.04E+11 & 5.21E+10 & 0.1718 \\
103 & 6.04E+06 & 1.12E+11 & 5.28E+10 & 0.1692 \\
104 & 5.95E+06 & 1.22E+11 & 5.36E+10 & 0.1667 \\
105 & 5.86E+06 & 1.33E+11 & 5.45E+10 & 0.1641 \\
106 & 5.77E+06 & 1.44E+11 & 5.53E+10 & 0.1615 \\
107 & 5.68E+06 & 1.57E+11 & 5.62E+10 & 0.1589 \\
108 & 5.59E+06 & 1.70E+11 & 5.71E+10 & 0.1573 \\
109 & 5.51E+06 & 1.85E+11 & 5.81E+10 & 0.1557 \\
110 & 5.42E+06 & 2.01E+11 & 5.91E+10 & 0.1540 \\
111 & 5.34E+06 & 2.18E+11 & 6.02E+10 & 0.1524 \\
112 & 5.25E+06 & 2.38E+11 & 6.13E+10 & 0.1507 \\
113 & 5.17E+06 & 2.58E+11 & 6.24E+10 & 0.1489 \\
114 & 5.08E+06 & 2.81E+11 & 6.36E+10 & 0.1473 \\
115 & 5.00E+06 & 3.05E+11 & 6.48E+10 & 0.1464 \\
116 & 4.92E+06 & 3.31E+11 & 6.61E+10 & 0.1453 \\
117 & 4.84E+06 & 3.60E+11 & 6.75E+10 & 0.1441 \\
118 & 4.76E+06 & 3.91E+11 & 6.89E+10 & 0.1429 \\
119 & 4.69E+06 & 4.25E+11 & 7.03E+10 & 0.1416 \\
120 & 4.61E+06 & 4.62E+11 & 7.18E+10 & 0.1408 \\
121 & 4.53E+06 & 5.01E+11 & 7.34E+10 & 0.1402 \\
122 & 4.46E+06 & 5.44E+11 & 7.50E+10 & 0.1397 \\
123 & 4.38E+06 & 5.90E+11 & 7.66E+10 & 0.1390 \\
124 & 4.31E+06 & 6.41E+11 & 7.83E+10 & 0.1382 \\
125 & 4.23E+06 & 6.97E+11 & 8.01E+10 & 0.1375 \\
126 & 4.16E+06 & 7.56E+11 & 8.19E+10 & 0.1374 \\
127 & 4.09E+06 & 8.20E+11 & 8.38E+10 & 0.1370 \\
128 & 4.02E+06 & 8.91E+11 & 8.57E+10 & 0.1365 \\
129 & 3.95E+06 & 9.68E+11 & 8.77E+10 & 0.1358 \\
130 & 3.88E+06 & 1.05E+12 & 8.97E+10 & 0.1352 \\
131 & 3.81E+06 & 1.14E+12 & 9.17E+10 & 0.1350 \\
132 & 3.74E+06 & 1.24E+12 & 9.38E+10 & 0.1347 \\
133 & 3.67E+06 & 1.35E+12 & 9.60E+10 & 0.1341 \\
134 & 3.61E+06 & 1.47E+12 & 9.82E+10 & 0.1333 \\
135 & 3.54E+06 & 1.60E+12 & 1.00E+11 & 0.1325 \\
136 & 3.48E+06 & 1.74E+12 & 1.03E+11 & 0.1322 \\
137 & 3.41E+06 & 1.90E+12 & 1.05E+11 & 0.1317 \\
138 & 3.35E+06 & 2.07E+12 & 1.08E+11 & 0.1310 \\
139 & 3.28E+06 & 2.26E+12 & 1.10E+11 & 0.1300 \\
140 & 3.22E+06 & 2.47E+12 & 1.12E+11 & 0.1290 \\
141 & 3.16E+06 & 2.70E+12 & 1.15E+11 & 0.1286 \\
142 & 3.10E+06 & 2.95E+12 & 1.18E+11 & 0.1281 \\
143 & 3.04E+06 & 3.23E+12 & 1.21E+11 & 0.1274 \\
144 & 2.98E+06 & 3.53E+12 & 1.24E+11 & 0.1267 \\
145 & 2.92E+06 & 3.86E+12 & 1.26E+11 & 0.1260 \\
146 & 2.86E+06 & 4.22E+12 & 1.30E+11 & 0.1260 \\
147 & 2.80E+06 & 4.61E+12 & 1.33E+11 & 0.1261 \\
148 & 2.74E+06 & 5.03E+12 & 1.36E+11 & 0.1261 \\
149 & 2.68E+06 & 5.49E+12 & 1.40E+11 & 0.1262 \\
150 & 2.63E+06 & 5.98E+12 & 1.44E+11 & 0.1269 \\
151 & 2.57E+06 & 6.50E+12 & 1.48E+11 & 0.1280 \\
152 & 2.52E+06 & 7.06E+12 & 1.53E+11 & 0.1291 \\
153 & 2.46E+06 & 7.67E+12 & 1.57E+11 & 0.1302 \\
154 & 2.41E+06 & 8.31E+12 & 1.62E+11 & 0.1321 \\
155 & 2.35E+06 & 8.98E+12 & 1.67E+11 & 0.1341 \\
156 & 2.30E+06 & 9.70E+12 & 1.72E+11 & 0.1361 \\
157 & 2.25E+06 & 1.05E+13 & 1.78E+11 & 0.1379 \\
158 & 2.20E+06 & 1.13E+13 & 1.84E+11 & 0.1407 \\
159 & 2.14E+06 & 1.22E+13 & 1.89E+11 & 0.1438 \\
160 & 2.09E+06 & 1.31E+13 & 1.95E+11 & 0.1468 \\
161 & 2.04E+06 & 1.41E+13 & 2.01E+11 & 0.1497 \\
162 & 1.99E+06 & 1.51E+13 & 2.07E+11 & 0.1535 \\
163 & 1.94E+06 & 1.62E+13 & 2.13E+11 & 0.1571 \\
164 & 1.89E+06 & 1.74E+13 & 2.20E+11 & 0.1606 \\
165 & 1.85E+06 & 1.87E+13 & 2.26E+11 & 0.1648 \\
166 & 1.80E+06 & 2.00E+13 & 2.32E+11 & 0.1696 \\
167 & 1.75E+06 & 2.15E+13 & 2.39E+11 & 0.1746 \\
168 & 1.70E+06 & 2.30E+13 & 2.45E+11 & 0.1794 \\
169 & 1.66E+06 & 2.47E+13 & 2.52E+11 & 0.1849 \\
170 & 1.61E+06 & 2.65E+13 & 2.58E+11 & 0.1914 \\
171 & 1.57E+06 & 2.84E+13 & 2.64E+11 & 0.1979 \\
172 & 1.52E+06 & 3.04E+13 & 2.70E+11 & 0.2054 \\
173 & 1.48E+06 & 3.26E+13 & 2.76E+11 & 0.2133 \\
174 & 1.43E+06 & 3.50E+13 & 2.82E+11 & 0.2218 \\
175 & 1.39E+06 & 3.76E+13 & 2.86E+11 & 0.2317 \\
176 & 1.35E+06 & 4.04E+13 & 2.91E+11 & 0.2422 \\
177 & 1.30E+06 & 4.34E+13 & 2.94E+11 & 0.2537 \\
178 & 1.26E+06 & 4.69E+13 & 2.97E+11 & 0.2656 \\
179 & 1.22E+06 & 5.09E+13 & 2.98E+11 & 0.2775 \\
180 & 1.18E+06 & 5.57E+13 & 2.98E+11 & 0.2888 \\
181 & 1.14E+06 & 6.19E+13 & 2.95E+11 & 0.2976 \\
182 & 1.10E+06 & 7.01E+13 & 2.90E+11 & 0.3023 \\
183 & 1.06E+06 & 8.19E+13 & 2.81E+11 & 0.3002 \\
184 & 1.02E+06 & 9.93E+13 & 2.68E+11 & 0.2902 \\
185 & 9.77E+05 & 1.23E+14 & 2.51E+11 & 0.2747 \\
186 & 9.39E+05 & 1.52E+14 & 2.30E+11 & 0.2599 \\
187 & 9.00E+05 & 1.81E+14 & 2.09E+11 & 0.2490 \\
188 & 8.62E+05 & 2.08E+14 & 1.88E+11 & 0.2422 \\
189 & 8.24E+05 & 2.32E+14 & 1.70E+11 & 0.2387 \\
190 & 7.87E+05 & 2.52E+14 & 1.56E+11 & 0.2376 \\
191 & 7.50E+05 & 2.70E+14 & 1.45E+11 & 0.2382 \\
192 & 7.14E+05 & 2.86E+14 & 1.38E+11 & 0.2397 \\
193 & 6.78E+05 & 2.99E+14 & 1.34E+11 & 0.2421 \\
194 & 6.42E+05 & 3.12E+14 & 1.32E+11 & 0.2446 \\
195 & 6.06E+05 & 3.22E+14 & 1.32E+11 & 0.2471 \\
196 & 5.72E+05 & 3.32E+14 & 1.33E+11 & 0.2493 \\
197 & 5.37E+05 & 3.42E+14 & 1.34E+11 & 0.2511 \\
198 & 5.03E+05 & 3.51E+14 & 1.35E+11 & 0.2522 \\
199 & 4.69E+05 & 3.59E+14 & 1.34E+11 & 0.2529 \\
200 & 4.35E+05 & 3.67E+14 & 1.32E+11 & 0.2533 \\
201 & 4.02E+05 & 3.74E+14 & 1.30E+11 & 0.2534 \\
202 & 3.69E+05 & 3.81E+14 & 1.29E+11 & 0.2533 \\
203 & 3.36E+05 & 3.87E+14 & 1.29E+11 & 0.2532 \\
204 & 3.04E+05 & 3.92E+14 & 1.28E+11 & 0.2530 \\
205 & 2.72E+05 & 3.97E+14 & 1.28E+11 & 0.2528 \\
206 & 2.41E+05 & 4.01E+14 & 1.28E+11 & 0.2526 \\
207 & 2.10E+05 & 4.04E+14 & 1.27E+11 & 0.2524 \\
208 & 1.79E+05 & 4.07E+14 & 1.27E+11 & 0.2522 \\
209 & 1.48E+05 & 4.10E+14 & 1.26E+11 & 0.2521 \\
210 & 1.18E+05 & 4.12E+14 & 1.26E+11 & 0.2520 \\
211 & 8.79E+04 & 4.13E+14 & 1.26E+11 & 0.2519 \\
212 & 5.83E+04 & 4.14E+14 & 1.26E+11 & 0.2518 \\
213 & 2.90E+04 & 4.15E+14 & 1.26E+11 & 0.2518 \\

\end{longtable}

\begin{table}[H]
\centering
\resizebox{\textwidth}{!}{%
\begin{tabular}{ccccc}
\hline
\textbf{Centers 1-10} & \textbf{Centers 11-20} & \textbf{Centers 21-30} & \textbf{Centers 31-40} & \textbf{Centers 41-50} \\
\hline
1 & 3.2028 & 10.258 & 32.855 & 105.23 \\
1.1234 & 3.5982 & 11.525 & 36.911 & 118.22 \\
1.2621 & 4.0424 & 12.947 & 41.468 & 132.81 \\
1.4179 & 4.5415 & 14.546 & 46.587 & 149.21 \\
1.593 & 5.1021 & 16.341 & 52.338 & 167.63 \\
1.7896 & 5.732 & 18.359 & 58.799 & 188.32 \\
2.0106 & 6.4396 & 20.625 & 66.058 & 211.57 \\
2.2588 & 7.2345 & 23.171 & 74.213 & 237.69 \\
2.5376 & 8.1276 & 26.031 & 83.374 & 267.03 \\
2.8509 & 9.131 & 29.245 & 93.667 & 300 \\
\hline
\end{tabular}%
}
\caption{Energy Grid for a 50-Species Network.}
\label{tab:energy_grid_segments}
\end{table}

\begin{table}[H]
\centering
\resizebox{\textwidth}{!}{%
\begin{tabular}{ccccc}
\hline
\textbf{Widths 1-10} & \textbf{Widths 11-20} & \textbf{Widths 21-30} & \textbf{Widths 31-40} & \textbf{Widths 41-50} \\
\hline
1 & 0.35194 & 1.1272 & 3.6103 & 11.563 \\
0.12345 & 0.39539 & 1.2664 & 4.056 & 12.991 \\
0.13869 & 0.4442 & 1.4227 & 4.5567 & 14.594 \\
0.15581 & 0.49903 & 1.5983 & 5.1192 & 16.396 \\
0.17504 & 0.56064 & 1.7956 & 5.7511 & 18.42 \\
0.19665 & 0.62985 & 2.0173 & 6.4611 & 20.694 \\
0.22093 & 0.70761 & 2.2663 & 7.2587 & 23.249 \\
0.2482 & 0.79496 & 2.5461 & 8.1548 & 26.119 \\
0.27885 & 0.8931 & 2.8604 & 9.1615 & 29.343 \\
0.31327 & 1.0033 & 3.2136 & 10.293 & 32.965 \\
\hline
\end{tabular}%
}
\caption{Energy Widths for a 50-Species Network.}
\label{tab:energy_widths_segments}
\end{table}
    \addToTOC{Vita}
    \chapter*{Vita} \label{ch:vita}
Raghav Chari was born in Charlottesville, Virginia, in September of 2003. He graduated from Panther Creek High School in 2021. During his high school years, he concurrently engaged in research at UNC Greensboro, focusing on ``Be'' stars. In 2021, he enrolled in the University of Tennessee, Knoxville’s Department of Physics \& Astronomy, where he began his pursuit of a Bachelor of Science in Physics. In his third year, he was accepted into the College Scholars program, through which he is also pursuing a Bachelor of Arts in the Philosophy of Physics. Throughout his undergraduate career, Raghav has been actively involved in computational astrophysics research. He has also served as a Teaching Assistant for undergraduate physics courses. Following his graduation, he plans to pursue a Ph.D. in Physics.

\end{document}